\documentclass[aas2pp4]{aastex}

\slugcomment{Preprint to be submitted to ApJ}
 
\ifdim\textheight>25.2cm\textheight=25.0cm\fi

\shorttitle{Red Galaxy Growth and the HOD}
\shortauthors{Brown {\it et al.}}

\begin{document}

\title{Red Galaxy Growth and the Halo Occupation Distribution}

\author{
Michael J. I. Brown\altaffilmark{1},
Zheng Zheng\altaffilmark{2,3,4},
Martin White\altaffilmark{5},
Arjun Dey\altaffilmark{6},
Buell T. Jannuzi\altaffilmark{6},
Andrew J. Benson\altaffilmark{7},
Kate Brand\altaffilmark{8,9},
Mark Brodwin\altaffilmark{6}, and
Darren J. Croton\altaffilmark{5}
}
\altaffiltext{1}{School of Physics, Monash University, Clayton, Victoria 3800, Australia,}
\altaffiltext{2}{Institute for Advanced Study, Einstein Drive, Princeton, NJ 08540}
\altaffiltext{3}{Hubble Fellow}
\altaffiltext{4}{John Bahcall Fellow}
\altaffiltext{5}{Departments of Physics and Astronomy, University of California, Berkeley, CA.}
\altaffiltext{6}{National Optical Astronomy Observatory, Tucson, AZ 85726-6732}
\altaffiltext{7}{California Institute of Technology, 1200 E. California Blvd., Pasadena, CA 91125}
\altaffiltext{8}{Space Telescope Science Institute, Baltimore, MD 21218}
\altaffiltext{9}{Giacconi Fellow}

\email{Michael.Brown@sci.monash.edu.au}

\begin{abstract}
We have traced the past $7~{\rm Gyr}$ of red galaxy stellar mass growth within dark matter halos.
We have determined the halo occupation distribution, which describes how galaxies reside 
within dark matter halos, using the observed luminosity function and clustering of  40,696 $0.2<z<1.0$ 
red galaxies in Bo\"otes. Half of $\simeq 10^{11.9}~h^{-1}~M_\odot$ halos host a red 
central galaxy, and this fraction increases with increasing halo mass.
We do not observe any evolution of the relationship between red galaxy stellar mass and host halo mass,
although we expect both galaxy stellar masses and halo masses to evolve over cosmic time.
We find that the stellar mass contained within the red population has doubled since $z=1$,
with the stellar mass within red satellite galaxies tripling over this redshift range.
In cluster mass halos ($>10^{14}~h^{-1}~M_\odot$) most of the stellar mass resides within satellite
galaxies and the intra-cluster light, with a minority of the stellar mass residing within central galaxies.
The stellar masses of the most luminous red central galaxies are 
proportional to halo mass to the power of $\simeq 0.35$.
We thus conclude that halo mergers do not always lead to rapid growth of central galaxies. 
While very massive halos often double in mass over the past $7~{\rm Gyr}$, 
the stellar masses of their central galaxies typically grow by only $\simeq 30\%$.
\end{abstract}


\section{INTRODUCTION}
\label{sec:intro}

In a universe where the bulk of the mass is in the form of collisionless cold dark matter (CDM), galaxies
reside within gravitationally bound halos of CDM particles \markcite{whi78}({White} \& {Rees} 1978). 
The motion of these particles is effectively governed by gravity alone, so both the mass function and 
clustering of dark matter halos are predictable, albeit non-trivial, functions of redshift and cosmological 
parameters \markcite{pre74,she99,jen01,whi02,tin05}(e.g., {Press} \& {Schechter} 1974; {Sheth} \& {Tormen} 1999; 
{Jenkins} {et~al.} 2001; {White} 2002; {Tinker} {et~al.} 2005). 
Gravitational collapse results in the growth of dark matter halos over cosmic time while dynamical friction leads 
to the orbital decay and merging of substructure within halos.
As a consequence, there is the expectation that the galaxies residing within these halos will also grow via mergers.

In a CDM cosmology, halos more massive than groups ($10^{13}~h^{-1}~M_\odot$) undergo rapid growth via merging 
between $z=1$ and the present day. If the stellar mass within every dark matter halo was always contained
within a single central galaxy,  we would expect the most massive galaxies to undergo rapid growth
via mergers at $z<1$. Until relatively recently, the rapid growth of massive galaxies at $z<1$ was
an almost universal prediction of CDM galaxy formation models \markcite{whi91,del06}(e.g., {White} \& {Frenk} 1991; {De Lucia} {et~al.} 2006). However,
it is difficult to predict the rate at which substructure merges within halos \markcite{tay01,ben03,taf03,boy08}(e.g., {Taylor} \& {Babul} 2001; {Benson} {et~al.} 2003; {Taffoni} {et~al.} 2003; {Boylan-Kolchin}, {Ma}, \&  {Quataert} 2008)
and at least some simulations overestimate the rate of galaxy growth via mergers.

As much of the stellar mass within groups and clusters resides within satellites \markcite{san85,lin04b}(e.g., {Sandage}, {Binggeli}, \& {Tammann} 1985; {Lin} \& {Mohr} 2004) 
and the diffuse intra-cluster light \markcite{kem91,gon00,arn02,fel04,zib05}(ICL; e.g., {Kemp} \& {Meaburn} 1991; {Gonzalez} {et~al.} 2000; {Arnaboldi} {et~al.} 2002; {Feldmeier} {et~al.} 2004; {Zibetti} {et~al.} 2005), not all halo mergers funnel 
stellar mass directly into central galaxies. It is thus plausible that massive galaxies do not grow as rapidly 
as their host halos. The current generation of CDM galaxy formation models predict a range of assembly histories 
for massive galaxies, with rates of $z<1$ stellar mass growth varying by up to a factor 
of 2 \markcite{bau05,bow06,del06,naa07}(e.g., {Baugh} {et~al.} 2005; {Bower} {et~al.} 2006; {De Lucia} {et~al.} 2006; {Naab} {et~al.} 2007). When plausible models produce widely varying predictions, there is 
a clear need for robust observations to test these predictions.

There is compelling observational evidence that galaxies grow via mergers at $z<1$.
There are thousands of examples of merging galaxies in the literature, spanning a broad 
range of galaxy types, mass and redshift \markcite{arp87,tom72,con03,van05,rin07,mci07,lot08}(e.g., {Arp} \& {Madore} 1987; {Toomre} \& {Toomre} 1972; {Conselice} {et~al.} 2003; {van Dokkum} 2005; {Rines}, {Finn}, \& {Vikhlinin} 2007; {McIntosh} {et~al.} 2007; {Lotz} {et~al.} 2008).
While there have been valiant attempts to measure the rate of galaxy growth using catalogs of 
merging galaxies or galaxy pairs, there are large uncertainties associated with the selection function
of such objects and the time-scales for merging. At low redshift, measured rates of 
red galaxy stellar mass growth via mergers span from $\simeq 2\%$ per ${\rm Gyr}$ 
\markcite{mas06,mas07}(e.g., {Masjedi} {et~al.} 2006; {Masjedi}, {Hogg}, \& {Blanton} 2007)
to $\simeq 10\%$ per ${\rm Gyr}$ \markcite{van05}(e.g., {van Dokkum} 2005). Although the exact merger rate remains
a matter of debate, clearly there is some assembly of red galaxies via merging at $z<1$.

The evolving space density of galaxies is a conceptually simple method for measuring the growth
of galaxy stellar masses. At the epoch where the most massive galaxies are assembled, their space
density will increase with decreasing redshift. Although conceptually simple,
in practice such measurements are difficult.  Very luminous galaxies are strongly clustered \markcite{nor02,zeh05a}(e.g., {Norberg} {et~al.} 2002; {Zehavi} {et~al.} 2005a), 
so uncertainties are dominated by the sample variance from large-scale structure rather than Poisson counting statistics. 
The most massive galaxies typically  possess strong 4000~\AA\ breaks and lack strong emission lines,
which complicates both optical photometry and spectroscopy of $z\gtrsim 0.8$ galaxies.
Despite these difficulties, there is increasing evidence that the stellar masses of the most
massive galaxies grow by 50\% or less at $z<1$ \markcite{bun06,cim06,wak06,bro07,sca07}(e.g., {Bundy} {et~al.} 2006; {Cimatti}, {Daddi}, \& {Renzini} 2006; {Wake} {et~al.} 2006; {Brown} {et~al.} 2007; {Scarlata} {et~al.} 2007).
As a consequence, pure passive evolution models without any mergers remain popular  \markcite{cim06,wak06,sca07}(e.g., {Cimatti} {et~al.} 2006; {Wake} {et~al.} 2006; {Scarlata} {et~al.} 2007), 
as they provide a better {\it approximation} of massive galaxy evolution than many simulations using CDM cosmologies.
However, since there is compelling evidence for galaxy mergers, the rate of stellar mass growth must be non-zero.

Measurements of the relationship between galaxy stellar mass and dark matter halo mass provide
important clues as to how galaxies grow over cosmic time. For example, if galaxy stellar mass
was directly proportional to host halo mass we would expect galaxies to grow as rapidly as their
host dark matter halos during mergers. The relationship between galaxy stellar mass and halo mass
has been explored, using various methods for estimating halo mass, including gravitational 
lensing \markcite{hoe04,man06a}(e.g., {Hoekstra}, {Yee}, \& {Gladders} 2004; {Mandelbaum} {et~al.} 2006b), 
satellite galaxy velocities \markcite{con07a}(e.g., {Conroy} {et~al.} 2007b), X-ray 
temperature \markcite{lin04b}(e.g., {Lin} \& {Mohr} 2004), X-ray luminosity \markcite{brou08}(e.g., {Brough} {et~al.} 2008), 
group velocity dispersion \markcite{brou06}(e.g., {Brough} {et~al.} 2006), galaxy 
clustering \markcite{vdb03,zhe07}(e.g., {van den Bosch}, {Yang}, \&  {Mo} 2003; {Zheng}, {Coil}, \& {Zehavi} 2007) 
and matching the galaxy luminosity function with the predicted halo mass function \markcite{val04}(e.g., {Vale} \& {Ostriker} 2004).
For the most massive galaxies, these studies find that galaxy stellar mass scales as halo mass to the power of roughly a third. 

The halo occupation distribution \markcite{pea00,sel00,ber02,coo02}(HOD; e.g., {Peacock} \& {Smith} 2000; {Seljak} 2000; {Berlind} \& {Weinberg} 2002; {Cooray} \& {Sheth} 2002) describes 
the number and distribution of galaxies within dark matter halos. As the clustering and space density of dark matter
halos are predictable functions of redshift, HOD models can be constrained with the observed clustering
and space density of galaxies. HOD models combine much of the relevant astrophysics (e.g., the halo mass function)
with empirical descriptions of how galaxies reside within dark matter halos. While such models cannot be considered
complete descriptions of galaxy evolution, they do provide key insights into how galaxies grow over cosmic time.
For example, if halo mergers efficiently funnel stellar mass into central galaxies, we would
expect relatively little stellar mass to reside within satellite galaxies.

In this paper, we determine the HOD using the measurements of the luminosity function
and clustering of $0.2<z<1.0$ red galaxies selected from the Bo\"otes field of the NOAO Deep Wide-Field 
Survey (NDWFS) and {\it Spitzer} IRAC Shallow Survey. This work builds upon our previous study of the 
HOD of the most massive  ($\geq 1.6L^*$) red galaxies in Bo\"otes \markcite{whi07}({White} {et~al.} 2007, see \S5.3 for a summary). 
Red galaxies are an ideal population for measuring the HOD and testing the predictions of galaxy formation models.
As red galaxies have low star-formation rates, the growth of individual galaxies at $z<1$ should
be dominated by merging (via dynamical friction) rather than star-formation, thus simplifying the 
comparison of CDM models and data. In CDM models the most massive halos undergo rapid growth via 
mergers at $z<1$, and if this results in rapid growth of central galaxies, then this should be 
observable within our sample. The spectral energy distributions of red galaxies are dominated by old
stellar populations, so the optical luminosities of $z<1$ red galaxies are 
tightly correlated with stellar mass. Although our halo masses rely upon CDM models of the space density
and clustering of halos, we can compare our halo masses with those derived with other techniques (e.g., 
X-ray temperature, weak lensing). In several galaxy formation models
virial shock heating  \markcite{bir03,dek06,kho07}(e.g., {Birnboim} \& {Dekel} 2003; {Dekel} \& {Birnboim} 2006; {Khochfar} \& {Ostriker} 2007) 
or feedback from Active Galactic Nuclei \markcite{sil98,wyi03,cro06,hop06}(AGNs; e.g., {Silk} \& {Rees} 1998; {Wyithe} \& {Loeb} 2003; {Croton} {et~al.} 2006; {Hopkins} {et~al.} 2006)
truncates star-formation in halos above a critical mass, and this mass scale should be observable with a well constrained HOD.

The structure of this paper is as follows. In \S\ref{sec:catalogs} we describe the NDWFS and IRAC Shallow Survey, the object
catalogs and our red galaxy sample. We present measurements of the evolving luminosity function and luminosity density of red
galaxies in \S\ref{sec:lf}. In \S\ref{sec:angular} we discuss the correlation function of red galaxies as a function
of both luminosity and redshift, including comparisons with the literature. We discuss the HOD modeling of the clustering
and space density of red galaxies in \S\ref{sec:hod}, including a discussion of the evolving relationship between galaxy luminosity and halo 
mass. In \S\ref{sec:analytic} we introduce an illustrative analytic model of the red galaxy HOD, and discuss how 
stellar mass evolves within the central and satellite red galaxy populations.
We summarize our key results in \S\ref{sec:summary}. Appendix A provides an overview of the analytic method used to 
determine preliminary angular correlation function covariance matrices. Appendix B describes the mock galaxy catalogs, which 
we used to determine the uncertainties of the luminosity function and our final estimates of the correlation function
covariance matrices.

Throughout this paper we use Vega magnitudes and a flat cosmology with $\Omega_m=0.25$, $\Omega_b=0.043$, 
$H_0=72~{\rm km~s^{-1}~Mpc^{-1}}$, $\sigma_8=0.8$ and $n_s=0.97$.
Our choice of cosmology is similar to the WMAP-3 cosmology \markcite{spe07}({Spergel} {et~al.} 2007) and 
matches the $1~h^{-3}~{\rm Gpc^3}$ dark matter simulation we used to generate mock galaxy catalogs.
For the analytic calculation of the HOD (\S\ref{sec:hod}), we define halos as spherical objects where the mean
density within the sphere is 200 times that of the background. For the mock catalogs, halos are identified 
using a Friends-of-Friends algorithm which is discussed in detail in Appendix B.
We denote base ten logarithms with ${\rm log}$ and natural logarithms with ${\rm ln}$.

\section{IMAGING, CATALOGS AND THE RED GALAXY SAMPLE}
\label{sec:catalogs}

We selected our red galaxy sample from imaging of the Bo\"otes field by NDWFS and IRAC Shallow Survey. 
Our sample is extremely similar to that of \markcite{bro07}{Brown} {et~al.} (2007), and we
refer the reader to that paper for a more thorough description of the surveys, source detection, 
photometry, photometric redshifts, rest-frame properties, catalog completeness and sanity checks. 
The only significant difference between this work and \markcite{bro07}{Brown} {et~al.} (2007) is that we adopt a flat cosmology 
with $\Omega_m=0.25$ and $H_0=72 {\rm km~s^{-1}~Mpc^{-1}}$ rather than $\Omega_m=0.30$ and $H_0=70 {\rm km~s^{-1}~Mpc^{-1}}$.
This cosmology matches that of the $1~h^{-3}~{\rm Gpc}^3$ dark matter simulation we use to generate
mock galaxy catalogs  (Appendix~\ref{sec:mock}) and is similar to the WMAP-3 cosmology \markcite{spe07}({Spergel} {et~al.} 2007).

\subsection{IMAGING}
\label{sec:surveys}

In this paper we utilize optical and infrared imaging of Bo\"otes from the 
NDWFS \markcite{jan99}({Jannuzi} \& {Dey} 1999) and {\it Spitzer} IRAC Shallow Survey \markcite{eis04}({Eisenhardt} {et~al.} 2004).
The NDWFS is an optical  ($B_WRI$) and near-infrared ($K$) imaging survey of two $\approx 9.3~{\rm deg}^2$
fields with telescopes of the National Optical Astronomy Observatory.
We utilize the third NDWFS data release\footnote{Available from the NOAO Science 
Archive at http://www.archive.noao.edu/ndwfs/} of optical imaging with the MOSAIC-I camera on the Kitt
Peak 4-m telescope. To obtain accurate optical colors with fixed aperture photometry across the Bo\"otes field, we have
smoothed copies of the released images to a common Point Spread Function (PSF) with a full width at half maximum of $1.35^{\prime\prime}$.
The IRAC Shallow Survey is $8.5~{\rm deg^2}$ of contiguous imaging with the Infrared Array Camera \markcite{faz04}(IRAC; {Fazio} {et~al.} 2004) of 
{\it Spitzer Space Telescope}. The Bo\"otes field was imaged at wavelengths of  $3.6$, $4.5$, $5.8$, and $8.0~\mu {\rm m}$
with a typical exposure time of $90~{\rm sec}$. We utilize the $3.6$ and $4.5~\mu {\rm m}$ imaging
to remove contaminants (e.g., stars) from our galaxy sample and (when combined with the NDWFS photometry) for photometric redshifts . 

\subsection{SOURCE DETECTION AND PHOTOMETRY}

We detected sources using SExtractor $2.3.2$ \markcite{ber96}({Bertin} \& {Arnouts} 1996), run on the 
$I$-band images of the NDWFS third data release.
For this paper, we only include objects which are detected within nominal subfield boundaries.
The subfields have small overlaps and we remove duplicate object detections from these regions.
To minimize contamination of the catalog, we exclude regions surrounding
very extended galaxies and saturated stars. Regions without good coverage from both the 
NDWFS and the IRAC Shallow Survey are also excluded. The final sample area
is $6.96~{\rm deg}^2$ over a $2.9^\circ \times 3.4^\circ$ field-of-view.

We measured aperture photometry for each object using our own code.
SExtractor segmentation maps were used to exclude flux associated with 
neighboring objects. We corrected the photometry for missing
pixels (e.g., bad pixels) using the mean flux per pixel measured in a series
of annuli surrounding each object. Uncertainties were determined by 
measuring photometry at $\simeq 100$ positions within $2^\prime$ of the object
position and finding the uncertainty which encompassed 68\% of the
measurements. The accuracy of the photometry was verified
by adding artificial galaxies to our data, recovering them with SExtractor
and measuring their photometry with our code \markcite{bro07}({Brown} {et~al.} 2007).

\subsection{PHOTOMETRIC REDSHIFTS}

We determined redshifts for our galaxies using the empirical ANNz 
photometric redshift code \markcite{fir03,col04}({Firth}, {Lahav}, \& {Somerville} 2003; {Collister} \& {Lahav} 2004).
ANNz uses artificial neural networks to determine the relationship between measured galaxy properties and redshift.
It does not use any prior assumptions about the shape of galaxy spectral energy distributions (SEDs), though it
does assume the relationship between observed galaxy properties and redshift is a relatively smooth function. 

The basis of our training set is galaxies with spectroscopic redshifts in Bo\"otes.
The AGN and Galaxy Evolution Survey (AGES, C.~S.~Kochanek~et~al.~in preparation) has 
obtained spectroscopic redshifts of $\simeq 16000$ $I\lesssim 20$ galaxies, while several
hundred additional redshifts are available from a variety of programs with 4 and 8~m-class telescopes.
At $z>0.6$ and $I>20.5$ there are relatively few galaxies with spectroscopic redshifts, so we added artificial 
galaxies to the training set by (at fixed redshift) interpolating in color and extrapolating in luminosity.
We trained and measured photometric redshifts using the $4^{\prime\prime}$ aperture photometry and the 2nd order moments of
the $I$-band light distribution.  Using 4319 red galaxies with spectroscopic redshifts, we find the 
$1\sigma$ uncertainties of our photometric redshifts are $\simeq 0.1$ in redshift at $I=22$, decreasing to $\simeq 0.03$ at $I=19.5$.

\subsection{REST-FRAME PROPERTIES}
\label{sec:rest}

To measure absolute magnitudes and rest-frame colors, 
we used maximum likelihood fits of \markcite{bru03}{Bruzual} \& {Charlot} (2003) SED models 
to the optical photometry.  Throughout this paper we use solar
metallicity models with a \markcite{sal55}{Salpeter} (1955) initial mass function, a 
formation redshift of $z=4$, and a broad range of exponentially declining star formation rates.  
As with all SED models and templates, these models do have errors but these have 
little impact upon our principal conclusions \markcite{bro07}({Brown} {et~al.} 2007, \S6.5).

The $4^{\prime\prime}$ aperture photometry captures 86\% or less of the total flux. 
We corrected for the flux outside this aperture by assuming galaxies within our
sample have \markcite{dev48}{de Vaucouleurs} (1948) profiles truncated at 7 half-light radii.
We adopt the same size-luminosity relation as \markcite{bro07}{Brown} {et~al.} (2007) and assume this relation undergoes
luminosity evolution described by a \markcite{bru03}{Bruzual} \& {Charlot} (2003) $\tau=0.65~{\rm Gyr}$ stellar synthesis model with 
a formation redshift of $z=4$. While the size-luminosity relation of red galaxies has some scatter,
this should have little impact upon our results as $4^{\prime\prime}$ aperture 
photometry captures most of the flux from $z>0.4$ red galaxies \markcite{bro07}(see Figure~5 of {Brown} {et~al.} 2007).

\subsection{THE RED GALAXY SAMPLE}
\label{sec:sample}

The distribution of galaxy colors is bimodal \markcite{hog04}(e.g., {Hogg} {et~al.} 2004), and selection criteria
for red galaxies typically fall near the minimum between the red and blue galaxy populations \markcite{mad02,bel04,wil06}(e.g., {Madgwick} {et~al.} 2002; {Bell} {et~al.} 2004; {Willmer} {et~al.} 2006).
We apply a similar approach here, and use the following rest-frame color selection criterion:
\begin{eqnarray}
U-V & > & 1.40-0.25-0.08\times(M_V-5~{\rm log}~h+20.0) \nonumber \\
 & & - 0.42\times(z-0.05)+0.07\times(z-0.05)^2.
\end{eqnarray}
Our criterion selects galaxies with rest-frame $U-V$ colors within
$0.25$ magnitudes of the evolving color-magnitude relation of red
galaxies.  This criterion allows comparison with the recent literature
and is similar to the criterion of \markcite{bel04}{Bell} {et~al.} (2004).
We also applied apparent color cuts to remove contaminants while not significantly
reducing sample completeness, and these are discussed in detail in \markcite{bro07}{Brown} {et~al.} (2007).

The completeness of the catalogs was verified by adding artificial 
galaxies with \markcite{dev48}{de Vaucouleurs} (1948) profiles  to copies of our data and recovering them with SExtractor. 
Our catalogs are more than 85\% complete for $I<23.5$ galaxies with
half-light radii of $0.5^{\prime\prime}$ or less.
We limit the absolute magnitude range in each of our redshift bins so
we can determine accurate redshifts and have a sample completeness of 85\% or more.
Our final sample contains 40,696 red galaxies with photometric redshifts between $z=0.2$ and $z=1.0$.

\section{THE RED GALAXY LUMINOSITY FUNCTION}
\label{sec:lf}


Our $0.2<z<1.0$ red galaxy sample is very similar to that of \markcite{bro07}{Brown} {et~al.} (2007), except  
we adopt a slightly different flat cosmology with $\Omega_m=0.25$ and $H_0=72~{\rm km~s^{-1}~Mpc^{-1}}$. 
This increases the $z=1$ luminosity distance by 3\% and the $z<1$ comoving volume by 10\%.
We select our sample using absolute magnitude criteria, so our sample size is 4\%
larger than that of \markcite{bro07}{Brown} {et~al.} (2007). While these changes are small, they are comparable to our
random uncertainties, so we present revised measurements of the red galaxy luminosity function
and luminosity density here.

Our methodology for measuring the luminosity function is very similar to that of \markcite{bro07}{Brown} {et~al.} (2007).
We measure the evolving luminosity function of red galaxies using both the binned 
$1/V_{\rm max}$ estimator \markcite{sch68}({Schmidt} 1968) and maximum-likelihood fits of \markcite{sch76}{Schechter} (1976) luminosity functions.
Our photometric redshift errors are relatively small and their impact upon the measured luminosity functions
is estimated to be negligible \markcite{bro07}({Brown} {et~al.} 2007).
We split the sample into four photometric redshift slices: $0.2<z<0.4$, $0.4<z<0.6$, $0.6<z<0.8$ and $0.8<z<1.0$, which have comoving volumes of 
$8.1\times 10^{-4}$, $1.8\times 10^{-3}$, $2.9\times 10^{-3}$ and $3.9\times 10^{-3}$ $h^{-3}~{\rm Gpc}^3$ respectively. 
The total comoving volume of $0.01~h^{-3} {\rm Gpc}^3$ is comparable to that of the $z<0.15$ 2dF Galaxy Redshift Survey (2dFGRS).

Unlike \markcite{bro07}{Brown} {et~al.} (2007), we estimate the uncertainties for the luminosity function using multiple mock galaxy catalogs, 
whose construction is described in Appendix~\ref{sec:mock}. We expect the uncertainties derived from 
mock catalogs to be more robust than those determined with subsamples of the data, as  
individual large-scale structures can span multiple subsamples. Fractional uncertainties derived from mocks are on the 
order of 10\%, and can be 50\% larger than those determined with subsamples.

To quantify the evolution of stellar mass within the red population, we measure the luminosity density 
(the luminosity weighted integral of the luminosity function).
To determine the $B$-band luminosity density ($j_B$) we sum over the galaxy 
catalog ($\sum_{i=1}^n L_i$) at each redshift bin, apply corrections for sample incompleteness, and
divide by the volume of the relevant redshift slice. Galaxies fainter than our magnitude limits contribute up to $\sim 15\%$ of the luminosity density,
and we include their contribution using an analytic approximation of the HOD (\S\ref{sec:analytic}).  
Our measurements of the luminosity function and luminosity density are summarized in Tables~\ref{table:vmax} and ~\ref{table:mlf}.

The overall evolution of the luminosity function is very similar to that reported by \markcite{bro07}{Brown} {et~al.} (2007) and our 
principal conclusions remain unchanged. The luminosity density decreases by only $27 \pm 20\%$ between $z=1$ and $z=0$, 
while the $B$-band luminosity of an aging stellar population fades by a factor of 3 over the same redshift range.
We thus conclude that the stellar mass within the red galaxy population has increased by $118\pm 45\%$ since $z=1$.
Mergers of red galaxies (without accompanying star-formation) do not increase the stellar mass contained within the red
population. As there is little star-formation within the red population, the increase in stellar mass must result from
stellar mass being transferred from the blue galaxy population to the red galaxy 
population \markcite{bel04,bro07,fab07}(e.g., {Bell} {et~al.} 2004; {Brown} {et~al.} 2007; {Faber} {et~al.} 2007). 

The transfer of stellar mass from the blue galaxy population into the red galaxy population will have little impact on the space
density of the most luminous red galaxies unless significant merging takes place, as very massive blue galaxies are rare
at $z<1$ \markcite{bel04}(e.g., {Bell} {et~al.} 2004). The evolving space density of luminous red galaxies thus provides
strong constraints on the rate of galaxy stellar mass growth via galaxy mergers.
The bright end ($M_B-5{\rm log}h\lesssim -21$) of the red galaxy luminosity function fades by $0.9$ $B$-band magnitudes between $z=1$ and $z=0$. 
We would not expect to see any fading if massive galaxies were being rapidly assembled via mergers at $z<1$. 
However, the bright end of the luminosity function does not fade by the $\simeq 1.2$ $B$-band magnitudes 
predicted by passive stellar population synthesis models \markcite{bru03}(e.g., {Bruzual} \& {Charlot} 2003) and the luminosity evolution of the 
fundamental plane  \markcite{van03,tre05}(e.g., {van Dokkum} \& {Stanford} 2003; {Treu} {et~al.} 2005). We thus conclude that there is ongoing 
assembly of massive galaxies at $z<1$, albeit at a rate that only increases their stellar masses by $\simeq 30\%$ between $z=1$ and $z=0$.

\section{MEASURING THE RED GALAXY CORRELATION FUNCTION}
\label{sec:angular}

The spatial correlation function of galaxies is one of the principal constraints on the HOD. 
In particular, the clustering of galaxies on scales less than $1~h^{-1}~{\rm Mpc}$ is a strong function of the number
of satellite galaxies residing within halos. We measured the clustering of red galaxies with the angular correlation function,
and then used the \markcite{lim54}{Limber} (1954) equation to compare models of the spatial clustering 
with our observations. Angular correlation functions were determined using subsamples of red galaxies
selected as a function of photometric redshift, luminosity and 
space density. A summary of the subsample properties, including galaxy
counts and photometric redshift ranges, is provided in Table~\ref{table:r0}.

\subsection{THE ANGULAR CORRELATION FUNCTION}

We measured the angular correlation function with the \markcite{lan93}{Landy} \& {Szalay} (1993) estimator:
\begin{equation}
 \hat{\omega}(\theta)=\frac{DD-2DR+RR}{RR}
\end{equation}
where $DD$, $DR$, and $RR$ are the number of galaxy-galaxy, galaxy-random
and random-random pairs at angular separation $\theta\pm\delta\theta/2$.
The pair counts were determined in bins spaced by $0.2~{\rm dex}$ between
$10^{\prime\prime}$ and $0.44^\circ$. 

For a field of finite size, estimators of the correlation function are subject to the 
integral constraint where
\begin{equation}
\int \int \hat{\omega}(\theta_{12}) d\Omega_1 d\Omega_2 \simeq 0
\end{equation}
\markcite{gro77}({Groth} \& {Peebles} 1977), where $\theta_{12}$ is the angle separating solid angle
elements $d\Omega_1$ and $d\Omega_2$. The integral constraint results in 
a systematic underestimate of the true clustering.
If the number density fluctuations in the volume are small, and the angular
correlations are smaller than the variance within the volume, then to first
order the correlation function is simply biased low by a constant equal to
the fractional variance of the number counts. Often, to remove this bias, the term
\begin{equation}
  \omega_\Omega=\frac{1}{\Omega^2}\int\int \omega(\theta_{12})d\Omega_1 d\Omega_2
\end{equation}
is added to $\hat{\omega}(\theta)$ where $\Omega$ is the survey area.
The value of $\bar{n}^2\omega_\Omega$, where $\bar{n}$ is an estimate of the mean
number of galaxies per unit area, is the contribution of clustering to
the variance of the galaxy number counts \markcite{gro77,efs91}({Groth} \& {Peebles} 1977; {Efstathiou} {et~al.} 1991). 

Our final estimates of the integral constraint use the variance of the galaxy number
counts measured from multiple mock galaxy catalogs (Appendix \ref{sec:mock}). 
Before constructing the mock catalogs, preliminary estimates of the integral constraint were 
derived from power-law fits to the angular correlation function. The integral constraint
depends on the value of $\omega(\theta)$ on large angular scales. On such scales, the angular correlation
function of mock catalog galaxies is constrained by the clustering of dark matter halos while 
power-law extrapolations of the correlation function are strongly dependent on the measured value of
the power-law index. Both mock catalogs and power-law fits (where the power-law index of the spatial correlation function is $\simeq -2$) 
provide comparable estimates of the integral constraint, with $\omega_\Omega \simeq 0.02 w(1^\prime)$.
The significance of the integral constraint reaches a maximum of $\simeq 0.5 \sigma$ for the largest
angular scales of our highest redshift bins. On the angular scales where most of our constraining power comes
from, the integral constraint is negligible.

\subsection{THE COVARIANCE MATRIX}

To fit a functional form to the measured angular correlation function,
we need a model covariance matrix to estimate uncertainties of our
data points and the correlations between them. Commonly used methods for 
estimating the covariance matrix include mock catalogs, subsampling or 
resampling of the data, and analytic approximations. 

For our preliminary measurements, we estimated covariance matrices using 
the analytic approximations described in Appendix \ref{sec:anacov}.
Analytic covariance matrices can be computed quickly and do not 
include random noise, but can have large systematic errors.
The final results presented in this paper use covariance matrices 
determined with mock catalogs, whose construction we discuss in 
Appendix~\ref{sec:mock}.  The mock covariance matrices naturally 
include the effects of sample variance and shot-noise without making 
assumptions about the linearity of the underlying clustering or special relations of higher
order functions. The mock catalogs are generated using a $1~h^{-3}~{\rm Gpc}^3$ dark 
matter simulation and an analytic approximation of how red galaxies reside within 
dark matter halos (\S\ref{sec:analytic}). The analytic approximation of the HOD was 
initially constrained with angular correlation functions determined with the analytic covariance 
matrices, and was then verified using mock catalog covariance matrices.

\subsection{THE GALAXY REDSHIFT DISTRIBUTION}
\label{sec:modeldndz}

\begin{figure*}
\plotone{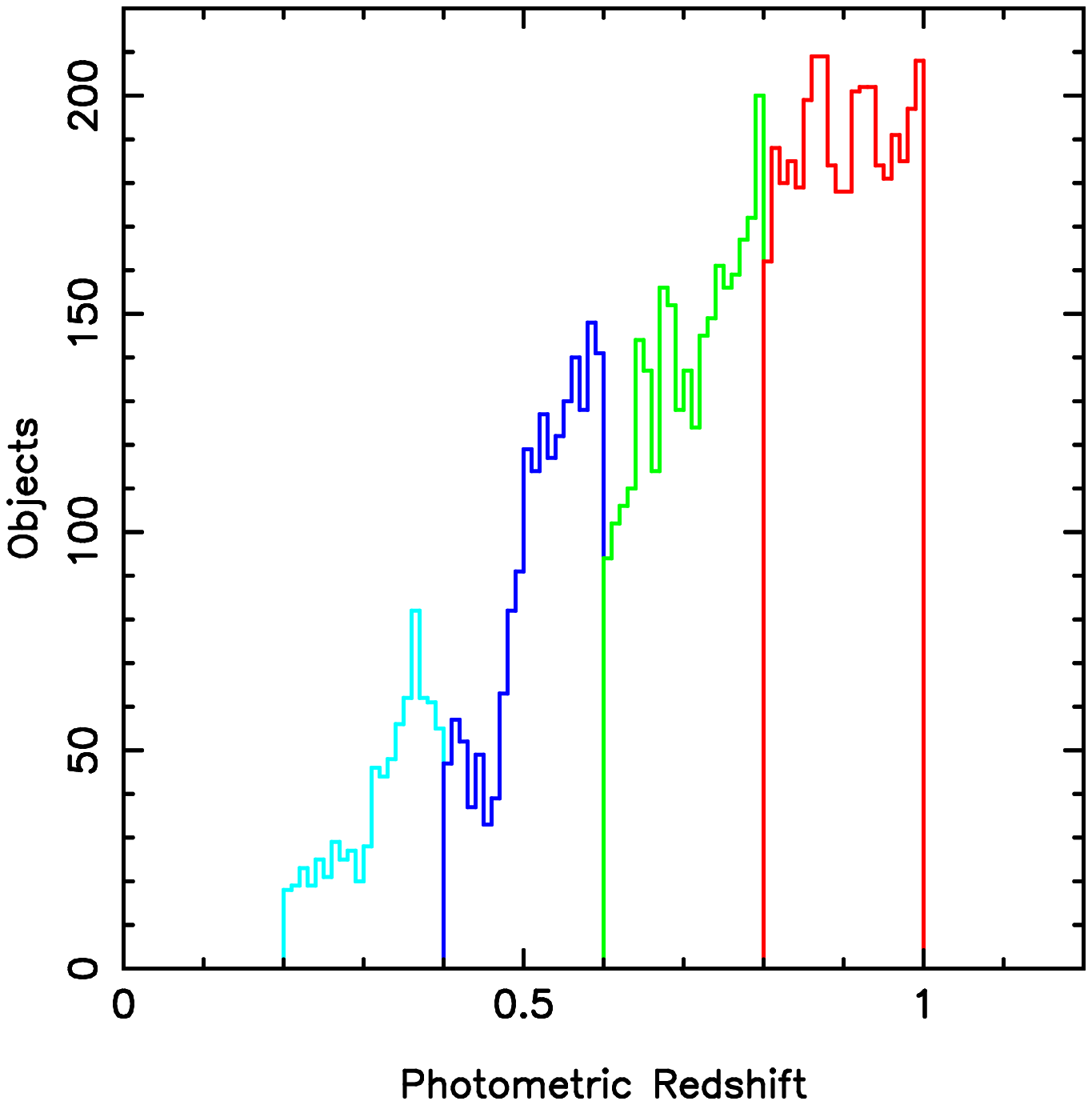}\plotone{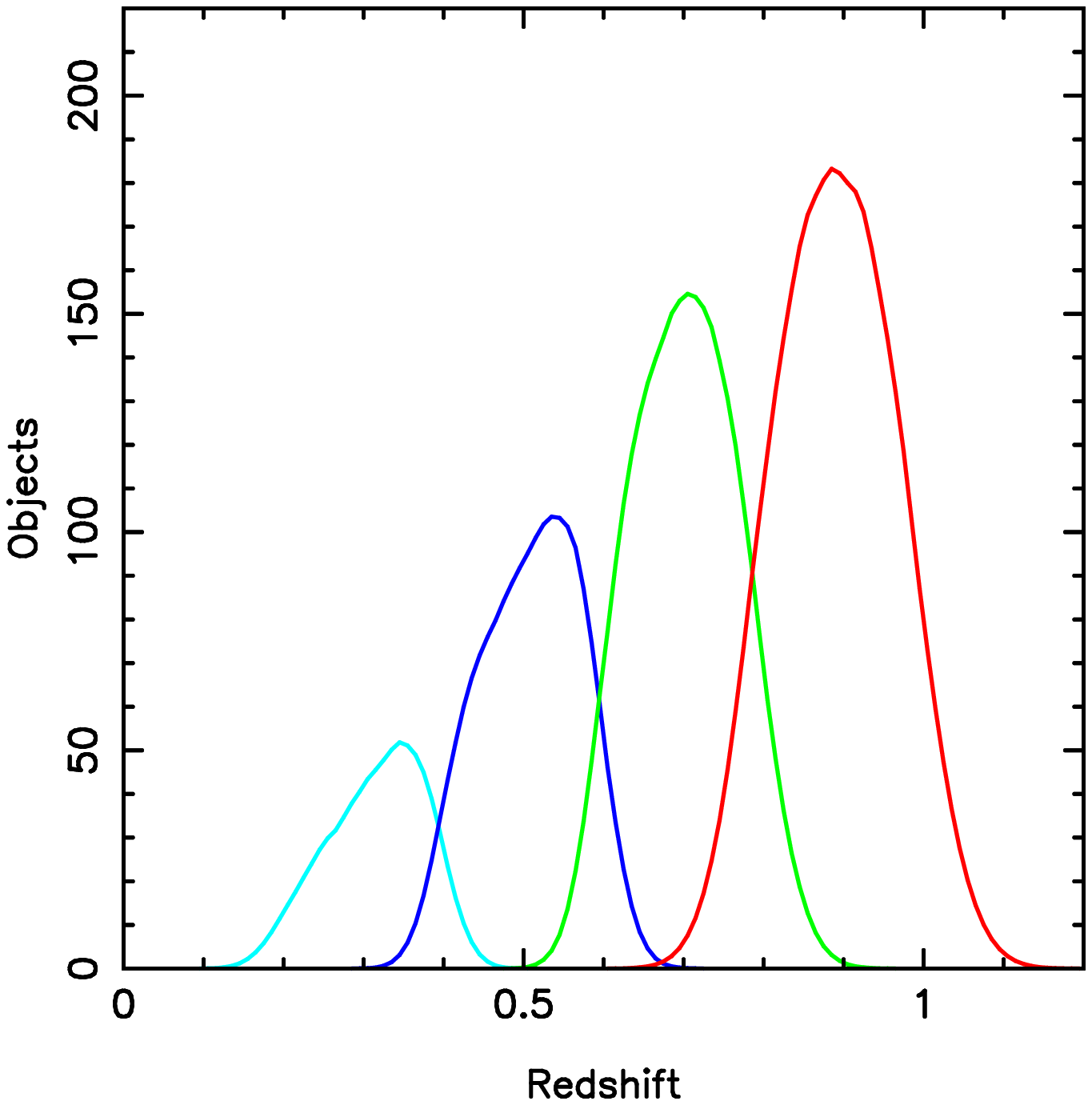}
\caption{The photometric (left) and model (right) redshift distributions for the $n$ brightest
red galaxies selected down to a space density threshold of $10^{-3} ~h^{3}~ {\rm Mpc}^{-3}$.
To model the galaxy redshift distribution, we used the luminosity functions from Table~\ref{table:mlf}
and convolved them with the measured photometric redshift uncertainties.
The mean redshift of the model galaxy redshift distribution is strongly constrained by the upper and lower limits
of the photometric redshift bins and has a weak dependence on Schechter function parameters.
If we did not account for photometric redshift errors when modeling the redshift distribution, we would overestimate the number
of galaxy pairs with small physical separations and underestimate the spatial clustering of galaxies by $\simeq 20\%$.
\label{fig:dndz}}
\end{figure*}

The angular correlation function, $\omega(\theta)$, can be derived from the spatial correlation function
$\xi(r,z)$, using the \markcite{lim54}{Limber} (1954) equation:
\begin{eqnarray}
w(\theta) & = & \int^\infty_0  \frac{dN}{dz}
\left[ \int^\infty_0 \xi (r(\theta,z,z^\prime),z)
\frac{dN}{dz^\prime} dz^\prime \right] dz \nonumber \\
& & \left/
\left( \int^\infty_0 \frac{dN}{dz} dz \right)^2 \right.
\label{eq:lim}
\end{eqnarray}
where $dN/dz$ is the spectroscopic redshift distribution, and $r(\theta,z,z^\prime)$ is the
comoving distance between two objects at redshifts $z$ and $z^\prime$
separated by angle $\theta$ on the sky. If the functional form of the
spatial correlation function is known, the spatial correlation
function can be determined with accurate measurements of the angular
clustering and a robust model of the galaxy redshift distribution.

To model the redshift distributions of our subsamples, we use the
Schechter luminosity functions listed in Table~\ref{table:mlf}.
Photometric redshift uncertainties result in objects being scattered
into and out of our subsamples. To model this, we use the measured
photometric redshift uncertainties from Table~4 of \markcite{bro07}{Brown} {et~al.} (2007), 
which are a function of both galaxy magnitude and redshift, and 
assume Gaussian random errors. We have not attempted to model gross photometric redshift
errors as they are rare in our sample (see Figure~3 of  \markcite{bro07}{Brown} {et~al.} 2007).
If we did not include these uncertainties,
we would underestimate the mean physical separation of galaxy pairs and 
underestimate $\xi(r)$ by approximately 20\%.
The photometric redshift and model spectroscopic 
redshift distributions for four of our red galaxy
samples are shown in Figure~\ref{fig:dndz}. As the spectroscopic
redshift distribution of neighboring photometric redshift bins
overlap, neighboring bins are not completely independent of each other.  

Our model redshift distributions are a function of the measured photometric redshift uncertainties.
As a consequence, if our photometric redshift uncertainties are in error our measurement of $\xi(r)$ 
will also be in error. To verify this is not the case, we checked that our estimate of $\xi(r)$
did not depend on the width of our redshift bins \markcite{bro03}({Brown} {et~al.} 2003). When a redshift bin
is very narrow, the width of the model redshift distribution is dominated by the photometric
redshift uncertainties. Conversely, the model redshift distribution for a very 
wide redshift bin has relatively little dependence on the photometric redshift uncertainties.
We found that our clustering estimates did not vary when we modified the width of our redshift 
bins, consistent with there being no gross errors in our model redshift distributions.

\subsection{POWER-LAW MODELS OF THE CORRELATION FUNCTION}

Power-laws provide a simple empirical approximation of the spatial correlation function, 
enabling quick comparison with the literature. Also, unlike the HOD, power-law models
of the correlation function do not depend on models of the space density and clustering
of dark matter halos. However, spatial correlation functions measured with large redshift surveys
do show significant departures from power-laws  \markcite{zeh04,zhe07}(e.g., {Zehavi} {et~al.} 2004; {Zheng} {et~al.} 2007). Also, power-laws are
a purely empirical description of the galaxy clustering and are not physically motivated.
Power-law approximations of the correlation function are thus useful but should be treated with due caution.

We approximate the spatial correlation function with the standard power-law parameterization:
\begin{equation}
\xi(r) = \left(r/r_0\right)^{-\gamma}
\end{equation}
where $r$ is the galaxy pair separation in comoving coordinates and $r_0$ is the spatial scale
where $\xi(r)=1$.
As a consequence, the angular correlation function is also parameterized by a power-law:
\begin{equation}
w(\theta) = w(1^\prime) \left(\frac{\theta}{1^\prime}\right)^{1-\gamma}.
\end{equation}
We use $\omega(1^\prime)$ rather than $\omega(1^\circ)$ to parameterize the strength of the angular clustering, 
as the measured value $\omega(1^\circ)$ strongly depends on the measured value $\gamma$.

\begin{figure*}
\plotone{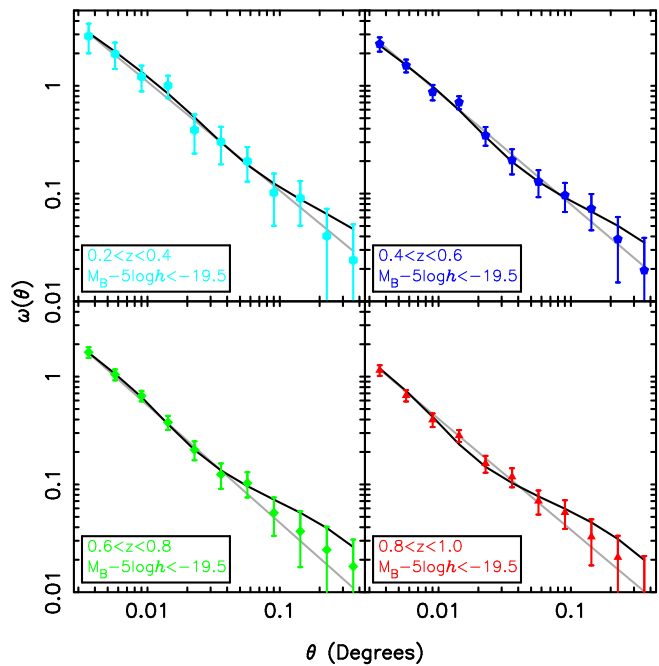}
\caption{The angular clustering of $M_B-5{\rm log}h<-19.5$ red galaxies in~Bo\"otes as a function
of redshift. Power-law fits to the data are shown with the grey line while the best-fit HOD models are shown
with the black curves. The datapoints at large angular scales are highly correlated with each other, and contribute
little to the power-law or HOD fits to the data. If we exclude the last three data-points when fitting HOD models
to the data, the resulting HOD parameters change only marginally.
\label{fig:ang1}}
\end{figure*}

\begin{figure*}
\plotone{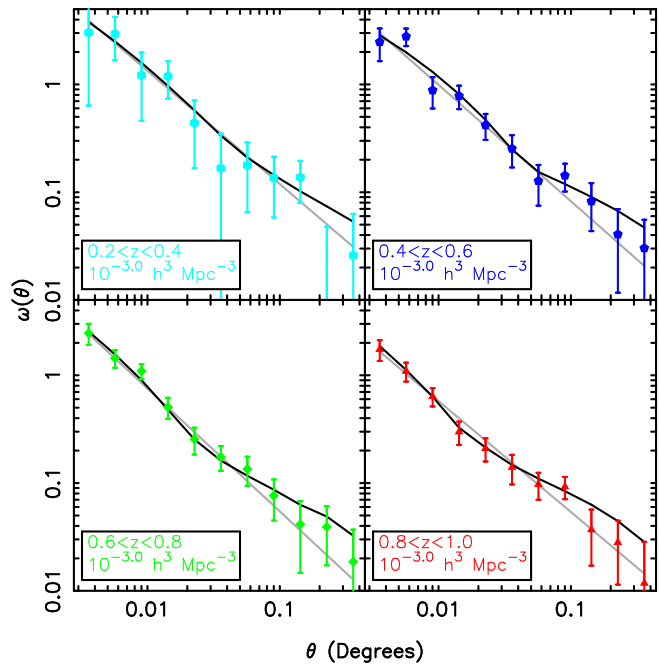}
\caption{The angular clustering of red galaxies as a function of redshift, determined using samples of
$n$ brightest galaxies per unit volume down to a space density threshold of $10^{-3}~h^{-3}~{\rm Mpc}$.
Power-law fits to the data are shown with the grey line while the best-fit HOD models are shown
with the black curves.
\label{fig:ang2}}
\end{figure*}

We plot examples of power-law fits to the angular correlation function in Figures~\ref{fig:ang1} and ~\ref{fig:ang2}, along with 
HOD model fits (see \S\ref{sec:hod}) for comparison. While HOD models provide better fits to spatial correlation functions measured
with large redshift surveys \markcite{zeh04,zhe07}(e.g., {Zehavi} {et~al.} 2004; {Zheng} {et~al.} 2007), we find comparable $\chi^2$ values for HOD model
and power-law fits to our data. This is not altogether surprising, as our sample is much smaller than comparable 
SDSS samples and (for a given sample) angular correlation functions have lower signal-to-noise than spatial correlation functions.

\begin{figure*}
\plotone{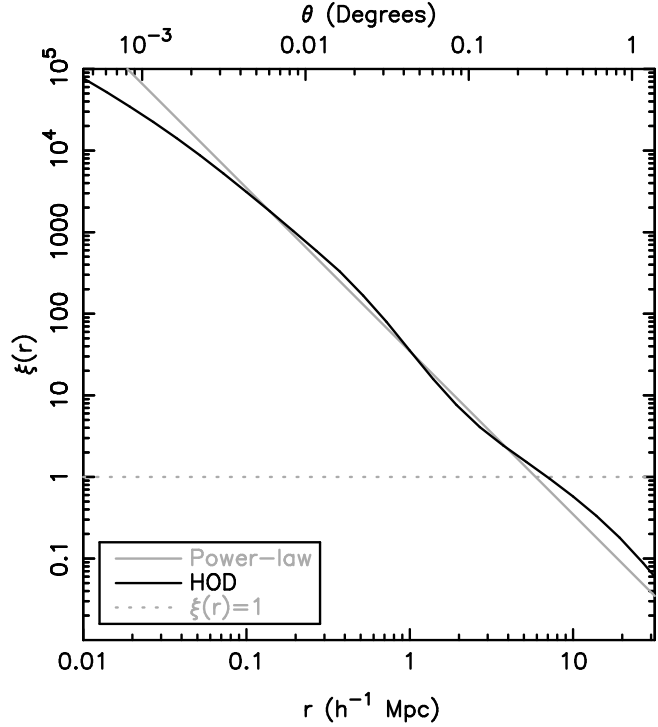}
\caption{Power-law and HOD models of the spatial correlation function, inferred from fits
to the angular clustering of $M_B-5{\rm log}h < -19.5$ red galaxies at $0.4<z<0.6$.
The angular scale corresponding to a transverse comoving distance of $r$ is shown at the top of the plot.
On scales of $\sim 1~h^{-1} {\rm Mpc}$, power-laws approximate HOD models of the correlation function.
On larger scales, power-law models derived from our data should be considered extrapolations.
Estimates of the spatial scale where $\xi(r)=1$ derived from HOD and power-law models can differ by $1~h^{-1}~{\rm Mpc}$.
\label{fig:xic}}
\end{figure*}

We determined the value of $r_0$ for each subsample using the measured values of  $\omega(1^\prime)$ and $\gamma$, models of the 
galaxy redshift distribution and the \markcite{lim54}{Limber} (1954) equation.  
We provide a complete list of our power-law correlation function parameters in Table~\ref{table:r0}.
We best constrain the spatial correlation function on scales of $\sim 1~h^{-1}~{\rm Mpc}$,
and on scales larger than $\sim 5~h^{-1}~{\rm Mpc}$ our power-law models should be considered extrapolations.
As we show in Figure~\ref{fig:xic}, the scale where power-law and HOD models of $\xi(r)$ equal unity often
differ by $\sim 1~h^{-1}~{\rm Mpc}$, so one should treat $r_0$ values from this work and the literature with caution.

\begin{figure*}
\plotone{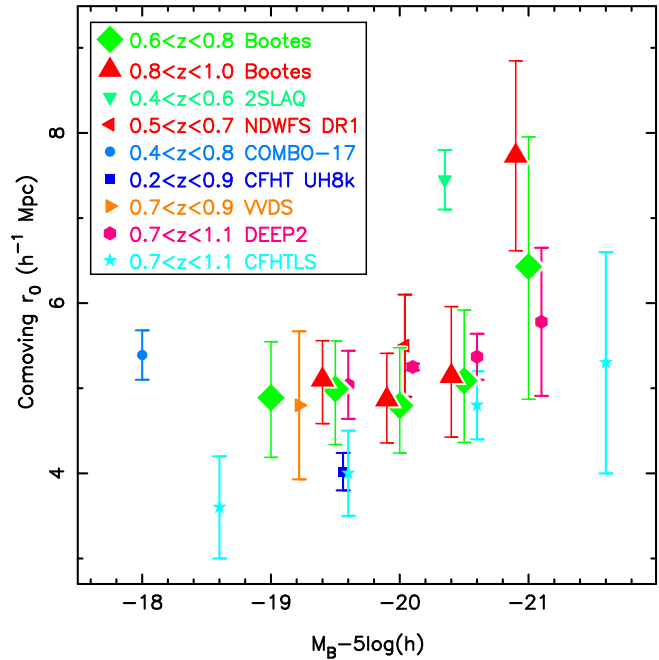}
\caption{The spatial clustering of red galaxies at $z\sim 0.8$, including our results and those from the
literature \markcite{bro03,wil03,men06,phl06,mcc07,ros07,coi08}({Brown} {et~al.} 2003; {Wilson} 2003; {Meneux} {et~al.} 2006; {Phleps} {et~al.} 2006; {McCracken} {et~al.} 2007; {Ross} {et~al.} 2007; {Coil} {et~al.} 2008). All the samples plotted here use
red galaxies brighter than an absolute magnitude threshold, with the exception of NDWFS DR1 and CFHTLS,
which use absolute magnitude bins which are $\pm 0.5~{\rm mag}$ wide. Spatial clustering measurements
from the CFHT UH8k survey and CFHT Legacy Survey are systematically low, as their model redshift distributions
do not account for photometric redshift errors. The clustering of red galaxies fainter than
$M_B-5{\rm log}h=-20$ is roughly constant with luminosity and redshift, with $r_0\simeq 5~h^{-1} {\rm Mpc}$.
\label{fig:r0hz}}
\end{figure*}

In Figure~\ref{fig:r0hz} we plot the spatial clustering of $z\sim 0.8$ red galaxies in Bo\"otes (parameterized by $r_0$) along 
with results from the recent literature at comparable redshifts \markcite{bro03,wil03,men06,phl06,mcc07,ros07,coi08}({Brown} {et~al.} 2003; {Wilson} 2003; {Meneux} {et~al.} 2006; {Phleps} {et~al.} 2006; {McCracken} {et~al.} 2007; {Ross} {et~al.} 2007; {Coil} {et~al.} 2008).
Measurements from the CFHT UH8k survey \markcite{wil03}({Wilson} 2003) and the CFHT Legacy Survey \markcite{mcc07}({McCracken} {et~al.} 2007) are systematically low,
as their model redshift distributions do not account for photometric redshift errors.
The clustering of red galaxies fainter than $M_B-5{\rm log}h=-20$ is roughly constant as a function 
of both luminosity and redshift, with $r_0\simeq 5~h^{-1} {\rm Mpc}$. There is some evidence that the 
spatial clustering of luminous red galaxies is correlated with luminosity at high redshift, as one would
expect if the most massive galaxies reside within the most massive (and strongly clustered) dark matter halos.
As halo masses are a product of HOD modeling, we will return to this point later in the paper.

\begin{figure*}
\plotone{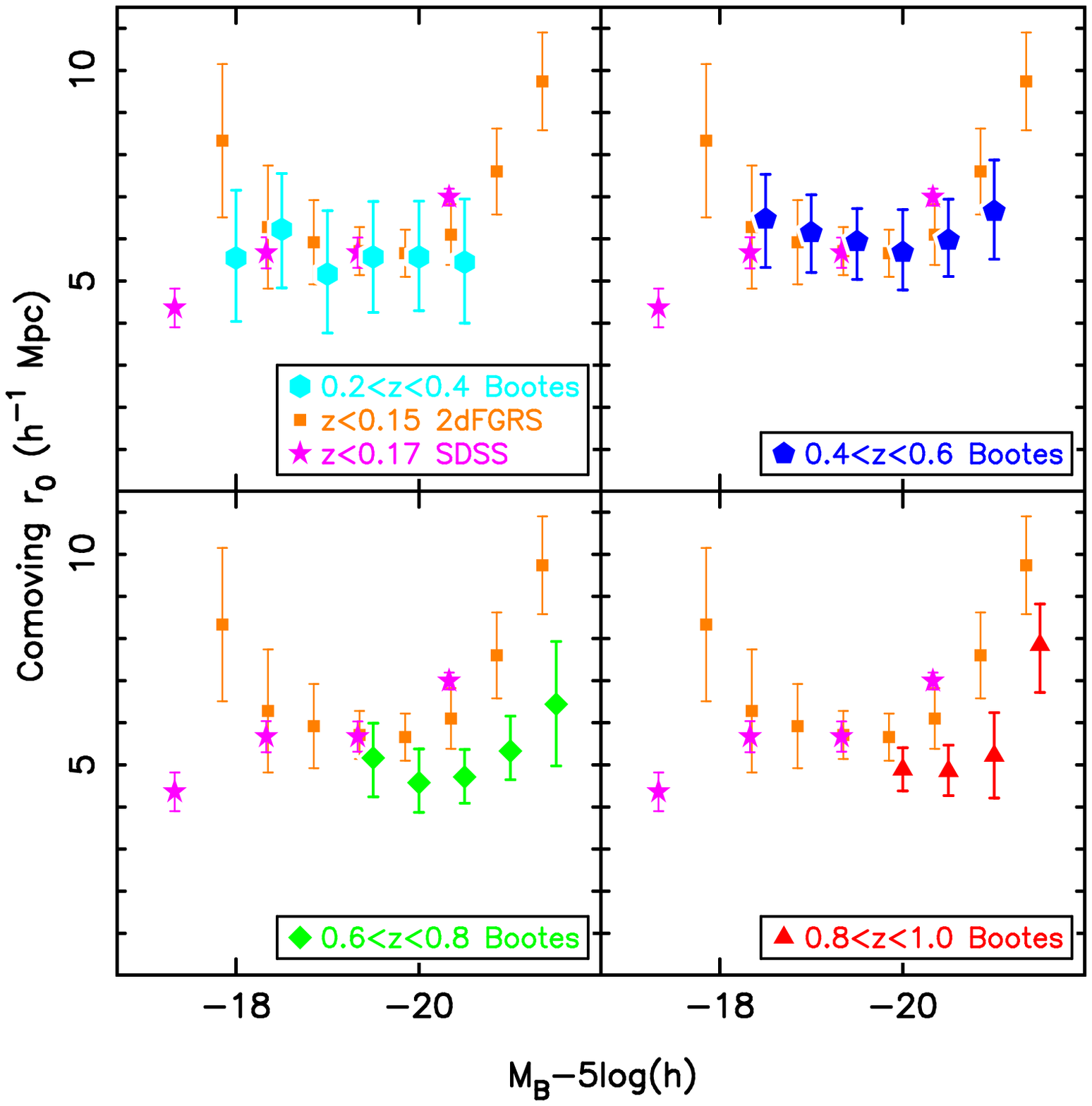}
\caption{The spatial clustering of red galaxies in Bo\"otes as a function of absolute magnitude and redshift.
For comparison, we also plot low redshift measurements from the 2dFGRS \markcite{nor02}({Norberg} {et~al.} 2002) and SDSS \markcite{zeh05b}({Zehavi} {et~al.} 2005b).
All the samples plotted here use bins $\pm 0.5~{\rm mag}$ wide.
The clustering of red galaxies brighter than $M_B-5{\rm log}h=-20$ is clearly a function of luminosity, and this
trends weakens at fainter magnitudes. Compared to the other measurements shown here, the power-law fit to the
faintest SDSS sample has an unusually high value of $\gamma$ (2.46) and a correspondingly low estimate of $r_0$.
The clustering of red galaxies as a function of luminosity evolves, but this is largely due to the fading
of aging stellar populations rather than evolution of large-scale structure.
\label{fig:r0lz}}
\end{figure*}

The clustering of red galaxies in Bo\"otes, as a function of both luminosity and redshift, is plotted in 
Figure~\ref{fig:r0lz}. For comparison we also plot low redshift measurements from the 2dFGRS \markcite{nor02}({Norberg} {et~al.} 2002) and SDSS \markcite{zeh05b}({Zehavi} {et~al.} 2005b). 
We have used the model S0 colors of \markcite{fuk95}{Fukugita}, {Shimasaku}, \&  {Ichikawa} (1995) to include the 2dFGRS and SDSS measurements in Figure~\ref{fig:r0lz}.
We find that the clustering of red galaxies is clearly a strong function of luminosity for galaxies brighter than $M_B-5{\rm log}h=-20$. 
This trend disappears or weakens for fainter galaxies. The faintest SDSS data-point suggests the correlation of $r_0$
with luminosity continues to fainter magnitudes, but the power-law approximation of this subsample has an unusually high
value of $\gamma$ (2.46) which may explain the unusually low estimate of $r_0$. The clustering of red galaxies as
a function of luminosity clearly evolves with redshift. However, as we expect aging stellar populations to fade by 
1.2 $B$-band magnitudes between $z=1$ and $z=0$, much of this evolution is due to luminosity evolution rather than the 
evolution of large-scale structure.

\section{HALO OCCUPATION DISTRIBUTION MODELING}
\label{sec:hod}

Halo occupation distribution \markcite{pea00,sel00,ber02,coo02}(HOD; e.g., {Peacock} \& {Smith} 2000; {Seljak} 2000; {Berlind} \& {Weinberg} 2002; {Cooray} \& {Sheth} 2002) modeling is a powerful
tool for understanding the clustering of galaxies and how galaxies reside within dark matter halos.
The principle of the HOD framework is to link galaxies to their host dark matter halos, whose formation
and properties can be predicted by both simulations and analytic methods. As the evolving 
space density and clustering of dark matter halos are predictable functions of redshift, one can 
determine the galaxy HOD with the observed space density and clustering of galaxies.
By measuring the HOD as a function of both galaxy luminosity and redshift, one 
obtains the evolving relationship between galaxy luminosity and host halo mass.
One can thus determine how galaxy stellar masses are growing with respect to the masses of the dark matter 
halos in which the galaxies reside.

\subsection{HOD MODELING OF THE CORRELATION FUNCTION}

The conceptually simplest method of implementing HOD modeling is to populate
halos identified in cosmological $N$-body simulations.
We use this approach to generate mock galaxy catalogs, which we discuss
in detail in Appendix~\ref{sec:mock}. The simulation provides halo masses and positions,
while the HOD specifies the mean number and spatial distribution of galaxies within halos.

The functional form of the HOD is typically motivated by the results of galaxy formation 
models, and it is physically reasonable to split the HOD into central and satellite galaxy 
components \markcite{kra04,zhe05}(e.g., {Kravtsov} {et~al.} 2004; {Zheng} {et~al.} 2005). 
By definition central galaxies are found at the center
of halos and there is only zero or one central galaxy per halo (although the mean number
of centrals per halo can lie between zero and one).
Multiple satellite galaxies can reside within a halo, and their spatial distribution
is often assumed to follow the dark matter distribution.

Once a set of HOD parameter values has been chosen, one can step through the halo catalog
and populate each halo with galaxies. The HOD provides the probabilities that a halo will contain
a central galaxy and a particular number of satellites. One can thus determine the number
of galaxies in a given halo using these probabilities and a random number generator.
The HOD also provides the probability of a satellite being at a particular radius from the halo
center, so a random number generator can be used to assign satellite positions within halos.
Once a catalog is populated with galaxies, one can mimic the observations and statistics
used to measure the properties of the galaxy population (in this case, the luminosity and angular correlation functions).
One can repeat the process, exploring a range of HOD parameter values, until a best-fit model is found.

While it is conceptually simple to populate a simulated halo catalog, it is computationally expensive.
Analytic methods of HOD modeling have been developed which allow quick computation of the galaxy 
space density and clustering 
statistics \markcite{pea00,sel00,ber02,coo02}(HOD; e.g., {Peacock} \& {Smith} 2000; {Seljak} 2000; {Berlind} \& {Weinberg} 2002; {Cooray} \& {Sheth} 2002).
These methods use descriptions of halo properties which have been calibrated against $N$-body simulations.
We employ an analytic methodology which is extremely similar to that of \markcite{whi07}{White} {et~al.} (2007) and \markcite{zhe07}{Zheng} {et~al.} (2007).
We utilize analytic approximations for the halo mass function \markcite{jen01}({Jenkins} {et~al.} 2001), 
the biased clustering of halos \markcite{tin05}({Tinker} {et~al.} 2005), the profile of dark matter within halos \markcite{NFW}(NFW; {Navarro}, {Frenk}, \& {White} 1996), 
the concentration of halo profiles \markcite{bul01}({Bullock} {et~al.} 2001) and the non-linear dark matter power-spectrum \markcite{smi03}({Smith} {et~al.} 2003).
The galaxy space density is the integral of the halo mass function multiplied by the mean number of galaxies per halo.
In the analytic calculation, the two-point correlation function is the sum of two terms, the one-halo term and
the two-halo term. 

The one-halo term results from pairs of galaxies which reside within the same halo (intra-halo pairs).
This term is dominant on small scales and, by definition, is sensitive to the fraction of galaxies which are satellites.
Given an HOD model, the one-halo term is just the distribution of intra-halo galaxy pair separations as a function of 
halo mass convolved with the halo mass function. The calculation of this term is usually decomposed into contributions 
by central-satellite and satellite-satellite pairs. We assume satellite galaxies follow the dark matter distribution within 
halos, which we model with NFW profiles. As a result, the spatial distribution of central-satellite pairs follows an NFW profile 
while that of satellite-satellite pairs follows an NFW profile convolved with itself.

The two-halo term results from pairs of galaxies which reside within 
different halos, and this term dominates the correlation function on large scales. 
On the largest scales it is equal to the mass correlation function times the 
mean galaxy-weighted halo bias squared. On scales smaller than $\sim 1~h^{-1}~{\rm Mpc}$
one must include prescriptions for halo exclusion (halos cannot reside within each other)
and scale-dependent bias. In practice it is easier to calculate the power spectrum of the two-halo term 
rather than directly calculate the spatial correlation function, and  we refer interested 
readers to \markcite{zhe04}{Zheng} (2004) and \markcite{tin05}{Tinker} {et~al.} (2005) for further details.

\subsection{HOD PARAMETERIZATION}

\begin{figure}
\plotone{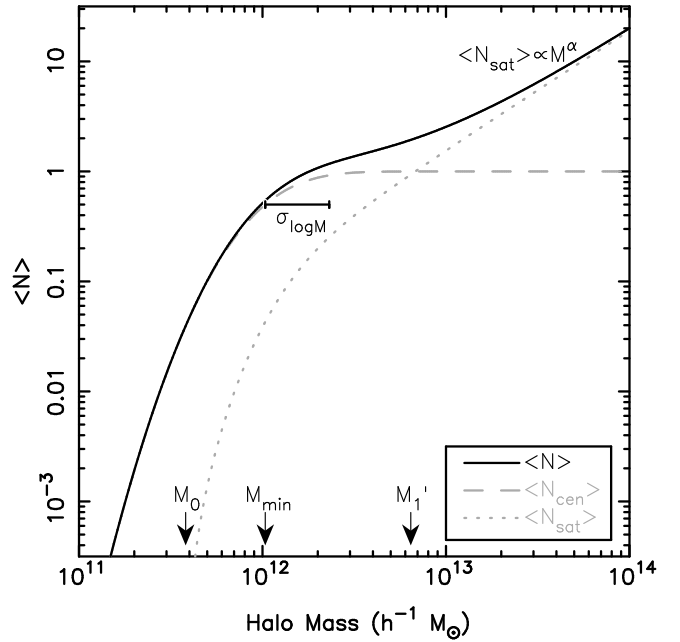}
\caption{The mean number of central and satellite galaxies per halo, as defined by equations~\ref{eq:cen} and ~\ref{eq:sat}.
Plotted is the HOD of $M_B-5{\rm log} h < -18$ red galaxies at $0.4<z<0.6$.
By definition halos host either zero or one central galaxy, with 50\% of halos of mass $M_{\rm min}$ hosting a central galaxy.
The parameter $\sigma_{{\rm log} M}$ quantifies the mass range where $\langle N_{\rm cen} \rangle$ transitions from $\simeq 0$ to $\simeq 1$.
Satellites can reside in halos more massive than $M_0$, and at mass of $M^\prime_1$ the mean number of satellites per halo is $\simeq 1$.
In very massive halos, the number of satellites is proportional to halo mass to the power of $\alpha$.
\label{fig:hodex}}
\end{figure}

We use a five parameter model to describe the mean number of central and satellite galaxies (brighter than some luminosity) per halo.
We plot an example of this model in Figure~\ref{fig:hodex} and discuss the model in detail below.
The functional form of this model is motivated by HODs observed in galaxy formation simulations \markcite{ber03,kra04,zhe05}({Berlind} {et~al.} 2003; {Kravtsov} {et~al.} 2004; {Zheng} {et~al.} 2005).
The mean number of central galaxies per halo is modeled with
\begin{equation}
 \langle N_{\rm cen}(M) \rangle = \frac{1}{2} \left[1 + {\rm erf}\left( \frac{{\rm log} M - {\rm log} M_{\rm min}}{\sigma_{{\rm log} M}} \right) \right],
\label{eq:cen}
\end{equation}
where erf is the error function
\begin{equation}
{\rm erf}(x) = \left. \frac{2}{\sqrt{\pi}} \int^x_0 e^{-t^2} dt \right.
\end{equation}
At a halo mass of $M_{\rm min}$, 50\% of halos host a central galaxy. If the relationship between galaxy luminosity and
halo mass had no scatter, $\langle N_{\rm cen} \rangle$ would be modeled by a step function. In reality this
relation must have some scatter, resulting in a gradual transition from $ \langle N_{\rm cen} \rangle \simeq 0$ to 
$ \langle N_{\rm cen} \rangle \simeq 1$, whose width we quantify with the parameter $\sigma_{{\rm log} M}$. 

We approximate the mean number of satellite galaxies per halo with a power-law truncated at a threshold mass of $M_0$:
\begin{equation}
 \langle N_{\rm sat}(M) \rangle =  \langle N_{\rm cen}(M) \rangle \left( \frac{M-M_0}{M_1^\prime} \right)^\alpha.
\label{eq:sat}
\end{equation}
The parameter $M_1^\prime$ corresponds to the halo mass where $\langle N_{\rm sat}(M) \rangle \simeq 1$ when  (as is the case here)
$M_1^{\prime}\gg M_0$ and $M_1^{\prime}\gg M_{\rm min}$.
When $\alpha=1$ and $M\gg M_0$, the mean number of satellites per halo is proportional to halo mass.
The number of satellites in halos of a given mass is assumed to follow a Poisson distribution, 
which is consistent with theoretical predictions \markcite{kra04,zhe05}({Kravtsov} {et~al.} 2004; {Zheng} {et~al.} 2005) and current observational constraints \markcite{yan05,ho07,yan07}({Yang} {et~al.} 2005; {Ho} {et~al.} 2007; {Yang}, {Mo}, \& {van den Bosch} 2007). 

For each subsample of the red galaxy catalog we tested a range of plausible HOD models.
For each HOD model we determined the galaxy space density and spatial correlation function using the analytic methodology
described above, and then determined the corresponding angular correlation function using the 
\markcite{lim54}{Limber} (1954) equation. For each HOD model we estimated $\chi^2$ values using the full covariance 
matrices and a space density prior including fractional uncertainties determined using mock catalogs. 

To rapidly explore the plausible range of HOD parameter space we applied the
Markov Chain Monte Carlo method \markcite{gil96}(MCMC; e.g., see {Gilks}, {Richardson}, \&  {Spiegelhalter} 1996). This method 
generates a list (or chain) of HOD parameters whose frequency in the chain traces 
the likelihood of that model fitting the data. It works by generating random 
HODs from a trial distribution and accepting or rejecting them based on the 
relative likelihood of the fit.  We choose new models by perturbing the HOD parameters from the last accepted chain
element by Gaussian offsets in the log of the relevant parameter.  The
step size and directions are determined from the covariance matrix of a previous run of the chain.
We further restricted the HOD parameter space to those models with $\sigma_{\log M} <0.6$.
Each chain provides the HOD distribution and a set of models which provide good fits to the observations. 

We provide a summary of our best-fit HOD parameter values in  Tables~\ref{table:hod1} and~\ref{table:hod2}.
While $M_{\rm min}$ and $M^\prime_1$ have small uncertainties, the other parameters are poorly constrained. 
As the measured HOD parameter values are correlated with each other, $M_{\rm min}$ and $M^\prime_1$ as a function of luminosity 
would show less scatter if the other HOD parameters had well determined values.
For this reason, in Table~\ref{table:hod2} we provide fits of HOD models where only $M_{\rm min}$ and $M^\prime_1$ are 
free parameters while the other parameters are fixed at $\sigma_{{\rm log} M}=0.3$, $\alpha=1$ and $M_0=M_{\rm min}$. 
The fixed parameters have values that are similar to those in Table~\ref{table:hod1} and the predictions of simulations \markcite{zhe05,seo07}(e.g., {Zheng} {et~al.} 2005; {Seo}, {Eisenstein}, \& {Zehavi} 2007).

\subsection{THE RED GALAXY HOD}

\begin{figure}
\plotone{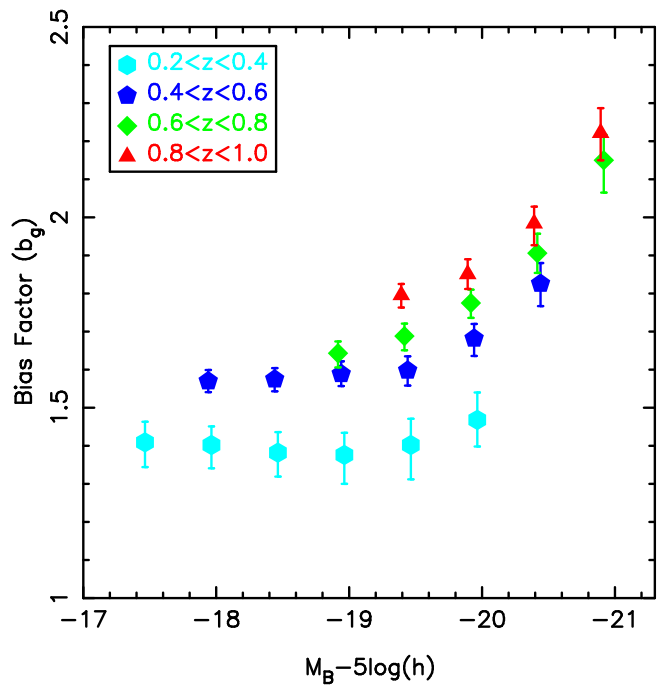}
\caption{The bias factor of red galaxies as a function of threshold absolute magnitude.
Data points at a given redshift are correlated with each other as bright and faint galaxies can reside within the
same large-scale structures and red galaxies selected by the brightest absolute magnitude thresholds are subsets
of the red galaxies selected with fainter absolute magnitude thresholds.
Red galaxies are highly biased tracers of the underlying dark matter distribution.
As the spatial clustering of red galaxies does not evolve as rapidly as the underlying
dark matter distribution, the bias increases with redshift.
\label{fig:bias}}
\end{figure}

Models of the correlation function derived from the HOD provide a robust estimate of the large-scale bias factor, $b_g$ (where $\xi_{g} \simeq b_g^2 \xi_{\rm DM}$),
which we plot as a function of absolute magnitude in Figure~\ref{fig:bias}. HOD models better constrain the bias factor
than power-law extrapolations, which depend on the measured value of $\gamma$ and result in a varying bias factor 
on large scales. The bias factor increases with redshift, as the spatial clustering of red galaxies evolves slowly
while the underlying dark matter distribution evolves rapidly. As noted previously, the bias (or clustering strength) is 
a strong function of luminosity for the brightest red galaxies while varying little for galaxies fainter than  $M_B-5{\rm log}h=-20$.

\begin{figure*}
\plotone{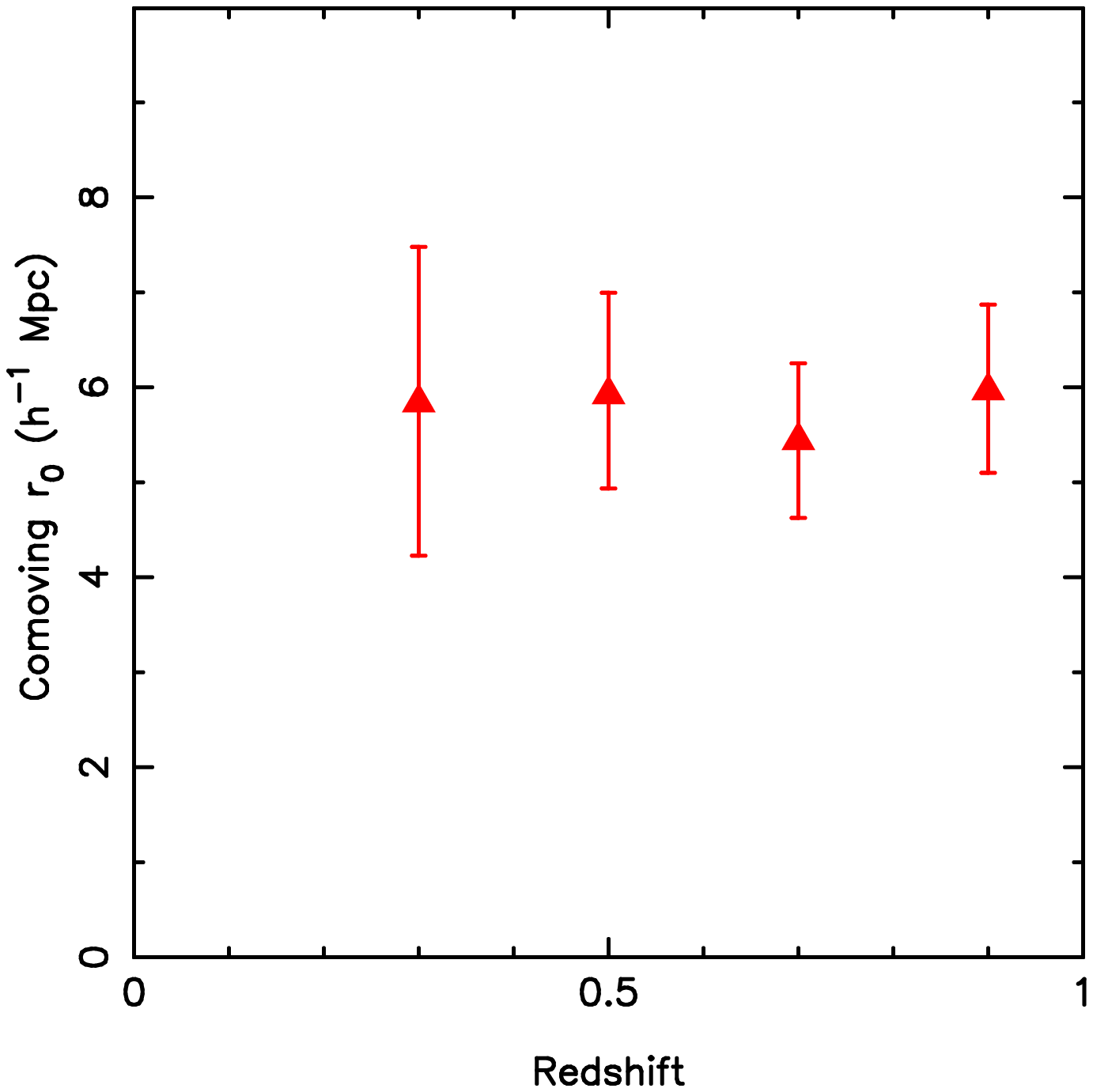}\plotone{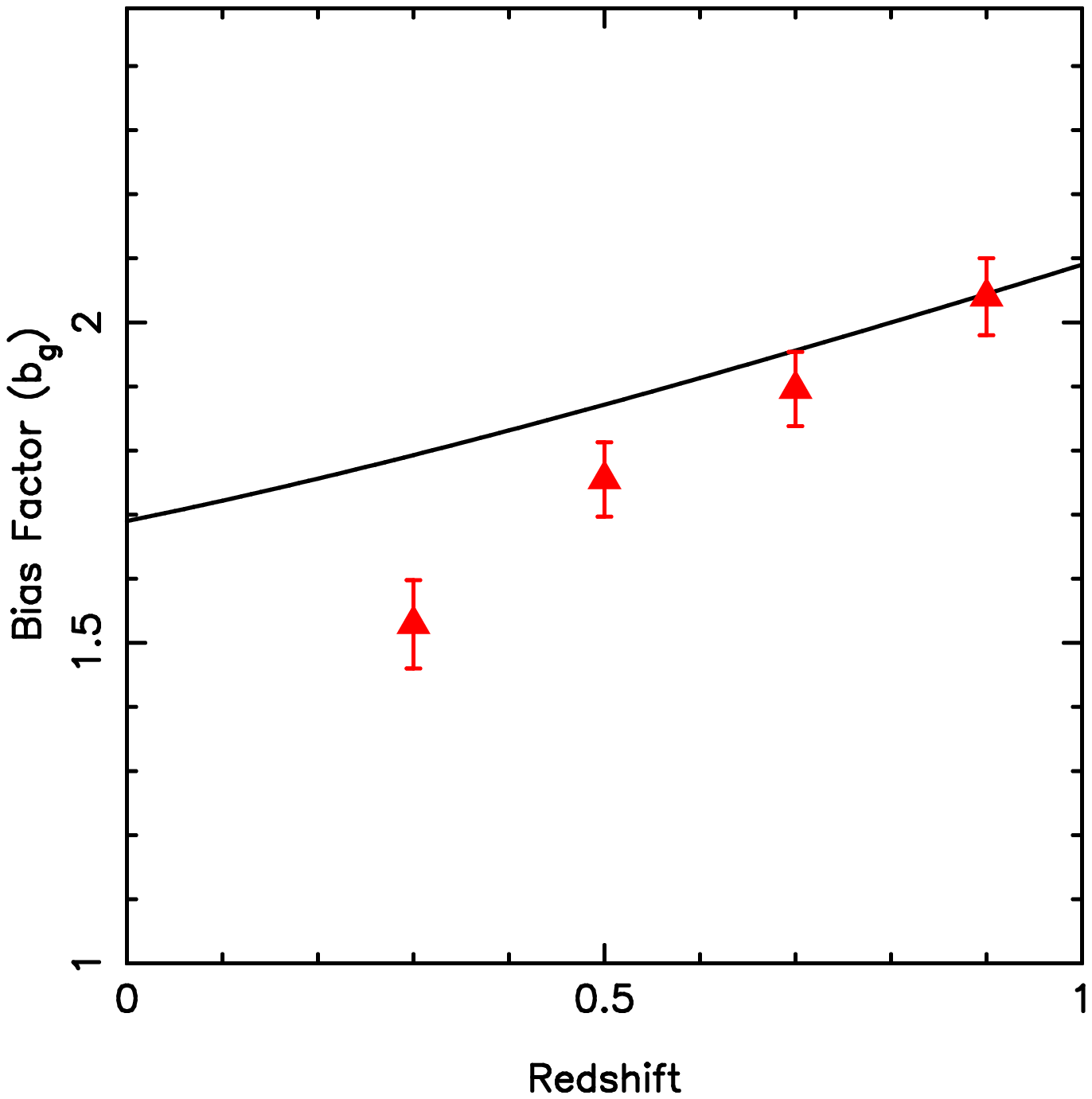}
\caption{The spatial clustering and bias factor of red galaxies as a function of redshift.
The most luminous red galaxies were selected down to a space density threshold of $10^{-3}~h^{3}~{\rm Mpc}^{-3}$.
The value of $r_0$ was derived from power-law fits to the data while the bias factor was determined using
HOD modeling of the clustering and space density of galaxies.
If red galaxies underwent pure luminosity evolution without mergers, gravitational collapse
would result in the spatial clustering increasing with decreasing redshift. A pure luminosity
evolution model without mergers \markcite{fry96}({Fry} 1996), which is shown with the solid line, does not match our observations.
\label{fig:space}}
\end{figure*}

In \markcite{whi07}{White} {et~al.} (2007), we discussed the HOD of space density selected samples of red galaxies, and 
we recap several key results of that study here. We selected the $n$ most luminous red galaxies per unit volume down 
to a space density threshold of $10^{-3.0}\,h^3{\rm Mpc}^{-3}$. 
If red galaxies underwent pure luminosity evolution (PLE) without galaxy mergers, a space density selected sample
would select the same (fading) galaxy population with redshift and the spatial clustering would increase
with decreasing redshift due to gravitational collapse.  As we show in Figure~\ref{fig:space}, 
we find little or no evolution of the spatial clustering of these galaxies with redshift, which is contrary to PLE. 
The evolving bias factor of these galaxies does not evolve in the same
manner as analytic approximations \markcite{fry96}({Fry} 1996) or simulations \markcite{whi07}({White} {et~al.} 2007) of PLE. The PLE
simulation, normalized to the $z=0.9$ observations, overestimates the number of satellite galaxies
at $z=0.5$ by roughly $50\%$. A simple solution is to remove a third of the satellites, by having them 
undergo merging or disruption between $z=0.9$ and $z=0.5$ \markcite{whi07}({White} {et~al.} 2007).

\begin{figure*}
\plotone{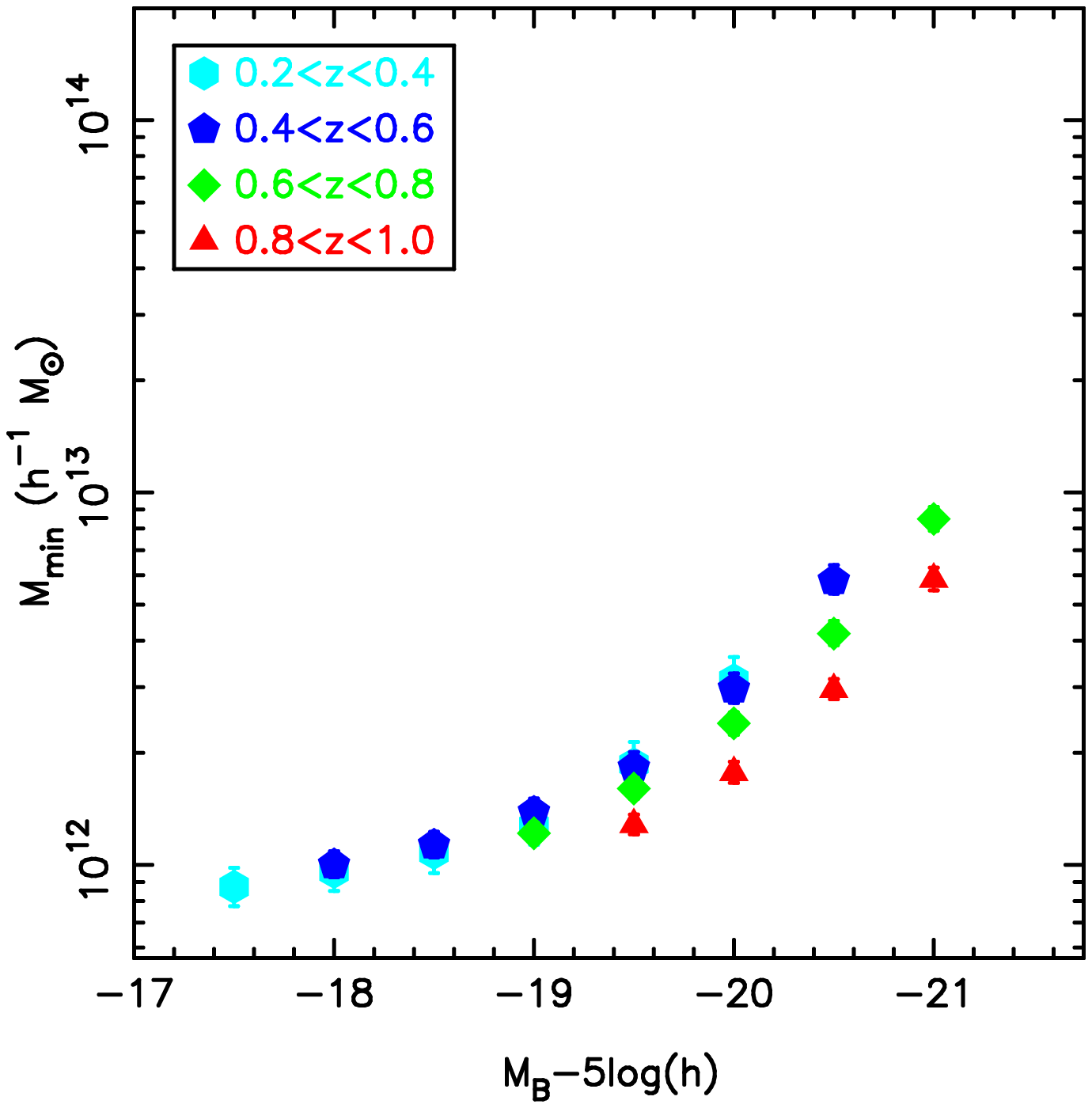}\plotone{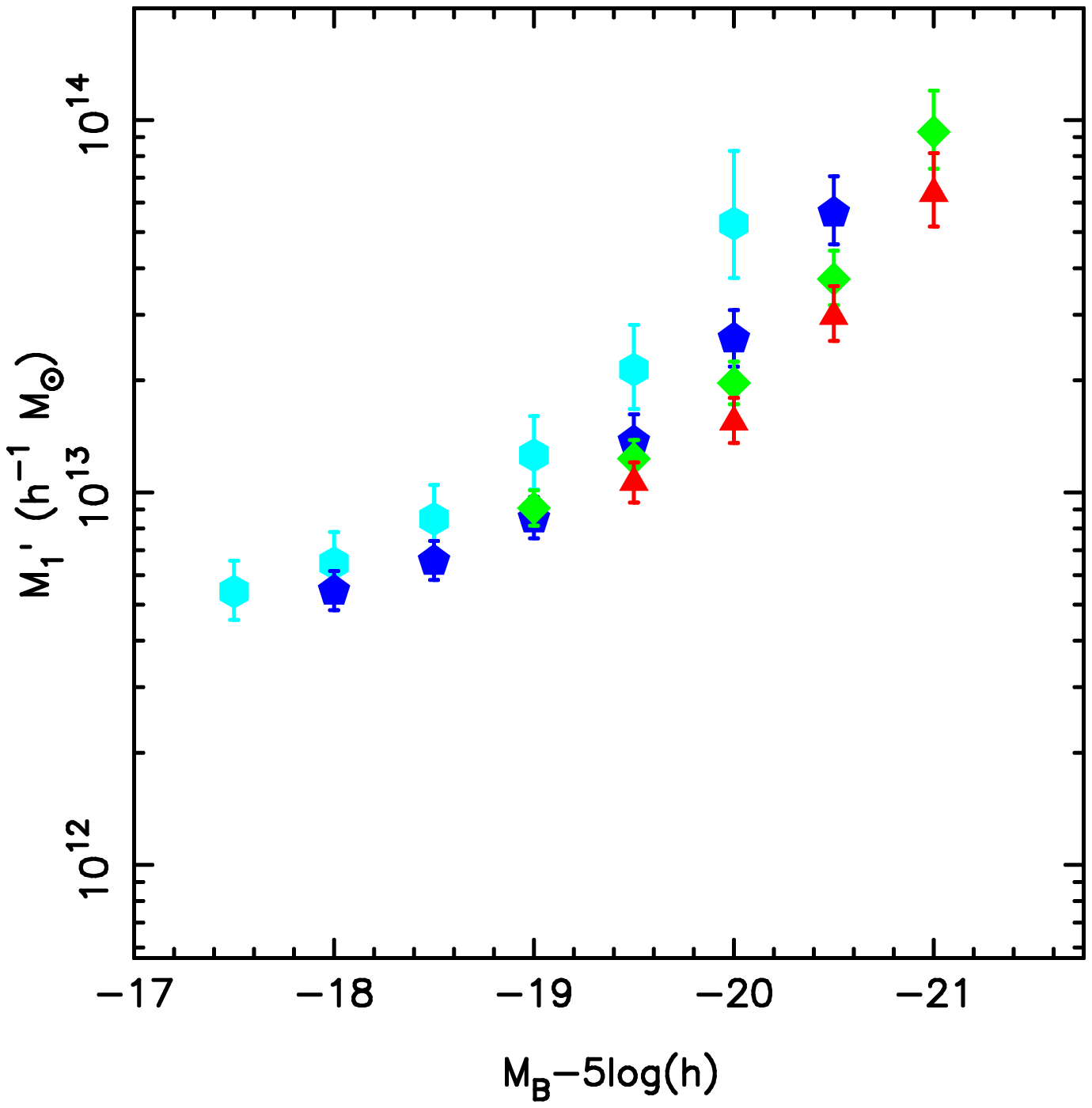}
\caption{The relationship between red galaxy threshold absolute magnitude and host halo mass,
parameterized with $M_{\rm min}$ (left) and $M_1^\prime$ (right).
$M_{\rm min}$ is the mass where 50\% of halos host a central galaxy while $M_1^\prime$
corresponds to the halo mass where the mean number of satellites is $\simeq 1$.
We determined $M_{\rm min}$ and $M_1^\prime$ using HOD modeling (with $\sigma_m$, $\alpha$ and $M_0$ fixed)
of the observed clustering and space density of red galaxies in Bo\"otes.
Both $M_{\rm min}$ and $M_1^\prime$ increase with increasing luminosity, as one would expect if
the most massive galaxies reside within the most massive dark matter halos.
\label{fig:mvsz}}
\end{figure*}

In Figure~\ref{fig:mvsz} we plot the values of $M_{\rm min}$ and $M_1^\prime$ as a function of 
$B$-band absolute magnitude. As one would expect, the most massive central and 
satellite galaxies can only reside in the most massive halos, so $M_{\rm min}$ and $M^\prime_1$ 
increase with luminosity. One can also see that the halo mass required to host a 
galaxy of fixed $B$-band luminosity decreases with increasing redshift. It is not 
obvious from Figure~\ref{fig:mvsz} if this trend results from evolution of the relationship 
between galaxy stellar mass and halo mass, or the evolution of galaxy stellar populations.

\begin{figure*}
\plotone{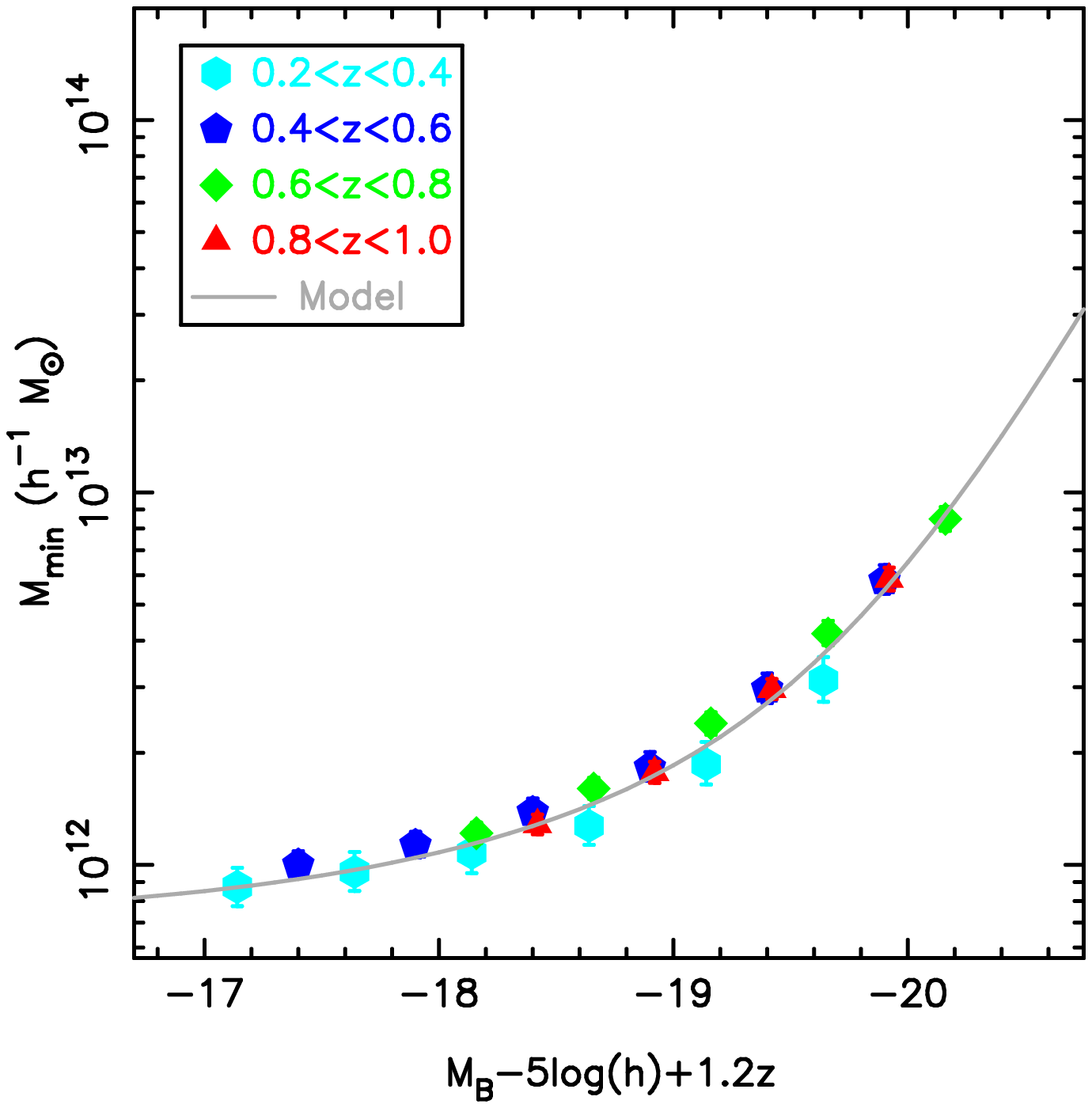}\plotone{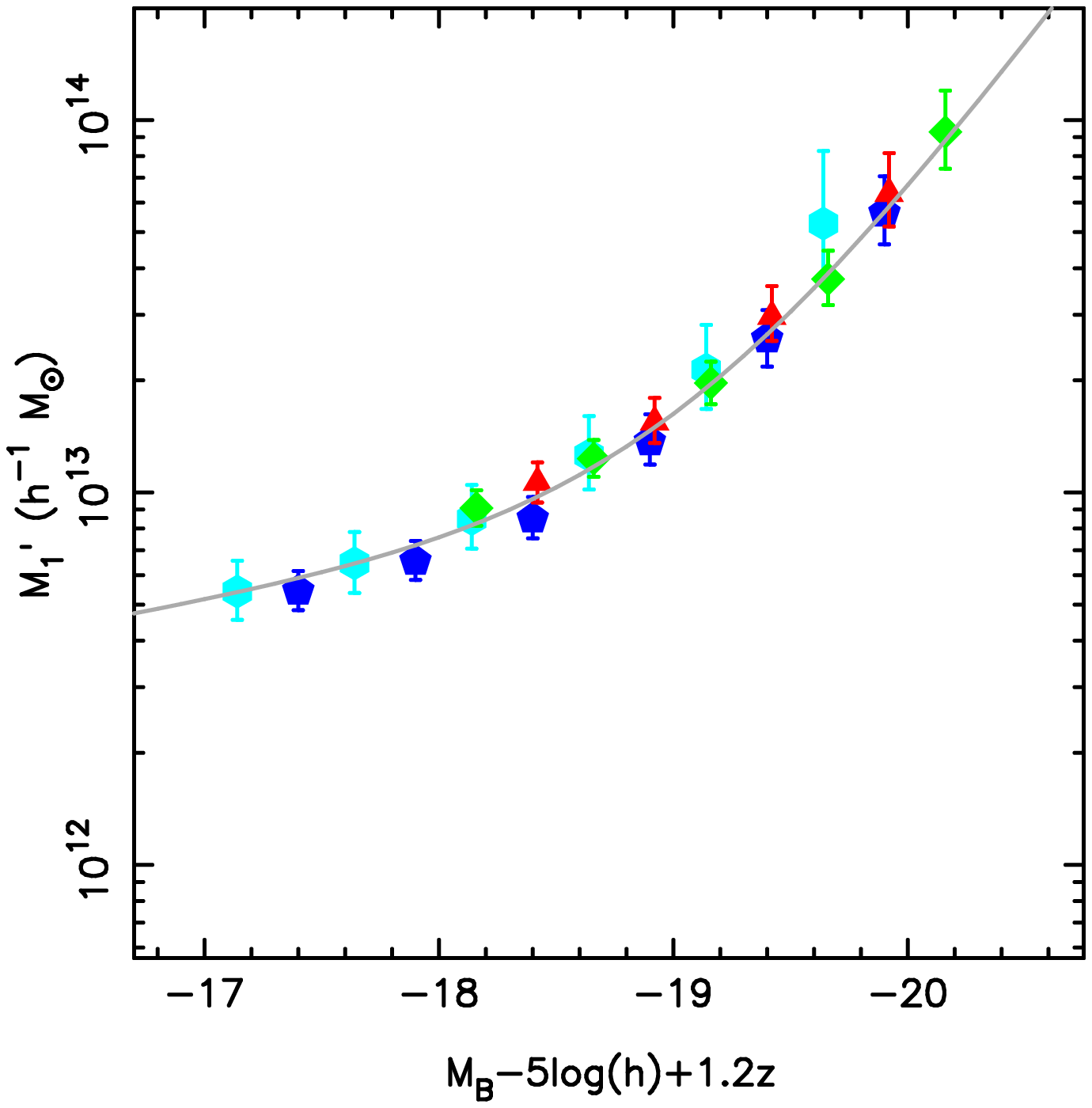}
\caption{HOD model parameters $M_{\rm min}$ (left) and $M_1^\prime$ (right) as a function
of threshold $B$-band absolute magnitude plus $1.2z$. By adding $1.2z$ to the absolute magnitudes,
we compensate for the fading of red galaxy stellar populations, so the x-axes of our plots are proxies for stellar mass.
Grey lines denote our analytic approximations of the HOD (given by Equations ~\ref{eq:mmin} and ~\ref{eq:m1pp}).
Both $M_{\rm min}$ and $M_1^\prime$ exhibit little or no evolution. We thus conclude that the relationship
between red galaxy stellar mass and host halo mass undergoes little or no evolution at $z<1$. As the
masses of dark matter halos increase with decreasing redshift, we expect galaxy stellar
masses to also increase with decreasing redshift.
\label{fig:mvsb}}
\end{figure*}

In Figure~\ref{fig:mvsb} we compensate for the fading of red galaxy stellar populations by 
adding $1.2z$ to the $B$-band absolute magnitudes. This correction is comparable to the observed
evolution of the fundamental plane  \markcite{van03,tre05}(e.g., {van Dokkum} \& {Stanford} 2003; {Treu} {et~al.} 2005) and is similar to the luminosity evolution 
of stellar populations models which reproduce the optical colors of red galaxies \markcite{bel04,bro07}(e.g., {Bell} {et~al.} 2004; {Brown} {et~al.} 2007). 
The value of $M_B+1.2z$ is effectively a proxy for stellar mass, although the exact conversion to stellar
mass will depend on details of the stellar initial mass function and the star formation history. 
One can see in Figure~\ref{fig:mvsb} that $M_{\rm min}$ and $M^\prime_1$ show little or no evolution as a 
function of red galaxy stellar mass. We thus conclude that the relationship between red galaxy stellar mass 
and host halo mass undergoes little or no evolution between $z=0.9$ and $z=0.3$.

\subsection{COMPARISON WITH HODs FROM THE LITERATURE}

An important consistency check is to compare our HODs with those from the literature.
Unfortunately, such a comparison is not straightforward due to differences in sample 
selection, adopted cosmology, and the assumed functional form of the HOD.
To simplify the comparison, we plot HODs as a function of galaxy space density rather than 
attempting to model the equivalent absolute magnitude threshold in the $B$-band.
As space density is strongly correlated with luminosity, this provides a simple and well defined way of comparing
samples selected using a variety of techniques.

HODs in the literature often use cosmologies or halo definitions that differ from the ones adopted here.
To shift halo masses to our $\Omega_m=0.25$ and $\sigma_8=0.8$ cosmology, we use the prescription of \markcite{zhe02}{Zheng} {et~al.} (2002).
For this comparison, we define halos to be spherical objects with a mean density 200 times that of the background, 
and we modify halo masses from the literature to be consistent with this definition.
We compare characteristic quantities which are not sensitive to the details of the HOD parameterization.
We use $M_{\rm min}$ and $M_1$, where $M_{\rm min}$ is the halo mass where $\langle N_{\rm cen} \rangle=0.5$ and $M_1$ is the halo mass 
where $\langle N_{\rm sat} \rangle =1$.

\begin{figure*}
\plotone{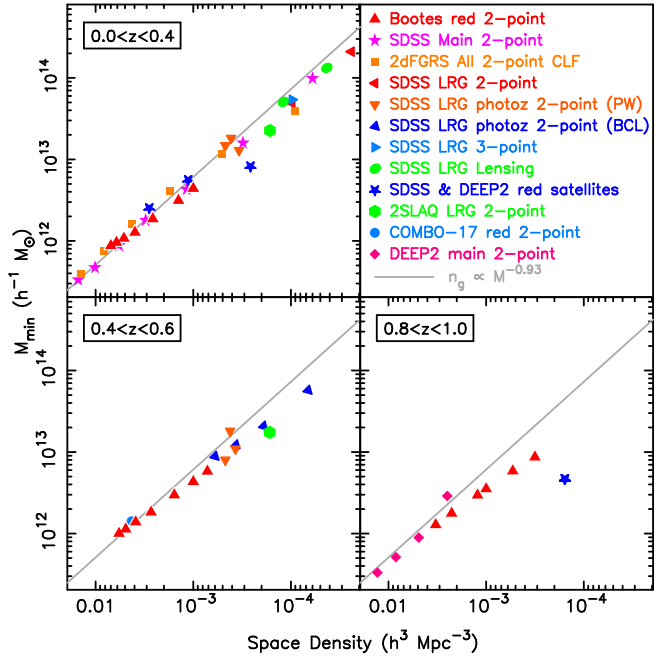}
\caption{$M_{\rm min}$ as a function of galaxy space density, where $M_{\rm min}$ is the halo mass where $\langle N_{\rm cen} \rangle=0.5$.
Random uncertainties are on the order of 10\%, with the exception of masses from satellite velocities, where uncertainties are $\sim 50\%$.
Satellite halo masses have been revised downwards by 30\% to compensate for spectroscopic incompleteness (as discussed by \markcite{con07a} {Conroy} {et~al.} 2007b).
Our measurements are in broad agreement with the recent literature, which is to be expected as many of the $M_{\rm min}$ measurements
are constrained with models of the halo mass function. A power-law fit to the SDSS main sample (grey line)
approximates $M_{\rm min}$ as a function of both space density and redshift. This is to be expected, as the space density of  $\sim 10^{12}~h^{-1}~M_\odot$
mass halos can be approximated by a slowly evolving power-law. Departures from this power-law are expected
for high mass and high redshift halos.
\label{fig:mmin_ng}}
\end{figure*}

\begin{figure*}
\plotone{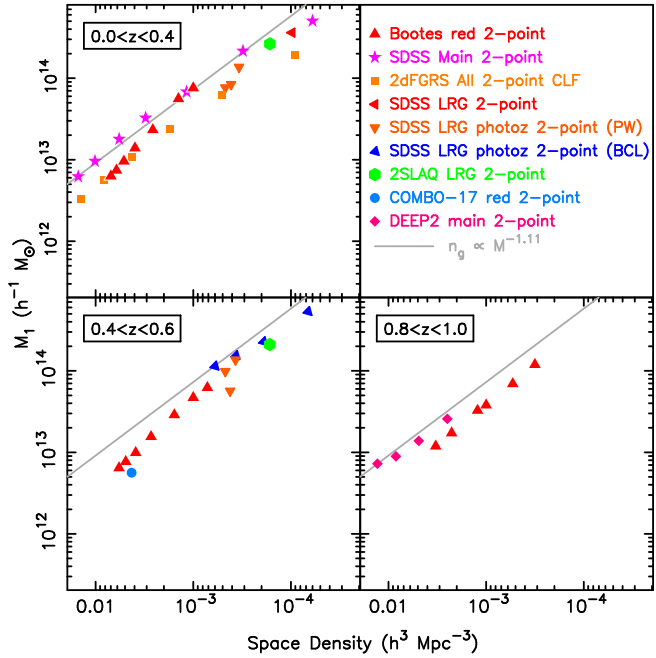}
\caption{$M_1$ as a function of galaxy space density, where $M_{1}$ is the halo mass where $\langle N_{\rm sat} \rangle =1$.
The grey line is a power-law fit to the HOD derived from the space density and 2-point correlation function of SDSS main sample galaxies.
There is broad agreement between the various studies, although the scatter is larger than in Figure~\ref{fig:mmin_ng}.
\label{fig:m1_ng}}
\end{figure*}

In Figures~\ref{fig:mmin_ng} and~\ref{fig:m1_ng} we compare our HOD estimates of $M_{\rm min}$ and $M_1$ with 
those from the literature. The HODs from the literature are mostly derived from red galaxy samples, although we have included
some HODs determined using samples including both red and blue galaxies (e.g., DEEP2, SDSS main).
The various studies are in broad agreement, including measurements derived from the
two-point correlation function \markcite{zeh05a,phl06,vdb07a,zhe07,bla08,wak08,pad08}({Zehavi} {et~al.} 2005a; {Phleps} {et~al.} 2006; {van den Bosch} {et~al.} 2007; {Zheng} {et~al.} 2007; {Blake}, {Collister}, \& {Lahav} 2008; {Wake} {et~al.} 2008; {Padmanabhan} {et~al.} 2008, Z. Zheng et al., in prep.),
three-point correlation function \markcite{kul07}({Kulkarni} {et~al.} 2007), satellite galaxy velocities \markcite{con07a}({Conroy} {et~al.} 2007b) and lensing \markcite{man06b}({Mandelbaum} {et~al.} 2006a).
The agreement between our measurements and those derived from lensing and satellite galaxy velocities is encouraging, 
as those measurements of halo masses do not depend on models of the halo mass function.

The relationship between galaxy space density and $M_{\rm min}$ at $z=0$ can be approximated by a power-law
\begin{equation}
n_g \simeq 5.3\times 10^{-3} \left( \frac{M_{\rm min}}{10^{12}~ h^{-1}~ M_\odot}\right)^{-0.93} ~h^{3}~ {\rm Mpc}^{-3},
\end{equation}
and this relationship appears to evolve slowly with redshift. 
Again, this is to be expected, as the mass function of dark matter halos less massive than 
$10^{13}~h^{-1}~M_\odot$ evolves slowly and can be approximated by $n_h(>M)\propto M^{-1}$.
This relationship should not hold for the most massive galaxies at high redshift, since the top end
of the halo mass function is not a power-law and rapidly evolves.

While the results from the literature are in broad agreement, there is no consensus
on how the relationship between galaxy stellar mass and host halo mass evolves. 
Some studies find this relationship evolves  \markcite{yan03,con07a,zhe07}({Yan}, {Madgwick}, \& {White} 2003; {Conroy} {et~al.} 2007b; {Zheng} {et~al.} 2007)\footnote{\markcite{yan03}{Yan} {et~al.} (2003) find that the relationship between $L^*$ and host halo mass doesn't evolve, 
but $L^*$ does not correspond to a fixed stellar mass. \markcite{con07a}{Conroy} {et~al.} (2007b) 
find that the relationship between stellar and host halo mass does not evolve for all but the most massive galaxies.}, while others find little or 
no evolution \markcite{lin06,phl06}({Lin} {et~al.} 2006; {Phleps} {et~al.} 2006). 
Measuring HOD evolution is complicated by several factors, which may explain the varying conclusions in the literature.
Redshift dependent selection effects can produce errors in the observed rate of HOD evolution.
Studies of the evolving HOD may also be comparing (somewhat) different galaxy populations at different epochs.
Finally, the relationship between blue galaxy stellar mass and host halo mass may evolve at $z<1$, while the
same relation for $\gtrsim L^*$ red galaxies undergoes little or no evolution.

\section{RED GALAXY EVOLUTION FROM HOD MODELING}
\label{sec:analytic}

As we show in Figure~\ref{fig:mvsb}, there is little or no evolution of the relationship between red galaxy 
stellar mass and host halo mass at $z<1$. Motivated by this, we introduce an analytic approximation of the HOD 
where the mean number of galaxies (of a given stellar mass) residing within halos (of a given mass) does not evolve.
The simplicity of this model allows one to easily discern important relationships present in our data, including
the correlation of central galaxy luminosity with host halo mass (\S\ref{sec:cen}) 
and the contributions of central and satellite galaxies to the total stellar mass within a halo (\S\ref{sec:satcen}).
Our analytic approximation also predicts how the luminosity function of central and satellite galaxies evolves over
a broad redshift range (\S\ref{sec:jb} and \S\ref{sec:jbcomp}).

Our approximation does not include all the relevant astrophysics (e.g., dynamical friction) and has empirical components
(e.g., the functional form of $\langle N_{sat} \rangle$), so it should not be confused with a complete 
model of galaxy evolution. We expect departures from this approximation (particularly at high $z$)
and these will be informative. In this respect our approximation is somewhat analogous to the illustrative and 
simple pure luminosity evolution models. That said, our model does include many of the key components
required to describe the evolution of $z<1$ red galaxies, including the aging of stellar populations and the growth of dark matter halos.

\subsection{AN ANALYTIC APPROXIMATION OF THE HOD}

Our analytic approximation of the HOD is largely constrained by the $M_{\rm min}$ and $M^\prime_1$ measurements presented in Figure~\ref{fig:mvsb}. 
These measurements span a limited luminosity range, so we use other measurements to constrain the HOD of the most massive galaxies.
The satellite fraction decreases with luminosity, with at least 80\% of  $M_B-5{\rm log}h + 1.2z<-20$ red galaxies being central galaxies.
We thus constrain the HOD of the most luminous galaxies by assuming one very luminous galaxy per halo, assuming some scatter ($\sigma_{\rm log M}=0.3$) in the 
relation between central galaxy luminosity and halo mass, and matching the cumulative luminosity
function of these galaxies with the cumulative mass function of dark matter halos.

We approximate the HOD parameters $M_{\rm min}$ and $M_1^\prime$ (for galaxies brighter than an absolute magnitude
threshold $M_B$) with a series of power-laws,
\begin{eqnarray}
M_{\rm min} (h^{-1} M_\odot) & =  &   10^{11.85}  \nonumber \\
                         &    & + 10^{11.95} \times 10^{0.40 \times [-19 - (M_B - 5 {\rm log} h + 1.2 z)]} \nonumber \\
                         &    & + 10^{13.70} \times 10^{1.15 \times [-21 - (M_B - 5 {\rm log} h + 1.2 z)]}  \nonumber \\
\label{eq:mmin}
\end{eqnarray}
and
\begin{eqnarray}
M_1^\prime (h^{-1} M_\odot) & =   &   10^{12.70} \times 10^{0.11 \times [-17 - (M_B - 5 {\rm log} h + 1.2 z)]}  \nonumber \\
                            &     & + 10^{14.60} \times 10^{0.85 \times [-21 - (M_B - 5 {\rm log} h + 1.2 z)]} \nonumber \\ \label{eq:m1pp}.
\end{eqnarray}
Each power-law component provides the relationship between galaxy luminosity and halo mass for a particular luminosity range.
For example, the last term of Equation~\ref{eq:mmin} shows halo mass is proportional to central galaxy luminosity to the 
power of $1.15\times 2.5$ for the most luminous galaxies.
As we show in Figure~\ref{fig:mvsb}, this approximation reproduces the measured
values of $M_{\rm min}$ and $M_1^\prime$ with an accuracy (RMS) of 8\% and 7\% respectively.
As the other HOD parameters are poorly constrained by our observations, we fix their 
values to $M_0=M_{\rm min}$, $\sigma_{{\rm log}M}=0.3$ and $\alpha=1$. 

We caution that  Equations~\ref{eq:mmin} and ~\ref{eq:m1pp} are only approximations of the true HOD.
They also depend upon the assumed cosmology and the analytic approximations discussed earlier.
It is also plausible that the HOD of red galaxies evolves very slowly at $z<1$. 
That said, we do expect the broad brush-strokes of our model to be valid and it is these which we discuss below.

\begin{figure}
\plotone{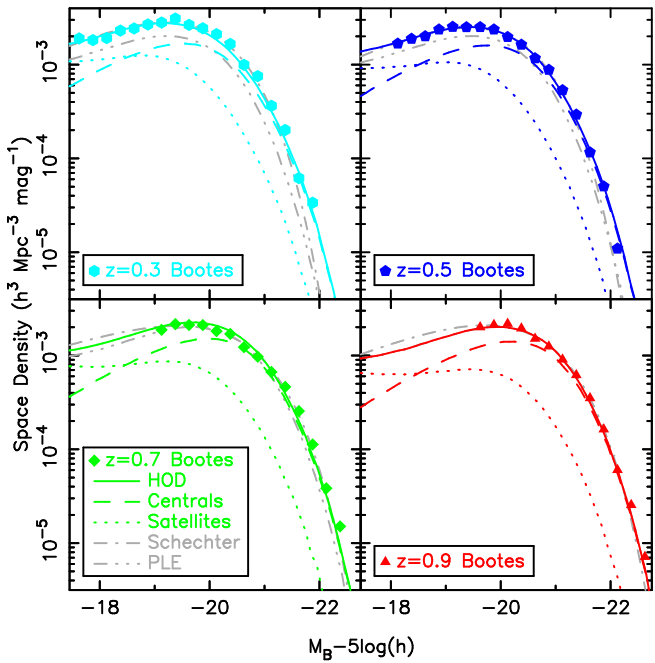}
\caption{Binned $1/V_{\rm max}$ red galaxy luminosity functions and our HOD model of the luminosity function (solid lines).
Random uncertainties for the $1/V_{\rm max}$ data points are on the order of 10\%.
A pure luminosity evolution model, which fades by $1.2$ $B$-band magnitudes per unit redshift and is
normalized to the $z=0.9$ luminosity function, does not match our observations.
The accuracy of our HOD model is comparable to Schechter functions individually fitted to the data at each redshift.
The contribution of central galaxies dominates the bright end of the luminosity function while the contribution
of satellite galaxies increases with decreasing luminosity and decreasing redshift.
\label{fig:lf}}
\end{figure}

Using the mean number of galaxies (brighter than a luminosity threshold) per halo and the halo mass function, 
one can determine the cumulative luminosity function by estimating the space density of galaxies for a series of absolute magnitudes.
Taking the derivative of this as a function of absolute magnitude, one obtains the conventional luminosity function.
In Figure~\ref{fig:lf} we compare the observed red galaxy luminosity function with our HOD model and Schechter
function fits. The accuracy of the HOD model is comparable to that of Schechter functions which were 
individually fitted to the data in each redshift bin. The HOD model has a consistently higher space 
density of very luminous galaxies than the Schechter function fits, particularly at $z<0.6$.
While the faint-end of Schechter luminosity functions is a power-law with a constant index,
the power-law index of the HOD model luminosity function varies at faint magnitudes.

\begin{figure}
\plotone{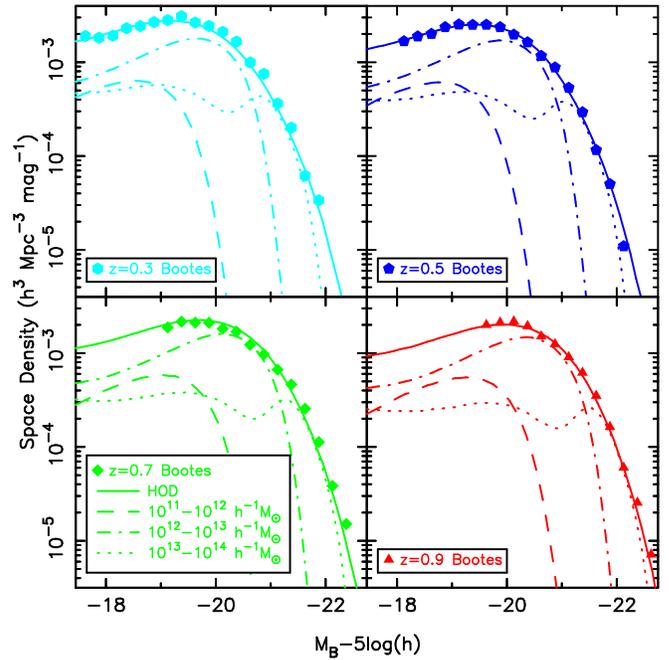}
\caption{Models of the luminosity function, split by halo mass, derived from our analytic approximation of the HOD.
Clearly the most luminous red galaxies reside within the most massive halos.
The least luminous red galaxies can be central galaxies in $\sim 10^{12}~h^{-1}~M_\odot$
halos or satellites in more massive halos. The luminosity function of galaxies in high mass halos has a wiggle at $\sim 2L^*$,
which is predicted by theory \markcite{zhe05}({Zheng} {et~al.} 2005) and has been observed in galaxy groups \markcite{yan07}({Yang} {et~al.} 2007). However, the
amplitude and location of the wiggle are functions $\sigma_{{\rm log} M}$ and $M_0$, which are poorly constrained by our observations.
\label{fig:lfm}}
\end{figure}

In Figure~\ref{fig:lfm} we plot the red galaxy luminosity function and the HOD model split 
by host halo mass. As one would expect, only the most massive halos can host the most massive galaxies.
However, while some of the faintest red galaxies are central galaxies in lower mass 
halos ($\sim 10^{12} ~h^{-1}~ M_\odot$), a large fraction of low luminosity red galaxies reside (as satellites)
in relatively massive halos. Galaxies in cluster mass halos ($>10^{14} ~h^{-1}~ M_\odot$) represent
a small fraction of the overall red galaxy population.

As we show in Figure~\ref{fig:lfm}, the shape of the red galaxy luminosity function varies with halo mass. 
In low mass halos, the shape of the galaxy luminosity function can be approximated by a 
Schechter function with a $\alpha\gtrsim 0$ while in high mass halos $\alpha\sim -1$. The value of 
$\alpha$ in high mass halos is very similar to the measured luminosity function of red cluster 
galaxies \markcite{dep03}(e.g., {De Propris} {et~al.} 2003).

The luminosity function of group and cluster galaxies ($>10^{13} ~h^{-1}~ M_\odot$ mass halos) has a wiggle at $\sim 2L^*$, 
corresponding to the transition from a satellite dominated luminosity function to a central dominated luminosity function.
Although predicted by theory  \markcite{zhe05}({Zheng} {et~al.} 2005) and observed in groups \markcite{yan07}({Yang} {et~al.} 2007), the size 
and location of the wiggle depends on both $\sigma_{{\rm log} M}$ and $M_0$, which are poorly constrained  by our data.
The luminosity function of red field galaxies is often approximated by a Schechter function with 
$\alpha=- 0.5$, and in our HOD model this results from faint red galaxies residing in halos spanning a broad mass range. 

\subsection{THE GROWTH OF CENTRAL GALAXIES}
\label{sec:cen}

In Figure~\ref{fig:lf} we plot the central and satellite components of the red galaxy luminosity function, 
determined using our analytic approximation of the HOD.
Clearly the top end of the luminosity function is dominated by central galaxies.
As we noted previously, halo mass scales as central galaxy luminosity
to the power of $1.15\times 2.5$, so the stellar masses of central galaxies scale as
halo mass to the power of $0.35$. We constrain the luminosity-mass relation for the brightest red 
galaxies using the observed luminosity function (\S\ref{sec:lf})
and the predicted halo mass function, so our luminosity-mass relation for these galaxies is similar to 
the $z\sim 0$ empirical model of \markcite{val04,val06}{Vale} \& {Ostriker} (2004, 2006) by construction. Our luminosity-mass relation 
for the most massive red galaxies is broadly similar to others in the literature, where halo mass 
was determined using X-ray temperature \markcite{lin04b,lin06}({Lin} \& {Mohr} 2004; {Lin} {et~al.} 2006), X-ray luminosity \markcite{brou08}({Brough} {et~al.} 2008), and 
weak lensing \markcite{man06a}({Mandelbaum} {et~al.} 2006b).

As the relationship between central galaxy stellar mass and halo mass is approximated by a non-evolving power-law with an 
index of $0.35$, the most massive galaxies do not grow as rapidly as their host halos.
While a merger of comparable mass halos presumably results in a merger of central galaxies, 
it is plausible that such  mergers are not 100\% efficient and dump stellar mass into the diffuse 
ICL \markcite{gon00,arn02,fel04,zib05,mon06,whi07,con07b,con07c}(e.g., {Gonzalez} {et~al.} 2000; {Arnaboldi} {et~al.} 2002; {Feldmeier} {et~al.} 2004; {Zibetti} {et~al.} 2005; {Monaco} {et~al.} 2006; {White} {et~al.} 2007; {Conroy}, {Wechsler}, \&  {Kravtsov} 2007c; {Conroy}, {Ho}, \&  {White} 2007a).
Minor halo mergers may produce satellite galaxies instead of directly funneling stellar mass into the central galaxy.
As a result, while mergers may increase the mass of a halo by 100\% at $z<1$, the stellar mass of the central galaxy 
typically grows by only 30\%. This is consistent with recent measurements of the luminosity function, where the most 
massive galaxies do not grow as rapidly as the most massive dark matter halos \markcite{bro07}(e.g., {Brown} {et~al.} 2007).
Semi-analytic models and simulations with rapid growth of massive galaxies may be 
overestimating the ability of halo mergers to funnel stellar mass into central galaxies  \markcite{tay01,ben03,taf03,boy08}(e.g., {Taylor} \& {Babul} 2001; {Benson} {et~al.} 2003; {Taffoni} {et~al.} 2003; {Boylan-Kolchin} {et~al.} 2008).

Our approximation of the HOD and the observed luminosity function of red galaxies (Figure~\ref{fig:lf}) 
are inconsistent with pure passive evolution without mergers. However, it remains plausible that the efficiency of central
galaxy growth via halo mergers continues to decline with increasing halo mass \markcite{coo05}(e.g., {Cooray} \& {Milosavljevi{\'c}} 2005). 
As the rate of major mergers with $>10^{14}~h^{-1}~M_\odot$ halos declines with decreasing redshift, the rate of massive 
galaxy growth will taper off. It is thus plausible that the space density of the most massive galaxies at low redshift 
can be {\it approximated} by pure passive evolution above some mass threshold \markcite{bun06,cim06,wak06}(e.g., {Bundy} {et~al.} 2006; {Cimatti} {et~al.} 2006; {Wake} {et~al.} 2006). 

Our analytic approximation of the HOD has a lower mass limit, so the cumulative luminosity function
of central red galaxies only marginally increases with decreasing luminosity at $L<0.2L^*$.
Halos far less massive than $10^{11.85}~h^{-1}~M_\odot$ presumably host blue central galaxies.
The color bimodality of galaxies also shows evidence for a transition mass \markcite{kau03}(e.g., {Kauffmann} {et~al.} 2003), 
which is broadly consistent with our results. If star formation were truncated as function of galaxy stellar
mass, one would not expect to observe red dwarf spheroidals. Thus, the bimodality
of galaxy colors may be better described by a truncation of central galaxy star-formation in halos
above a critical mass. Presumably there is a mechanism which prevents gas from cooling in high mass halos,
thus suppressing central galaxy star-formation.
The most plausible mechanisms in the current literature are virial shock 
heating \markcite{bir03,kho07,dek06}(e.g., {Birnboim} \& {Dekel} 2003; {Khochfar} \& {Ostriker} 2007; {Dekel} \& {Birnboim} 2006) and feedback from AGNs  \markcite{sil98,wyi03,cro06,hop06}(e.g., {Silk} \& {Rees} 1998; {Wyithe} \& {Loeb} 2003; {Croton} {et~al.} 2006; {Hopkins} {et~al.} 2006).

\subsection{STELLAR MASS IN CENTRAL AND SATELLITE GALAXIES}
\label{sec:satcen}

\begin{figure}
\plotone{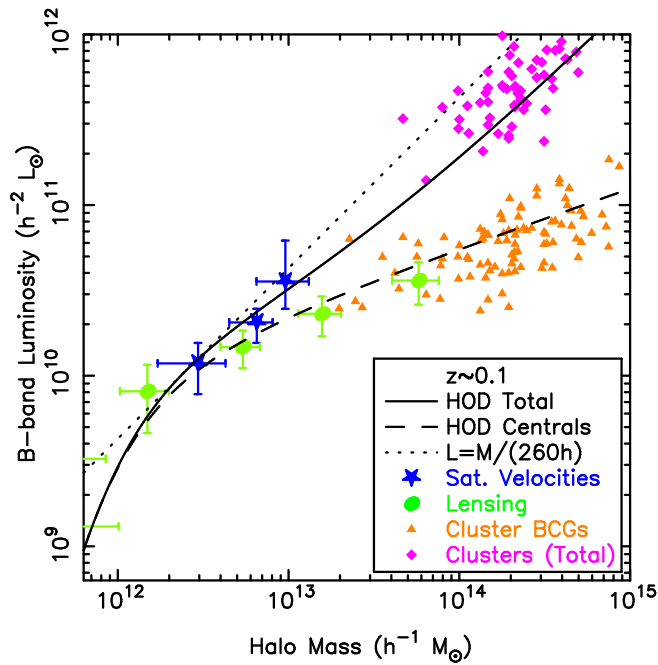}
\caption{The $B$-band luminosity per halo contributed by all and central red galaxies at $z=0.1$.
Measurements of the relationship between galaxy luminosity and halo mass, derived from weak lensing \markcite{man06a}({Mandelbaum} {et~al.} 2006b),
satellite galaxy velocities \markcite{con07a}({Conroy} {et~al.} 2007b) and galaxy clusters \markcite{lin04a,lin04b}({Lin} {et~al.} 2004; {Lin} \& {Mohr} 2004), are in broad agreement
with our analytic approximation of the HOD. (For clarity we do not show the uncertainties for individual clusters.)
The $B$-band luminosity of all red galaxies is not directly proportional to halo mass (i.e., $L=M/(260h)$), which may
indicate that stellar mass is lost from red galaxies during mergers with very massive halos \markcite{mon06,con07b,whi07}({Monaco} {et~al.} 2006; {Conroy} {et~al.} 2007c; {White} {et~al.} 2007).
\label{fig:ml}}
\end{figure}

In Figure~\ref{fig:ml} we plot the $B$-band luminosity contributed by red galaxies as a function
of halo mass at $z\sim 0.1$, determined using our analytic approximation of the HOD.
We also plot measurements of $B$-band luminosity versus mass, where the halo mass 
was determined with weak lensing \markcite{man06a}({Mandelbaum} {et~al.} 2006b), satellite galaxies \markcite{con07a}({Conroy} {et~al.} 2007b) and X-ray 
temperature \markcite{lin04a,lin04b}({Lin}, {Mohr}, \& {Stanford} 2004; {Lin} \& {Mohr} 2004). To include these measurements on our plot we assume
$B_{\rm Vega}-r_{\rm AB}=1.32$ and $B_{\rm Vega}-K_{\rm Vega}=4$. There is broad agreement between 
these measurements and our analytic approximation of the HOD, which we constrained with measurements 
of the space density and clustering of red galaxies. 

As we show in Figure~\ref{fig:ml}, the fraction of $B$-band light contributed by satellite
galaxies rapidly increases with halo mass. This is not unexpected, as the bulk of stellar
mass within clusters does not reside within the brightest cluster galaxy \markcite{san85,lin04a,lin04b}(e.g., {Sandage} {et~al.} 1985; {Lin} {et~al.} 2004; {Lin} \& {Mohr} 2004).
Clearly the growth of these halos did not result in stellar mass being efficiently funneled into the central galaxy.

In high mass halos the total $B$-band luminosity is proportional to halo mass to the power of $\simeq 0.9$ rather than $1$. 
This may indicate that mergers with very massive halos shift stellar mass out of the red galaxy population, perhaps into the 
ICL \markcite{gon00,arn02,fel04,zib05,mon06,whi07,con07c,con07b}(e.g., {Gonzalez} {et~al.} 2000; {Arnaboldi} {et~al.} 2002; {Feldmeier} {et~al.} 2004; {Zibetti} {et~al.} 2005; {Monaco} {et~al.} 2006; {White} {et~al.} 2007; {Conroy} {et~al.} 2007a, 2007c).

\subsection{THE EVOLVING LUMINOSITY DENSITY OF RED GALAXIES}
\label{sec:jb}

\begin{figure*}
\plotone{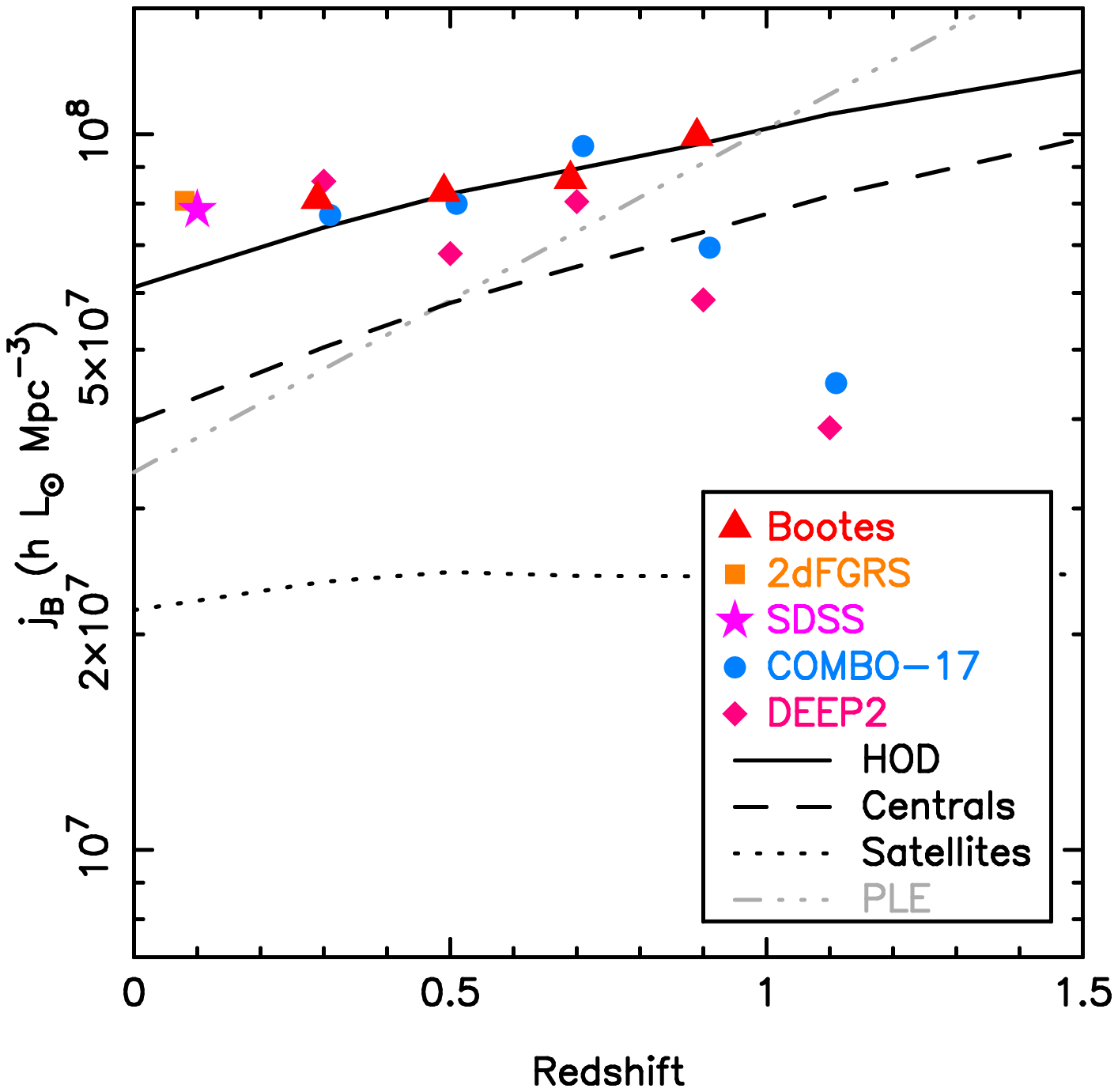}\plotone{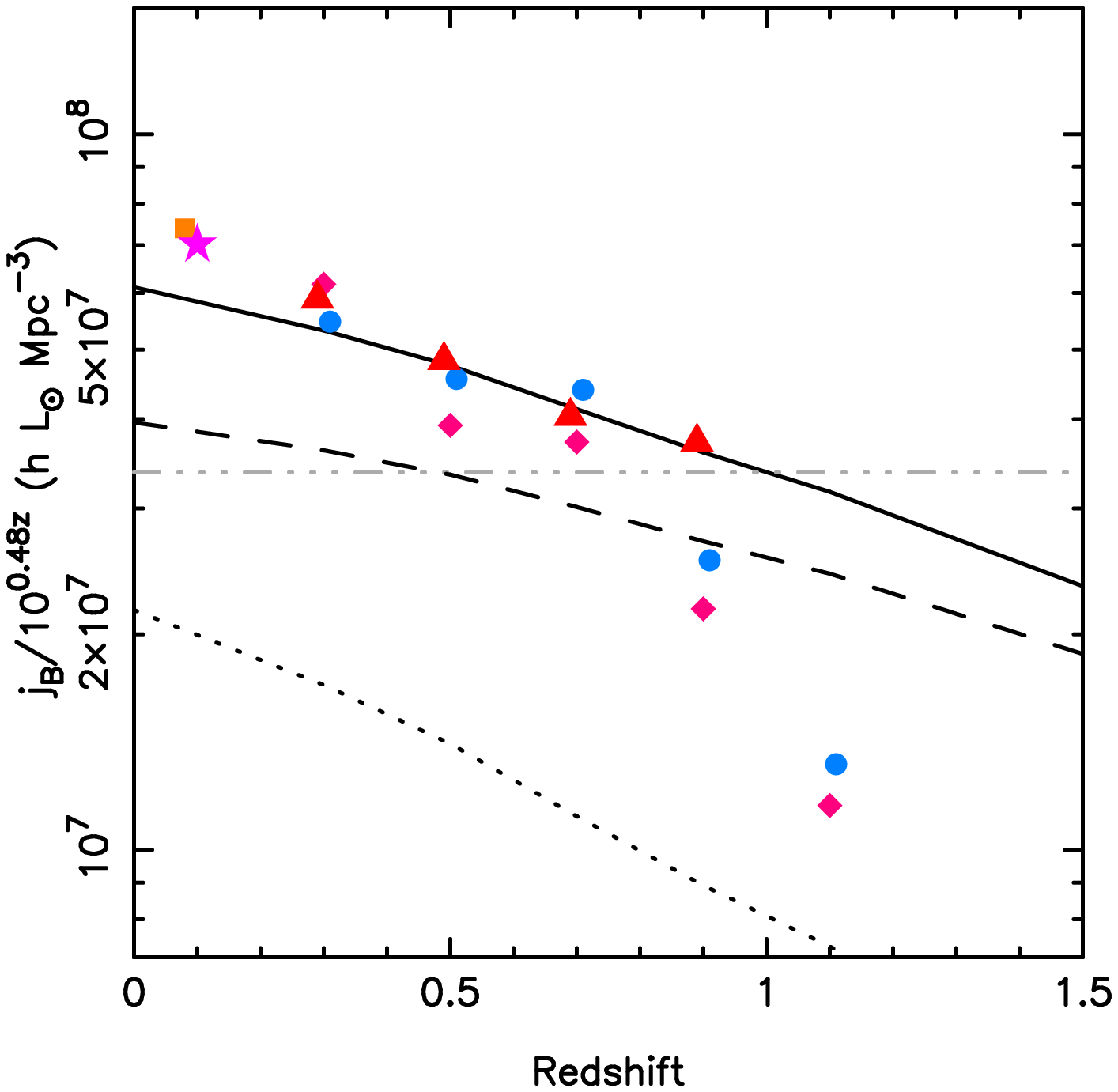}
\caption{The $B$-band luminosity density of red galaxies. The luminosity density values in the
right panel have been divided by $10^{0.4\times 1.2z}$, so a fixed stellar mass corresponds
to a horizontal line. Random uncertainties are on the order of 10\% while systematic errors
could be as high as 20\% at $z<0.8$. Pure luminosity evolution is ruled out by our observations
and those from the literature. Our approximation of the HOD is shown with the black solid line,
along with the contributions by central and satellite galaxies.
The stellar mass contained within the satellite population rapidly increases with time.
If the luminosity density of red galaxies rapidly declines at $z>0.8$ \markcite{bel04,wil06,fab07}({Bell} {et~al.} 2004; {Willmer} {et~al.} 2006; {Faber} {et~al.} 2007), the HOD must rapidly evolve.
\label{fig:jb}}
\end{figure*}

The luminosity weighted integral of the luminosity function, the luminosity density, is 
often used to estimate the evolution of the stellar mass within the red galaxy population.
In Figure~\ref{fig:jb} we plot $B$-band luminosity density measurements from Bo\"otes and the
literature \markcite{mad02,bel04,bla06,wil06,fab07}({Madgwick} {et~al.} 2002; {Bell} {et~al.} 2004; {Blanton} 2006; {Willmer} {et~al.} 2006; {Faber} {et~al.} 2007). Roughly 15\% of the luminosity density 
is contributed by galaxies fainter than our magnitude limits, and we have applied corrections
to the Bo\"otes measurements using our HOD approximation of the luminosity function.
While there are some discrepancies between the various surveys, which we discuss below, there 
is a broad consensus that the $B$-band luminosity density of red galaxies evolves slowly at $z<1$  \markcite{bel04,bro07,fab07}({Bell} {et~al.} 2004; {Brown} {et~al.} 2007; {Faber} {et~al.} 2007).

In Figure~\ref{fig:jb} we over-plot the evolving luminosity density derived from our analytic approximation of the HOD.
As the stellar populations of red galaxies fade between $z=1$ and $z=0$, a non-evolving luminosity density does not
correspond to a fixed stellar mass density. To simplify the interpretation of the luminosity density, in the 
right-hand panel of Figure~\ref{fig:jb} we divide the $B$-band luminosity density by $10^{0.4\times 1.2z}$, 
so the y-axis is proportional to stellar mass. One can clearly see that the stellar mass 
contained within the red population doubles between $z=1$ and $z=0$.
As red galaxies have low rates of star-formation, presumably this mass is being transferred from the
blue galaxies after a decline in star-formation \markcite{bel04,bro07,fab07}(e.g., {Bell} {et~al.} 2004; {Brown} {et~al.} 2007; {Faber} {et~al.} 2007).

In Figure~\ref{fig:jb} we show the contributions of central and satellite galaxies to the evolving 
luminosity density. If the stellar mass contained within the satellite and central populations remained fixed at 
their $z=1$ values, the $z=0$ luminosity density of both populations would be $\simeq 1.5 \times 10^7~h~L_\odot~{\rm Mpc}^{-3}$ lower 
than what we derive from HOD modeling. The stellar mass contained within the red satellite population triples between 
$z=1$ and $z=0$. The bulk of $z=0$ red satellites were either central or blue galaxies within the past $7~{\rm Gyr}$.

A naive interpretation of our results is that equal portions of new stellar mass are being transfered from the
blue population to the red central and satellite populations. Such an interpretation is risky, since satellites
can merge with central galaxies, and central galaxies can become satellites in groups
or clusters. One can imagine extreme scenarios where star-formation is
only truncated in central or satellite galaxies. More sophisticated models, 
which track the evolution of individual halos and galaxies over cosmic time will address
these issues, and we will discuss such a model in an upcoming paper (M.~White et~al. in prep.).

\subsection{COMPARISON WITH OTHER SURVEYS}
\label{sec:jbcomp}

\begin{figure*}
\plotone{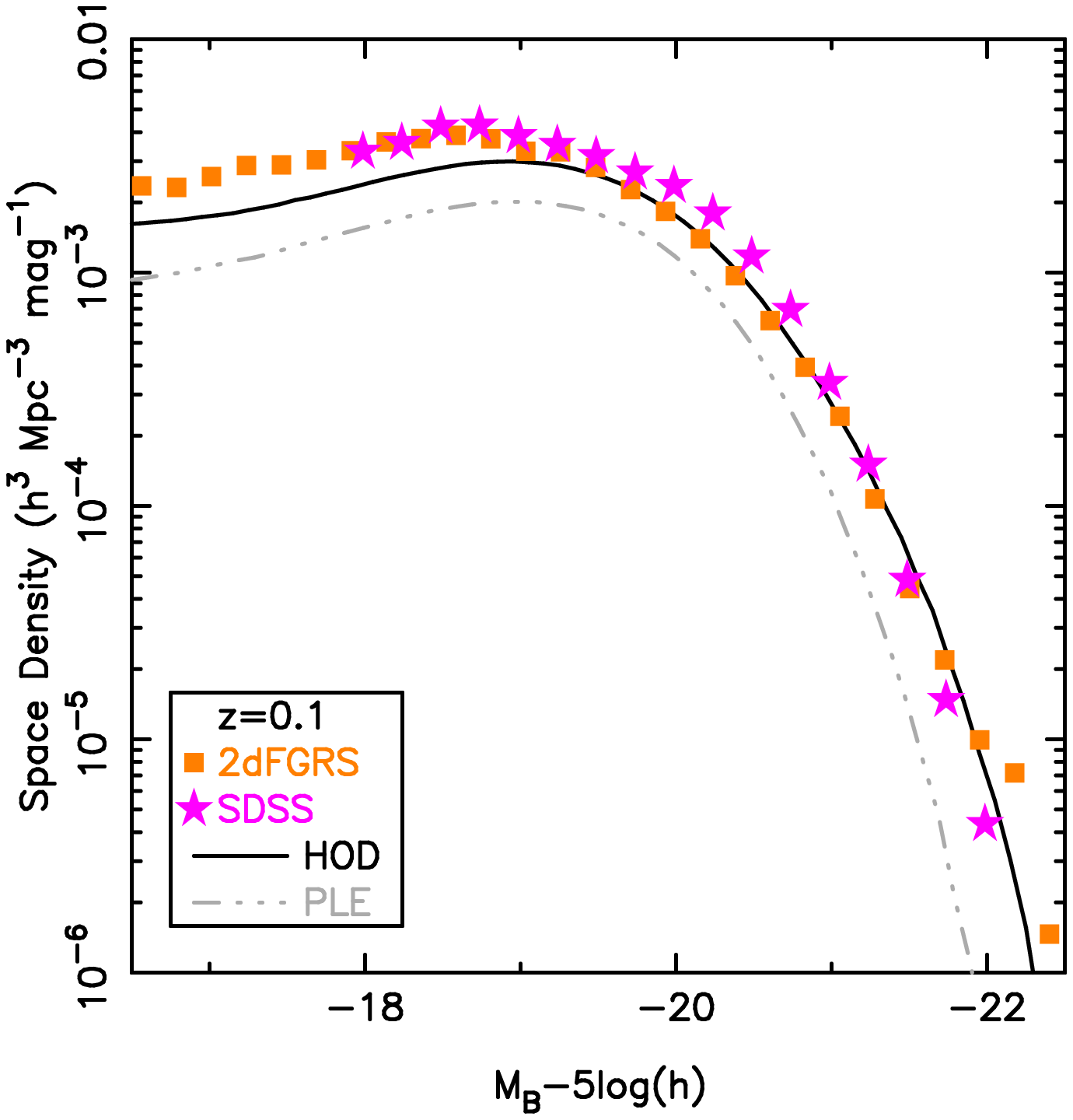}\plotone{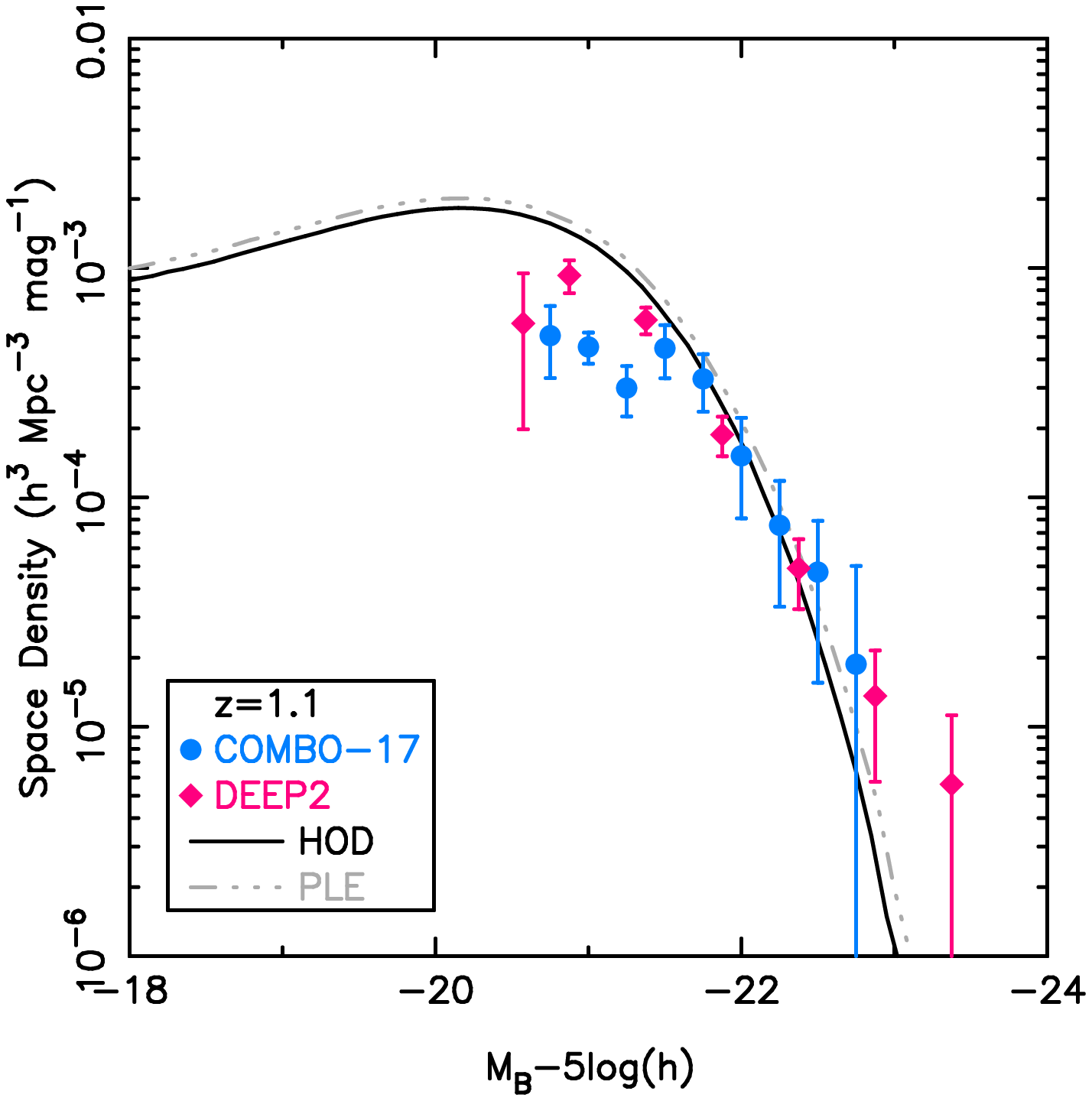}
\caption{Red galaxy luminosity functions for redshifts $z\simeq 0.1$ and $z\sim 1.1$ \markcite{mad02,bel04,bla06,wil06,fab07}({Madgwick} {et~al.} 2002; {Bell} {et~al.} 2004; {Blanton} 2006; {Willmer} {et~al.} 2006; {Faber} {et~al.} 2007).
For comparison, we also plot luminosity functions derived from our approximation of the HOD and a PLE model, which is normalized to
our $z=0.9$ data. For clarity we do not show the 2dFGRS and SDSS uncertainties, and we have
increased the DEEP2 luminosities by 15\% to correct for flux falling beyond the DEEP2 aperture \markcite{bro07}({Brown} {et~al.} 2007).
While our HOD model reproduces the bright end of the luminosity function, it fails for fainter galaxies where the
measured space density is sensitive to details of the galaxy selection criteria.
\label{fig:lzhz}}
\end{figure*}

In Figure~\ref{fig:lzhz} we plot measurements of the red galaxy luminosity function at low
and high redshift \markcite{mad02,bla06,bel04,wil06,fab07}({Madgwick} {et~al.} 2002; {Blanton} 2006; {Bell} {et~al.} 2004; {Willmer} {et~al.} 2006; {Faber} {et~al.} 2007) along with the predictions of our  HOD approximation 
and PLE. The PLE model, normalized to our $z\sim 0.9$ measurements, does not reproduce the red galaxy luminosity function.
The PLE model underestimates the space density of $\sim L^*$ red galaxies and underestimates the luminosities of the 
most massive galaxies at low redshift.

Our HOD model better approximates the galaxy luminosity function than a PLE model.
However, while our model reproduces the bright end of the luminosity function, there are systematic 
offsets at low luminosities. As the most luminous red galaxies lie along the red sequence, their measured space density is 
insensitive to details of the galaxy selection criteria. In contrast, there is a significant
population of $\lesssim L^*$ galaxies with colors falling between the red and blue galaxy populations.
By shifting our rest-frame $U-V$ selection criterion blueward by just 0.1 magnitudes, the measured space 
density of $L^*$ red galaxies increases by 25\% while the measured $r_0$ values decrease by 10\%.
Such a shift would remove the offset between our HOD approximation and the 2dFGRS and SDSS measurements.

\begin{figure}
\plotone{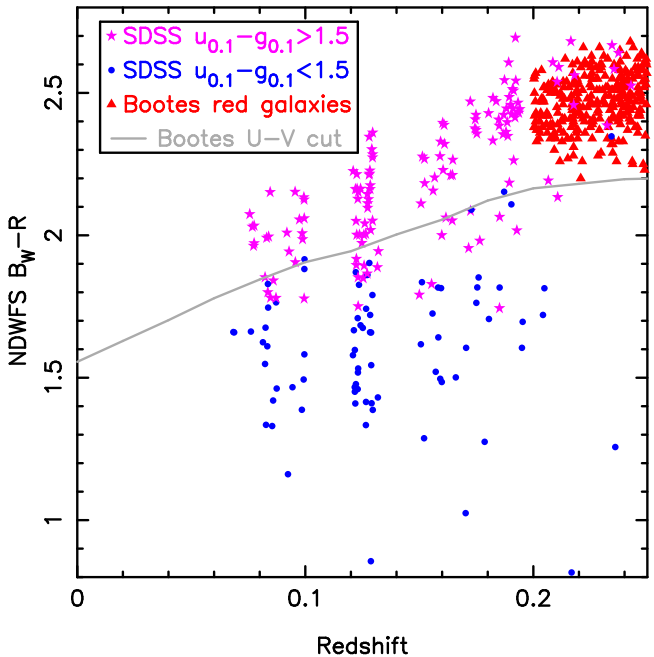}
\caption{NDWFS $B_W-R$ colors of $L>L^*$ red galaxies from the SDSS and this work.
We use a $8^{\prime\prime}$ and $4^{\prime\prime}$ diameter apertures to measure the $B_W-R$
colors of the SDSS and Bo\"otes galaxies respectively, so the physical aperture size is comparable for both surveys.
The SDSS $0.05<z<0.15$ red galaxy sample of \markcite{bla06}{Blanton} (2006), selected using a rest-frame $u_{0.1}-g_{0.1}>1.5$ (AB) criterion,
is systematically bluer than the evolving rest-frame $U-V$ criterion used to select $z>0.2$ red galaxies in ~Bo\"otes.
We thus expect the SDSS to measure a systematically higher space density of $\sim L^*$ red galaxies than ~Bo\"otes.
\label{fig:bwr}}
\end{figure}

As the SDSS includes the Bo\"otes field, we can directly compare our red galaxy sample with that of 
\markcite{bla06}{Blanton} (2006).  As we show in Figure~\ref{fig:bwr}, the \markcite{bla06}{Blanton} (2006) red galaxy sample 
includes systematically bluer galaxies than our sample. This explains why \markcite{bla06}{Blanton} (2006) measures
a higher space density of $<L^*$ red galaxies than Bo\"otes.
The rest-frame $u_{0.1}-g_{0.1}$ criterion of \markcite{bla06}{Blanton} (2006) may be bluer than our $U-V$ criterion,
although other factors could be significant including the choice of galaxy SED templates and the 
method used to measure galaxy photometry.

It is unclear if the HOD undergoes rapid evolution at $z\sim 1$. We find a 
significantly higher space density of $z\sim 0.9$ red galaxies than either COMBO-17 or DEEP2. 
If the DEEP2 or COMBO-17 measurements of the $z\sim 1$ luminosity function are valid,
the HOD must rapidly evolve at $z\sim 1$. As we noted previously, we do expect some 
evolution of the HOD as there is residual star-formation within the red 
galaxy population and the properties of dark matter halos of a given mass evolve.
Why this evolution should rapidly accelerate at $z\sim 1$ is unclear.

It is plausible that COMBO-17 and DEEP2 are underestimating the space density of $z\sim 1$ red galaxies.
DEEP2 uses a non-evolving selection criterion for red galaxies, and this criterion must
be redder than the evolving colors of an aging stellar population at high redshift.
We thus expect DEEP2 to underestimate the space density of red galaxies beyond some (as yet undetermined) redshift. 
There is an ongoing collaboration between the Bo\"otes  and DEEP2 surveys to resolve this discrepancy, and
we have obtained imaging of the DEEP2 Extended Groth Strip with the NDWFS filter-set to determine the 
impact of photometric errors on current luminosity function measurements.
Unfortunately, COMBO-17 optical photometric redshifts have gross errors for galaxies beyond
$z=1$ (E.~F. Bell, N. Taylor - private communication), resulting in a spuriously rapid decline in the space 
density of red galaxies. This systematic error is being addressed with the addition of near-infrared photometry 
to the COMBO-17 photometric redshifts. At present, it remains plausible that the relationship between galaxy stellar
mass and host halo mass evolves slowly out to $z \sim 1.5$, and there is a clear need for robust space density and clustering measurements
to test this conjecture.

\section{SUMMARY}
\label{sec:summary}

We have measured the past $7~{\rm Gyr}$ of red galaxy growth within dark matter halos. To do this, we have measured the halo occupation
distribution, which describes how galaxies reside within dark matter halos (as a function of halo mass). We have
constrained HOD models using measurements of the evolving luminosity function and clustering of $0.2<z<1.0$
red galaxies. Our sample of 40,696 red galaxies, selected from $6.96~{\rm deg}^2$ of imaging in Bo\"otes, 
is far larger than comparable galaxy samples at these redshifts.

We have measured the evolving luminosity function of red galaxies, and obtain results that are very similar to \markcite{bro07}{Brown} {et~al.} (2007).
The bright end of the luminosity function fades with time, indicating that the most
massive galaxies are not rapidly growing via mergers. The bright end of the luminosity function
does not fade as rapidly as an aging stellar population, and we conclude that the stellar masses
of the most massive galaxies have grown by 30\% over the past $7~{\rm Gyr}$. We find that the luminosity density of 
red galaxies evolves slowly with redshift, and the stellar mass contained within the 
red galaxy population has doubled between $z=1$ and $z=0$. As there is little star-formation within the 
red galaxy population, stellar mass must have been transfered from the blue galaxy population at $z<1$.

We have measured the angular correlation function of $0.2<z<1.0$ red galaxies, using 
samples selected as a function of photometric redshift, luminosity and space density. 
We evaluated power-law and HOD models of the spatial correlation function by using
the \markcite{lim54}{Limber} (1954) equation to determine corresponding angular correlation functions.
To evaluate the quality of the models, we determined $\chi^2$ values using the complete 
covariance matrices for the angular correlation function. Preliminary covariance matrices
were determined using analytic approximations while our final results were determined
using covariance matrices derived from mock catalogs. Our mock catalogs reproduce the luminosity
and correlation functions of red galaxies, and include large-scale structures that are frequently observed
in deep galaxy surveys.

We find that the value of $r_0$ increases with luminosity for red galaxies brighter than 
$M_B-5{\rm log}h = -20$, while for fainter galaxies $r_0\simeq 5~h^{-1}~{\rm Mpc}$.
While the spatial clustering of red galaxies as a function of luminosity clearly evolves, 
this is largely due to the fading of aging stellar populations rather than the evolution
of large-scale structures.

We performed HOD modeling of the observed luminosity function and angular clustering of  Bo\"otes red galaxies.
The large-scale bias factor increases with luminosity for red galaxies brighter than  $M_B-5{\rm log}h = -20$.
We find that the bias of red galaxies increases with redshift, as the spatial clustering 
of galaxies does not evolve as rapidly as the underlying distribution of dark matter.
The evolution of the bias factor is inconsistent with pure luminosity evolution models 
without mergers \markcite{whi07}({White} {et~al.} 2007), where the spatial clustering increases with time due to gravitational collapse.

The relationship between red galaxy luminosity and host halo mass evolves, due to the fading of aging 
stellar populations. We do not observe any evolution of the relationship between red galaxy 
stellar mass and host halo mass with redshift. As halo masses grow at $z<1$, we expect the 
stellar masses of central galaxies to also grow at $z<1$.

We find that the HOD of red central galaxies has a lower limit, with 50\% of  $10^{11.9}~h^{-1}~M_\odot$ mass halos 
hosting a red central galaxy. As one moves down further in halo mass, one presumably observes an 
increasing fraction of blue central galaxies. What truncates star-formation in central galaxies 
is unclear, but it appears to be strongly correlated with host halo mass.

We can reproduce both the luminosity function and clustering of red galaxies in Bo\"otes with an 
analytic approximation of the HOD, which also simplifies the interpretation our of HOD
measurements. In our model, the fraction of stellar mass within the satellite population
increases with host halo mass. In cluster ($>10^{14}~h^{-1}~M_\odot$) mass halos, more mass resides
within satellites than central galaxies. We thus conclude that mergers with these halos
do not always (efficiently) funnel stellar mass into central galaxies. 
The luminosities of central galaxies are proportional to halo mass to the power of $\simeq 0.35$.
As a result, if halo masses increase by 100\% at $z<1$, the stellar mass of the typical
central galaxy grows by only 30\% over the same redshift range. Massive galaxy assembly continues
at $z<1$, but the bulk of their growth took place at higher redshifts.

The growth of stellar mass within the red population at $z<1$ is reproduced by our analytic approximation
of the HOD. The stellar mass contained within red satellite galaxies triples between $z=1$ and $z=0$, 
and they account for a third of the stellar mass within the red population by $z\sim 0$. Our analytic approximation of the 
HOD reproduces current measurements of the bright-end of the $z>1$ red galaxy luminosity 
function, but there are offsets at lower luminosities. Either measurements of the $z\gtrsim 1$ red galaxy 
luminosity function have large systematic errors, or the red galaxy HOD rapidly evolves at $z\gtrsim 1$.

\acknowledgments

We thank our colleagues on the NDWFS, IRAC Shallow Survey, and AGES teams,
in particular R.~J. Cool, P.~R. Eisenhardt, D.~J. Eisenstein, G.~G. Fazio, C.~S. Kochanek, and G.~P. Tiede.
This paper would not have been possible without the efforts of the  KPNO, {\it Spitzer}, MMT,
W.~M.~Keck and {\it Gemini} support staff.
We are grateful to the IRAF team for the majority of the packages used to process the NDWFS images.
We thank Alyson Ford, Lissa Miller, and Jennifer Claver, for processing much of the Bo\"otes optical imaging.
H. Spinrad, S. Dawson, D. Stern, J.~E. Rhoads, S Malhotra, B.~T. Soifer, C. Bian, S.~G. Djorgovski, S.~A. Stanford,
S.~Croft, W.~van~Breugel and the AGES collaboration generously shared their spectroscopic redshifts with us prior to publication.
Several of the key ideas presented in this paper were developed during summer workshops of the Aspen Center for Physics, 
who we thank for their hospitality.
This work is based in part on observations made with the {\it Spitzer} Space Telescope, which is operated
by the Jet Propulsion Laboratory, California Institute of Technology
under a contract with NASA. This research was supported by the National Optical Astronomy Observatory which is
operated by the Association of Universities for Research in Astronomy (AURA), Inc.
under a cooperative agreement with the National Science Foundation.
The simulations were performed on the supercomputers at the National
Energy Research Scientific Computing center.
At an early stage of this work, ZZ was supported by NASA through Hubble 
Fellowship grants HF-01181.01-A, awarded by the Space Telescope Science 
Institute, which is operated by the Association of Universities for Research 
in Astronomy, Inc., for NASA, under contract NAS 5-26555. ZZ gratefully
acknowledges support from the Institute for Advanced Study through a John
Bahcall Fellowship.
While writing this paper we had many productive discussions with other astronomers working upon 
galaxy assembly and evolution, including E.~F. Bell, M.~R.~Blanton, S.~Brough, A.~L.~Coil, C.~Conroy, S.~M.~Faber,
J.~E.~Gunn, T.~R.~Lauer, J.~A. Newman, P. Norberg, J.~P. Ostriker, N.~P. Ross, A.~E.~Schulz, R. Sheth,  F. van~den~Bosch, 
M.~S. Vogeley, D.~.A.~Wake, C.~N.~A. Willmer, C.~Wolf, and I.~Zehavi.


\appendix

\section{AN ANALYTIC APPROXIMATION FOR THE COVARIANCE MATRIX}
\label{sec:anacov}

Our final estimates of the angular correlation function covariance matrices
are derived from mock catalogs.  However, one needs estimates of the HOD to
generate such mocks, so preliminary fits of HOD models to the angular
correlation function utilized analytic approximations of the covariance
matrices.
The basis of our analytic covariance matrices is the Gaussian approximation
\markcite{eis01}(e.g., {Eisenstein} \& {Zaldarriaga} 2001):
\begin{equation}
  C_\omega(\theta_i, \theta_j) = \frac{1}{\pi \Omega^2}
  \int_0^\infty K\,dK\ P^2_2(K) J_0(K\theta_i) J_0(K\theta_j)
\label{eq:ez}
\end{equation}
where $J_0$ is a Bessel function and $P_2(K)$ is the (2D) angular power
spectrum,
\begin{equation}
  P_2(K) = 2\pi \int_0^\infty \theta\,d\theta\ w(\theta) J_0(K\theta)
\end{equation}
This approximation is best suited to correlation functions where
$\omega(\theta)\ll 1$ and underestimates the covariance of very strongly
clustered objects.
For a power-law power spectrum, $\omega(\theta)\propto\theta^{-a}$ or
$P_2(K)\propto K^{a-2}$, the integral can be evaluated analytically
in terms of $\bar{\theta}\equiv\sqrt{\theta_i\theta_j}$ and
$r^2=\theta_i/\theta_j\le 1$:
\begin{equation}
  C_\omega(\bar{\theta},r) = \frac{(\bar{\theta}/r)^{2(1-a)}}{\pi\Omega^2}
  2^{2a-3} \frac{\Gamma(a+1)}{\Gamma(2-a)}
  \ {}_2F_1\left(a-1,a-1,1,r^4\right)
\end{equation}
valid for $1<a<3/2$.  The power-series expansion of the (confluent)
hypergeometric function, ${}_2F_1$, converges rapidly.
For cases where the power-law is (nearly) divergent we truncate the integral
at an angular scale corresponding to a transverse comoving distance of
$150\,h^{-1}$Mpc.  In practice, fits to our data are only marginally affected
by the value of the angular correlation function on scales of more than
$1^\circ$.  If the angular correlation function bins have significant
width, Equation~\ref{eq:ez} is modified to
\begin{equation}
C_\omega(\theta_i, \theta_j) =
\left(\frac{2}{\theta_{i,2}^2-\theta_{i,1}^2}\right)
\left(\frac{2}{\theta_{j,2}^2-\theta_{j,1}^2}\right)
\int^{\theta_{i,2}}_{\theta_{i,1}}\theta\,d\theta
\int^{\theta_{j,2}}_{\theta_{j,1}}\theta^\prime\,d\theta^\prime
C_\omega(\theta,\theta^\prime)
\end{equation}
(D. Eisenstein 2003, private communication) where $\theta_1$ and $\theta_2$
are the inner and outer radii of the bins.  This can be rewritten as the
single integral
\begin{eqnarray}
C_\omega(\theta_i, \theta_j) =
\frac{4}{(\theta_{i,2}^2-\theta_{i,1}^2)
          (\theta_{j,2}^2-\theta_{j,1}^2)\pi\Omega^2}
\times \nonumber \\
\int^\infty_0  P_2^2(K)
\left[\theta_{i,2}J_1(K\theta_{i,2})-\theta_{i,1}J_1 (K\theta_{i,1})\right]
\left[\theta_{j,2}J_1(K\theta_{j,2})-\theta_{j,1}J_1(K\theta_{j,1})\right]
\frac{dK}{K}.
\end{eqnarray}
The contribution of shot noise to the estimate of the covariance was included
by adding the reciprocal of the sky surface density of galaxies (per steradian)
to $P_2(K)$.  However, the shot noise only dominates the covariance on scales
of $\lesssim 1^\prime$ for the red galaxy sample.

The analytic covariance matrix of \markcite{eis01}{Eisenstein} \& {Zaldarriaga} (2001) underestimates the uncertainties
when $\omega(\theta)\gtrsim 1$.  To mitigate this issue, we approximate the
covariance matrix with,
\begin{eqnarray}
C_\omega^\prime (\theta_i, \theta_j) =
  C_\omega(\theta_i, \theta_j) + 0.01 w(\theta_j)^{2.5} & {\rm where} j \geq i
\end{eqnarray}
where $C_\omega$ is determined using the method of \markcite{eis01}{Eisenstein} \& {Zaldarriaga} (2001) and $\omega(\theta_j)$
is determined using the truncated power-law discussed earlier.
The second term of this equation was determined by trial and error, and roughly approximates the near-diagonal elements of the Bo\"otes  covariance
matrices derived from jack-knife subsamples and mock catalogs. While this term is a hack and should be treated with caution (and perhaps contempt),
it does prevent highly correlated $\omega(\theta)$ measurements on small angular scales from skewing preliminary fits of power-laws and HODs to our data.

\begin{figure}[t]
\plotone{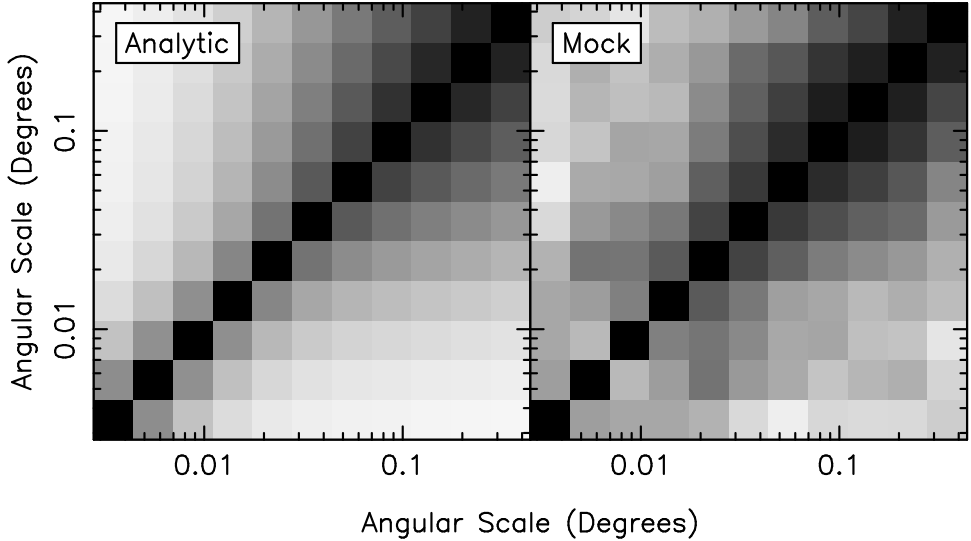}
\caption{Analytic and mock catalog correlation matrices for the $0.6<z<0.8$ $M_B-5{\rm log}h<-19.5$ red galaxy sample.
The correlation matrix is defined by $C_\omega(\theta_i, \theta_j)/\sqrt{C_\omega(\theta_i, \theta_i)C_\omega(\theta_j, \theta_j)}$, where
$C_\omega$ is the covariance matrix of the angular correlation function. The near-diagonal elements of the two covariance
matrices are in rough agreement, although the off diagonal elements show significant discrepancies.
Despite this, our principal results do not change significantly if we switch from one set of covariance matrices to the other.
\label{fig:cor}}
\end{figure}

In Figure~\ref{fig:cor} we plot the analytic and mock correlation matrices
for one of our red galaxy samples. While the diagonal elements of the
covariance matrices are comparable, the off-diagonal elements
of the analytic covariance matrix are systematically less than
those determined with mock galaxy catalogs. The uncertainties of
power-law fits to the binned angular correlation function increase by up to 100\% when
we switch from analytic to mock covariance matrices.
However, HOD parameters exhibit little change when we switch covariance matrices, 
as the HOD is constrained by the dark matter distribution, halo mass function and observed galaxy 
space density as well as the observed clustering of galaxies.

\section{MOCK CATALOGS} 
\label{sec:mock}

We used mock galaxy catalogs to estimate the covariance matrix of the correlation function
and the uncertainties for luminosity functions. We utilized a simulation which is very similar
to that of \markcite{whi07}{White} {et~al.} (2007), but with a larger volume and lower mass resolution.
The same cosmology is used as throughout this paper: 
$\Omega_m=0.25=1-\Omega_\Lambda$, $\Omega_b=0.043$, $h=0.72$, $n_s=0.97$ and $\sigma_8=0.8$.
As in \markcite{whi07}{White} {et~al.} (2007), the linear theory power spectrum for the initial
conditions was computed by evolution of the coupled Einstein, fluid and
Boltzmann equations using the code described in \markcite{whi96}{White} \& {Scott} (1996). 
\markcite{SSWZ}{Seljak} {et~al.} (2003)  find that this code agrees well with CMBfast \markcite{CMBfast}({Seljak} \& {Zaldarriaga} 1996).
The simulation employed $1024^3$ particles of mass $10^{11}\,h^{-1}~M_\odot$
in a periodic cube of side $1\,h^{-1}$Gpc using a {\sl TreePM\/} code
\markcite{whi02}({White} 2002) with a Plummer-equivalent softening length of $35\,h^{-1}~$kpc (comoving).
A detailed comparison of this {\sl TreePM\/} code with other codes can be found in \markcite{hei07}{Heitmann} {et~al.} (2007).

We used a series of simulation outputs at $z<1$ and for each output we generate a 
halo catalogs using the Friends-of-Friends (FoF) algorithm \markcite{DEFW}({Davis} {et~al.} 1985) with a 
linking length of 0.168 times the mean inter-particle spacing.
This procedure partitions the particles into equivalence classes, by linking
together all particle pairs separated by less than a distance $b$.
The halos correspond roughly to particles with
$\rho>3/(2\pi b^3)\simeq 100$ times the background density.
Our mass definition uses the sum of the particle masses in the halo,
however to obtain better correspondence between our definition of halo mass
and that implicitly defined by the mass functions of \markcite{she99}{Sheth} \& {Tormen} (1999) and
\markcite{jen01}{Jenkins} {et~al.} (2001) we rescaled the masses by
$M/M_{\rm fof}=1+0.01\left( \ln M_{\rm fof}-23.5 \right)$ where $M_{\rm fof}$
is the FoF mass in units of $h^{-1}M_\odot$.  With this redefinition the mass
function in the simulation lies between those of \markcite{she99}{Sheth} \& {Tormen} (1999) and
\markcite{jen01}{Jenkins} {et~al.} (2001), differing from them by less than 10\% in the mass range
of interest.

Mock catalogs were constructed by using an analytic approximation 
of the HOD (\S\ref{sec:analytic}) to assign central and satellite galaxies to halos.
We determined the mean number of central red galaxies as a function of halo mass and 
used a random number generator to determine if a central galaxy was assigned to 
a particular halo. We determined the mean number of satellite galaxies as a function
of halo mass, and used a random number generator and Poisson statistics to determine 
the number of satellites assigned to each halo. To determine the likelihood of a galaxy 
having a particular luminosity, we used conditional luminosity functions (number of galaxies
per magnitude per halo) for central and satellite galaxies. These luminosity functions 
are a function of host halo mass and were derived from the cumulative
luminosity functions provided by the analytic approximation of 
the HOD. Central galaxies were placed precisely at the center of dark matter halos 
while the distribution of satellite galaxies follows a spherical NFW profile.
Random number generators were used to assign galaxy luminosities and satellite positions
relative to halo centers.

\begin{figure}
\plotone{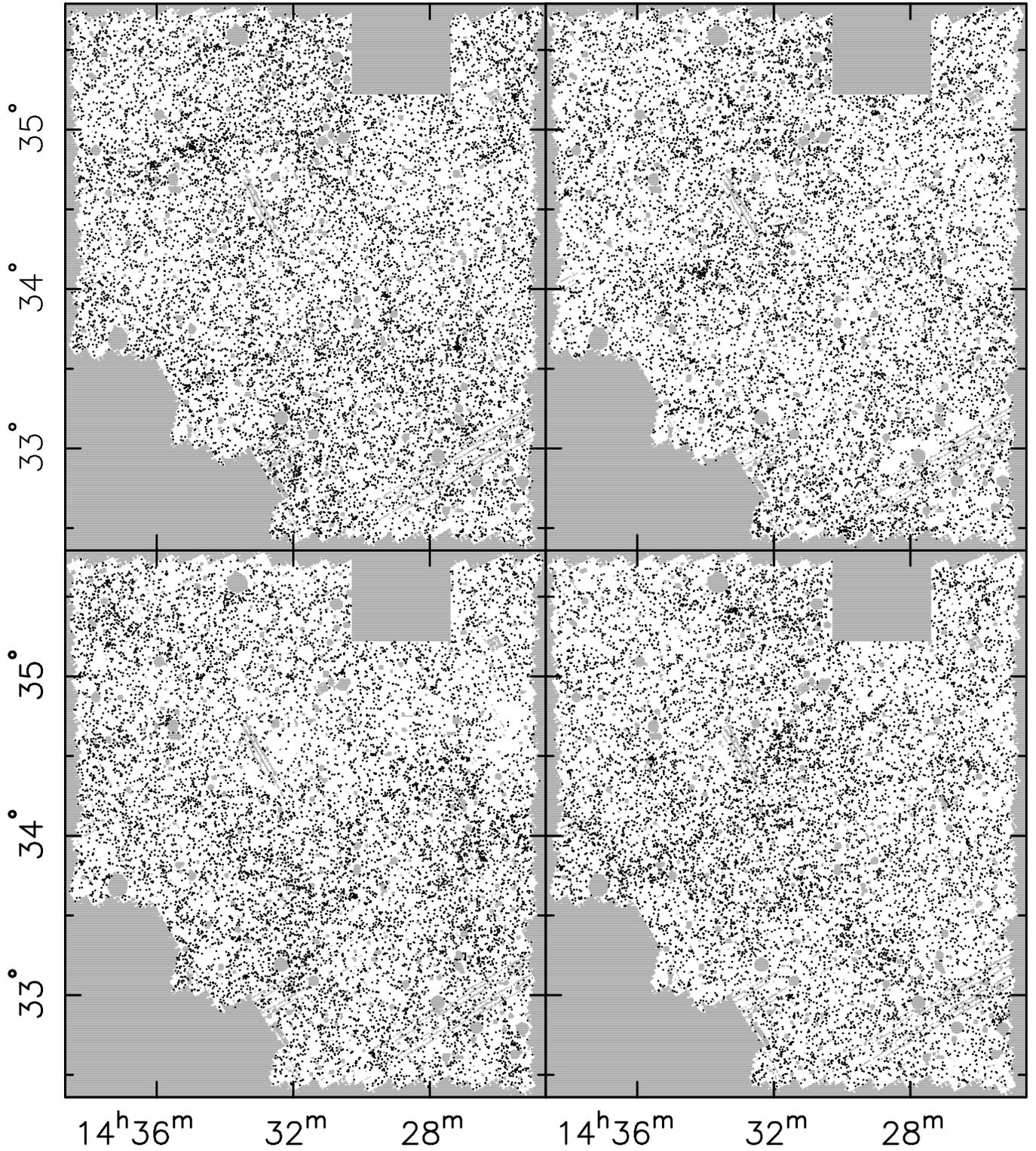}
\caption{The sky distribution of $0.8<z<1.0$ red galaxies in ~Bo\"otes  (top-left) and three mock catalogs.
Individual large-scale-structures are evident in both ~Bo\"otes and the mocks,
and are often found in deep galaxy surveys  \markcite{lub00,nak05,guz07}(e.g., {Lubin} {et~al.} 2000; {Nakata} {et~al.} 2005; {Guzzo} {et~al.} 2007).
While such high redshift structures were once thought to be rare, this assumed that massive galaxies
and large-scale structure rapidly evolved at $z<1$.
Although one can imagine these structures grossly bias both luminosity function and clustering measurements,
the number counts of the mock catalogs have a standard deviation of only 8\%.
\label{fig:skc}}
\end{figure}

The $1~h^{-3}~{\rm Gpc}^3$ simulation was used to produce between 56 ($z\sim 0.9$) and 400($z\sim 0.3$) 
~Bo\"otes sized samples. The simulation cube was viewed parallel to one of the Cartesian axes,
and each galaxy was assigned a spectroscopic redshift and corresponding apparent magnitude. 
Galaxies were then assigned photometric redshifts, assuming Gaussian random errors and the 
photometric redshift uncertainties reported by \markcite{bro07}{Brown} {et~al.} (2007), and the absolute 
magnitudes were scattered accordingly. In Figure~\ref{fig:skc} we show the 
sky distribution of three $0.8<z<1.0$ red galaxy mocks and the real $0.8<z<1.0$ red
galaxy catalog. Individual large-scale structures at $z\sim 0.9$ are seen in both the 
~Bo\"otes field and our mock catalogs, and are relatively common in other deep 
surveys  \markcite{lub00,nak05,guz07}(e.g., {Lubin} {et~al.} 2000; {Nakata} {et~al.} 2005; {Guzzo} {et~al.} 2007). While large structures of high redshift galaxies were 
once expected to be rare, this assumed rapid galaxy growth took place at $z<1$.

Angular correlation functions were determined for each ~Bo\"otes sized mock using the \markcite{lan93}{Landy} \& {Szalay} (1993) estimator,
and these were used to derive covariance matrices. These covariance matrices include 
both sample variance and shot-noise without needing to assume the underlying 
fluctuations are Gaussian or higher-order moments follow some particular pattern.
We also use the mocks to estimate the uncertainties for the galaxy luminosity 
functions.

\begin{figure}
\plotone{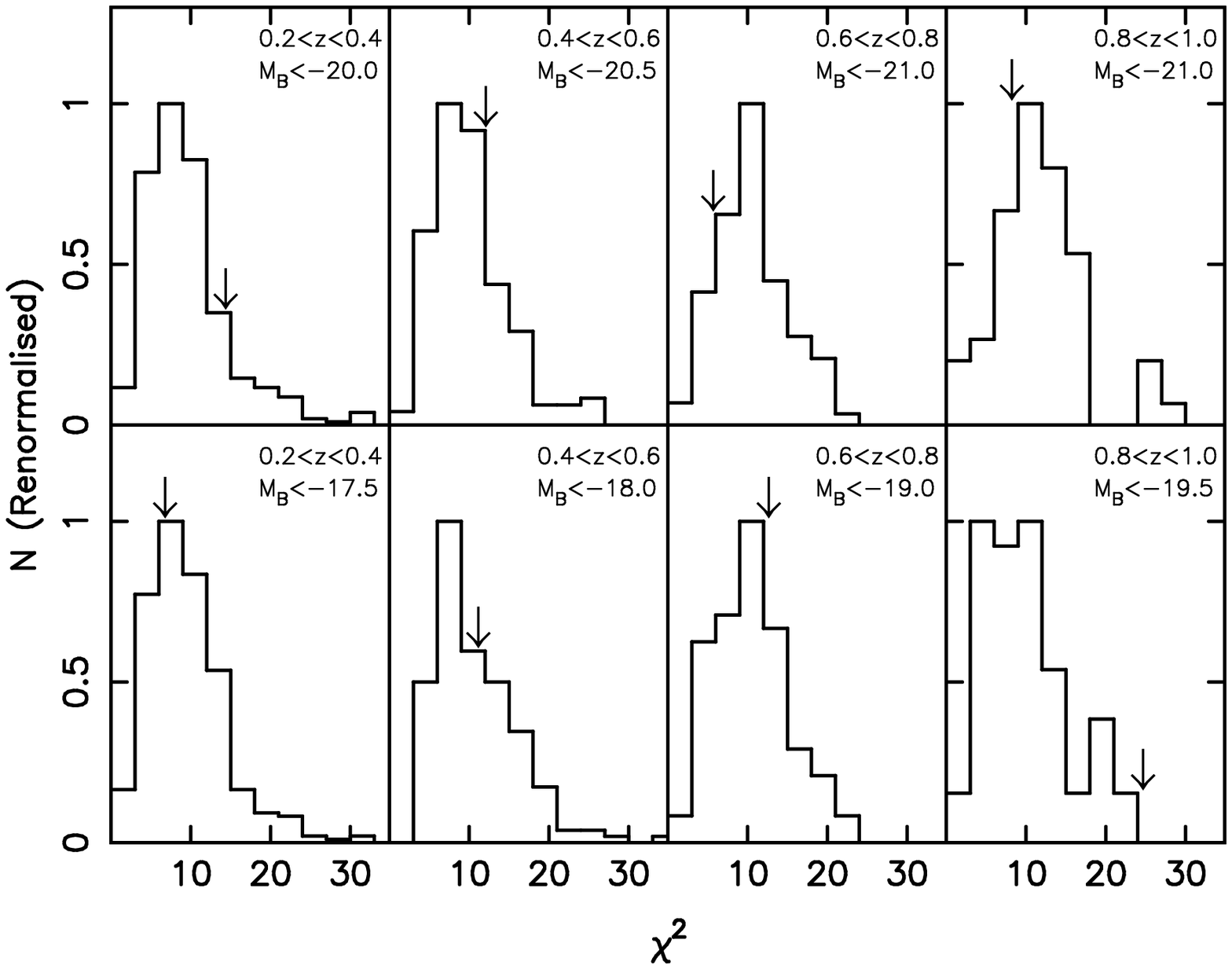}
\caption{The distribution of $\chi^2$ values for mock catalog correlation functions. The $\chi^2$ values were determined
using the angular correlation function measured from each mock and the analytic approximation of the HOD. Values of $\chi^2$
for the Bo\"otes  field were determined in the same manner, and are shown with arrows. With the possible exception
of the faintest red galaxies in our highest redshift bin, the Bo\"otes $\chi^2$ values are consistent with those derived
from mocks. As we discuss in Appendix~\ref{sec:mock}, the discrepancy between the $0.8<z<1.0$ data and the mocks may
be due to selection effects rather than an error in our approximation of the HOD.
\label{fig:wchi}}
\end{figure}

Are the mock catalogs consistent with the observed clustering of red galaxies in Bo\"otes?
To verify this we used the analytic approximation of the HOD to determine $\chi^2$ values for 
the angular correlation functions measured from each mock. We then determined
$\chi^2$ in the same manner using the real Bo\"otes field. In Figure~\ref{fig:wchi} we plot the 
distribution of $\chi^2$ values for the mocks and Bo\"otes. With the possible exception of the 
faintest red galaxies in our highest redshift bin, the Bo\"otes  $\chi^2$ values are comparable
to those derived from the mocks.
The analytic approximation of the HOD slightly underestimates the space density and overestimates the 
clustering of the faintest $0.8<z<1.0$ red galaxies. This discrepancy may be caused by selection effects, as a 
small shift in the rest-frame selection criteria can alter the measured space density and clustering 
of $<L^*$ red galaxies by $\sim 10\%$. As we discuss in \S\ref{sec:jb}, such selection effects may 
contribute to the large scatter of $z>0.8$ luminosity density measurements reported in the 
current literature. 

\bibliography{}

\clearpage

\begin{deluxetable}{ccccc}
\tablecolumns{5}
\tabletypesize{\scriptsize}
\tablecaption{Red Galaxy $1/V_{\rm max}$ Luminosity Function with mock catalog uncertainties.\label{table:vmax}}
\tablehead{ 
\colhead{Absolute}            & 
\multicolumn{4}{c}{Luminosity Function ($h^3~{\rm Mpc^{-3}~mag^{-1}}$)} 
\\  
  \colhead{Magnitude}           & 
  \colhead{$0.2<z<0.4$}         &
  \colhead{$0.4<z<0.6$}         &
  \colhead{$0.6<z<0.8$}         &
  \colhead{$0.8<z<1.0$}         
}
\startdata
$ -17.75 < M_B - 5 {\rm log} h < -17.50 $ & $ 1.90 \pm 0.27 \times 10^{-3} $ & -                                & -                                & -                                \\ 
$ -18.00 < M_B - 5 {\rm log} h < -17.75 $ & $ 1.82 \pm 0.26 \times 10^{-3} $ & -                                & -                                & -                                \\ 
$ -18.25 < M_B - 5 {\rm log} h < -18.00 $ & $ 1.92 \pm 0.27 \times 10^{-3} $ & $ 1.68 \pm 0.16 \times 10^{-3} $ & -                                & -                                \\ 
$ -18.50 < M_B - 5 {\rm log} h < -18.25 $ & $ 2.30 \pm 0.32 \times 10^{-3} $ & $ 1.89 \pm 0.19 \times 10^{-3} $ & -                                & -                                \\ 
$ -18.75 < M_B - 5 {\rm log} h < -18.50 $ & $ 2.43 \pm 0.34 \times 10^{-3} $ & $ 2.01 \pm 0.20 \times 10^{-3} $ & -                                & -                                \\ 
$ -19.00 < M_B - 5 {\rm log} h < -18.75 $ & $ 2.67 \pm 0.35 \times 10^{-3} $ & $ 2.37 \pm 0.25 \times 10^{-3} $ & -                                & -                                \\ 
$ -19.25 < M_B - 5 {\rm log} h < -19.00 $ & $ 2.81 \pm 0.36 \times 10^{-3} $ & $ 2.53 \pm 0.25 \times 10^{-3} $ & $ 1.88 \pm 0.16 \times 10^{-3} $ & -                                \\ 
$ -19.50 < M_B - 5 {\rm log} h < -19.25 $ & $ 3.09 \pm 0.42 \times 10^{-3} $ & $ 2.50 \pm 0.21 \times 10^{-3} $ & $ 2.16 \pm 0.17 \times 10^{-3} $ & -                                \\ 
$ -19.75 < M_B - 5 {\rm log} h < -19.50 $ & $ 2.65 \pm 0.36 \times 10^{-3} $ & $ 2.51 \pm 0.24 \times 10^{-3} $ & $ 2.11 \pm 0.15 \times 10^{-3} $ & $ 2.00 \pm 0.12 \times 10^{-3} $ \\ 
$ -20.00 < M_B - 5 {\rm log} h < -19.75 $ & $ 2.44 \pm 0.32 \times 10^{-3} $ & $ 2.37 \pm 0.23 \times 10^{-3} $ & $ 2.11 \pm 0.13 \times 10^{-3} $ & $ 2.12 \pm 0.15 \times 10^{-3} $ \\ 
$ -20.25 < M_B - 5 {\rm log} h < -20.00 $ & $ 2.11 \pm 0.29 \times 10^{-3} $ & $ 1.97 \pm 0.19 \times 10^{-3} $ & $ 1.82 \pm 0.14 \times 10^{-3} $ & $ 2.18 \pm 0.15 \times 10^{-3} $ \\ 
$ -20.50 < M_B - 5 {\rm log} h < -20.25 $ & $ 1.66 \pm 0.24 \times 10^{-3} $ & $ 1.63 \pm 0.19 \times 10^{-3} $ & $ 1.70 \pm 0.13 \times 10^{-3} $ & $ 1.94 \pm 0.16 \times 10^{-3} $ \\ 
$ -20.75 < M_B - 5 {\rm log} h < -20.50 $ & $ 9.94 \pm 1.53 \times 10^{-4} $ & $ 1.17 \pm 0.12 \times 10^{-3} $ & $ 1.22 \pm 0.09 \times 10^{-3} $ & $ 1.51 \pm 0.12 \times 10^{-3} $ \\ 
$ -21.00 < M_B - 5 {\rm log} h < -20.75 $ & $ 7.53 \pm 1.37 \times 10^{-4} $ & $ 8.80 \pm 1.03 \times 10^{-4} $ & $ 9.71 \pm 0.84 \times 10^{-4} $ & $ 1.25 \pm 0.09 \times 10^{-3} $ \\ 
$ -21.25 < M_B - 5 {\rm log} h < -21.00 $ & $ 3.64 \pm 0.78 \times 10^{-4} $ & $ 5.34 \pm 0.67 \times 10^{-4} $ & $ 6.66 \pm 0.65 \times 10^{-4} $ & $ 9.07 \pm 0.77 \times 10^{-4} $ \\ 
$ -21.50 < M_B - 5 {\rm log} h < -21.25 $ & $ 2.01 \pm 0.49 \times 10^{-4} $ & $ 2.93 \pm 0.51 \times 10^{-4} $ & $ 4.63 \pm 0.53 \times 10^{-4} $ & $ 6.19 \pm 0.57 \times 10^{-4} $ \\ 
$ -21.75 < M_B - 5 {\rm log} h < -21.50 $ & $ 6.17 \pm 2.03 \times 10^{-5} $ & $ 1.16 \pm 0.23 \times 10^{-4} $ & $ 2.55 \pm 0.35 \times 10^{-4} $ & $ 3.52 \pm 0.37 \times 10^{-4} $ \\ 
$ -22.00 < M_B - 5 {\rm log} h < -21.75 $ & $ 3.37 \pm 1.51 \times 10^{-5} $ & $ 5.04 \pm 1.39 \times 10^{-5} $ & $ 1.13 \pm 0.20 \times 10^{-4} $ & $ 1.64 \pm 0.18 \times 10^{-4} $ \\ 
$ -22.25 < M_B - 5 {\rm log} h < -22.00 $ & -                                & $ 1.09 \pm 0.45 \times 10^{-5} $ & $ 3.85 \pm 0.90 \times 10^{-5} $ & $ 6.03 \pm 0.94 \times 10^{-5} $ \\ 
$ -22.50 < M_B - 5 {\rm log} h < -22.25 $ & -                                & $ 2.47 \pm 1.68 \times 10^{-6} $ & $ 1.51 \pm 0.63 \times 10^{-5} $ & $ 2.56 \pm 0.65 \times 10^{-5} $ \\ 
$ -22.75 < M_B - 5 {\rm log} h < -22.50 $ & -                                & $ < 1.90 \times 10^{-5}        $ & $ 2.75 \pm 2.12 \times 10^{-6} $ & $ 7.16 \pm 3.35 \times 10^{-6} $ \\ 
$ -23.00 < M_B - 5 {\rm log} h < -22.75 $ & -                                & $ < 2.52 \times 10^{-5}        $ & $ 1.37 \pm 1.58 \times 10^{-6} $ & $ 1.02 \pm 1.03 \times 10^{-6} $ \\ 
\enddata
\end{deluxetable}

\begin{deluxetable}{cccccccc}
\tablecolumns{7}
\tabletypesize{\scriptsize}
\tablecaption{Schechter function fits and the red galaxy luminosity density.\label{table:mlf}}
\tablehead{
\colhead{$z$ range}                                                      &
\colhead{$N_{\rm galaxy}$}                                                 &
\colhead{Volume $(h^{-3} {\rm Mpc}^3)$}                                  &
\colhead{$M^*_B - 5 {\rm log} h$}                                        &
\colhead{$\phi^* (h^3 {\rm Mpc^{-3}})       $}                           &
\colhead{$\alpha$}                                                       &
\colhead{$j_B (10^7 h L_\odot {\rm Mpc^{-3}})$}
}
\startdata
$0.2<z<0.4$ &  6027  & $8.1\times 10^5$ & $ -19.60 \pm 0.05 $    & $ 7.86 \pm 0.71 \times 10^{-3} $   & $ -0.30 \pm  0.05 $   & $8.11\pm 0.85$ \\ 
$0.4<z<0.6$ & 11117  & $1.8\times 10^6$ & $ -19.78 \pm 0.04 $    & $ 6.98 \pm 0.60 \times 10^{-3} $   & $ -0.28 \pm  0.04 $   & $8.30\pm 0.79$ \\ 
$0.6<z<0.8$ & 11154  & $2.9\times 10^6$ & $ -20.22 \pm 0.05 $    & $ 5.16 \pm 0.36 \times 10^{-3} $   & $ -0.52 \pm  0.05 $   & $8.65\pm 0.62$ \\ 
$0.8<z<1.0$ & 12398  & $3.9\times 10^6$ & $ -20.29 \pm 0.05 $    & $ 5.61 \pm 0.39 \times 10^{-3} $   & $ -0.41 \pm  0.06 $   & $9.93\pm 0.95$ \\ 
\enddata
\end{deluxetable}

\begin{deluxetable}{ccccccc}
\tablecolumns{7}
\tabletypesize{\tiny}
\tablecaption{Red galaxy correlation function power-law fit parameters\label{table:r0}}
\tablehead{ 
  \colhead{Subsample Selection}     & 
  \colhead{Photometric}             & 
  \colhead{$N_{\rm galaxy}$}                  &
  \colhead{$\omega(1^\prime)$}           &
  \colhead{$\gamma$}                &
  \colhead{Comoving $r_0$}          &
  \colhead{$\chi^2/d.o.f.$}         
\\ 
  \colhead{Criterion}                          & 
  \colhead{Redshift}                           &
  \colhead{}                                   &
  \colhead{}                                   &
  \colhead{}                                   &
  \colhead{$(h^{-1}{\rm Mpc})$}                &
  \colhead{}                                   
}
\startdata
\\ 
\multicolumn{7}{l}{Narrow luminosity range}\\
\\
\hline
\\
 $-18.5<M_B-5{\rm log}h<-17.5$ & $0.2<z<0.4$    &  1608  & $ 0.60 \pm 0.16 $  & $ 2.04 \pm 0.16 $  & $  5.5 \pm  1.6 $  &  0.76 \\ 
 $-19.0<M_B-5{\rm log}h<-18.0$ & $0.2<z<0.4$    &  1890  & $ 0.73 \pm 0.14 $  & $ 2.02 \pm 0.12 $  & $  6.2 \pm  1.4 $  &  1.06 \\ 
 $-19.5<M_B-5{\rm log}h<-18.5$ & $0.2<z<0.4$    &  2230  & $ 0.54 \pm 0.13 $  & $ 2.04 \pm 0.16 $  & $  5.2 \pm  1.5 $  &  0.59 \\ 
 $-20.0<M_B-5{\rm log}h<-19.0$ & $0.2<z<0.4$    &  2229  & $ 0.46 \pm 0.11 $  & $ 1.89 \pm 0.13 $  & $  5.6 \pm  1.3 $  &  0.99 \\ 
 $-20.5<M_B-5{\rm log}h<-19.5$ & $0.2<z<0.4$    &  1788  & $ 0.57 \pm 0.11 $  & $ 1.97 \pm 0.14 $  & $  5.6 \pm  1.3 $  &  0.37 \\ 
 $-21.0<M_B-5{\rm log}h<-20.0$ & $0.2<z<0.4$    &  1071  & $ 0.57 \pm 0.14 $  & $ 1.98 \pm 0.19 $  & $  5.4 \pm  1.5 $  &  0.70 \\ 
\\
 $-19.0<M_B-5{\rm log}h<-18.0$ & $0.4<z<0.6$    &  3551  & $ 0.46 \pm 0.08 $  & $ 1.93 \pm 0.09 $  & $  6.5 \pm  1.1 $  &  1.22 \\ 
 $-19.5<M_B-5{\rm log}h<-18.5$ & $0.4<z<0.6$    &  4270  & $ 0.48 \pm 0.07 $  & $ 1.98 \pm 0.08 $  & $  6.2 \pm  0.9 $  &  0.85 \\ 
 $-20.0<M_B-5{\rm log}h<-19.0$ & $0.4<z<0.6$    &  4525  & $ 0.47 \pm 0.06 $  & $ 2.00 \pm 0.08 $  & $  5.9 \pm  0.8 $  &  0.66 \\ 
 $-20.5<M_B-5{\rm log}h<-19.5$ & $0.4<z<0.6$    &  3875  & $ 0.41 \pm 0.06 $  & $ 1.97 \pm 0.10 $  & $  5.7 \pm  1.0 $  &  1.58 \\ 
 $-21.0<M_B-5{\rm log}h<-20.0$ & $0.4<z<0.6$    &  2581  & $ 0.44 \pm 0.05 $  & $ 1.95 \pm 0.11 $  & $  6.0 \pm  0.9 $  &  0.67 \\ 
 $-21.5<M_B-5{\rm log}h<-20.5$ & $0.4<z<0.6$    &  1314  & $ 0.64 \pm 0.09 $  & $ 2.02 \pm 0.13 $  & $  6.7 \pm  1.2 $  &  0.96 \\ 
\\
 $-20.0<M_B-5{\rm log}h<-19.0$ & $0.6<z<0.8$    &  5864  & $ 0.24 \pm 0.05 $  & $ 1.91 \pm 0.11 $  & $  5.2 \pm  0.9 $  &  0.24 \\ 
 $-20.5<M_B-5{\rm log}h<-19.5$ & $0.6<z<0.8$    &  5590  & $ 0.25 \pm 0.04 $  & $ 2.08 \pm 0.12 $  & $  4.6 \pm  0.8 $  &  1.04 \\ 
 $-21.0<M_B-5{\rm log}h<-20.0$ & $0.6<z<0.8$    &  4158  & $ 0.28 \pm 0.04 $  & $ 2.11 \pm 0.10 $  & $  4.7 \pm  0.6 $  &  0.55 \\ 
 $-21.5<M_B-5{\rm log}h<-20.5$ & $0.6<z<0.8$    &  2421  & $ 0.42 \pm 0.05 $  & $ 2.17 \pm 0.12 $  & $  5.3 \pm  0.8 $  &  1.09 \\ 
\\
 $-22.0<M_B-5{\rm log}h<-21.0$ & $0.6<z<0.8$    &  1090  & $ 0.54 \pm 0.11 $  & $ 2.06 \pm 0.20 $  & $  6.4 \pm  1.5 $  &  0.40 \\ 
 $-20.5<M_B-5{\rm log}h<-19.5$ & $0.8<z<1.0$    &  7654  & $ 0.19 \pm 0.03 $  & $ 1.93 \pm 0.07 $  & $  4.9 \pm  0.5 $  &  0.36 \\ 
 $-21.0<M_B-5{\rm log}h<-20.0$ & $0.8<z<1.0$    &  6534  & $ 0.21 \pm 0.03 $  & $ 2.00 \pm 0.10 $  & $  4.9 \pm  0.6 $  &  0.84 \\ 
 $-21.5<M_B-5{\rm log}h<-20.5$ & $0.8<z<1.0$    &  4147  & $ 0.23 \pm 0.04 $  & $ 1.94 \pm 0.18 $  & $  5.2 \pm  1.0 $  &  1.14 \\ 
 $-22.0<M_B-5{\rm log}h<-21.0$ & $0.8<z<1.0$    &  1992  & $ 0.43 \pm 0.06 $  & $ 1.84 \pm 0.14 $  & $  7.8 \pm  1.1 $  &  1.40 \\ 
\\
\\ \hline
\\
\multicolumn{7}{l}{Wide luminosity range}\\
\\ \hline
\\
 $M_B-5{\rm log}h<-17.5$ & $0.2<z<0.4$    &  6027  & $ 0.66 \pm 0.11 $  & $ 1.97 \pm 0.10 $  & $  6.2 \pm  1.2 $  &  0.86 \\ 
 $M_B-5{\rm log}h<-18.0$ & $0.2<z<0.4$    &  5275  & $ 0.67 \pm 0.10 $  & $ 1.96 \pm 0.09 $  & $  6.3 \pm  1.1 $  &  0.87 \\ 
 $M_B-5{\rm log}h<-18.5$ & $0.2<z<0.4$    &  4419  & $ 0.63 \pm 0.11 $  & $ 1.98 \pm 0.11 $  & $  5.9 \pm  1.2 $  &  0.84 \\ 
 $M_B-5{\rm log}h<-19.0$ & $0.2<z<0.4$    &  3385  & $ 0.59 \pm 0.11 $  & $ 1.96 \pm 0.12 $  & $  5.8 \pm  1.2 $  &  0.82 \\ 
 $M_B-5{\rm log}h<-19.5$ & $0.2<z<0.4$    &  2189  & $ 0.65 \pm 0.13 $  & $ 2.01 \pm 0.13 $  & $  5.7 \pm  1.3 $  &  0.68 \\ 
 $M_B-5{\rm log}h<-20.0$ & $0.2<z<0.4$    &  1157  & $ 0.68 \pm 0.17 $  & $ 2.03 \pm 0.18 $  & $  5.6 \pm  1.5 $  &  0.87 \\ 
\\
 $M_B-5{\rm log}h<-18.0$ & $0.4<z<0.6$    & 11117  & $ 0.50 \pm 0.05 $  & $ 1.99 \pm 0.06 $  & $  6.2 \pm  0.7 $  &  1.29 \\ 
 $M_B-5{\rm log}h<-18.5$ & $0.4<z<0.6$    &  9541  & $ 0.51 \pm 0.06 $  & $ 2.01 \pm 0.06 $  & $  6.1 \pm  0.7 $  &  1.11 \\ 
 $M_B-5{\rm log}h<-19.0$ & $0.4<z<0.6$    &  7566  & $ 0.51 \pm 0.05 $  & $ 2.04 \pm 0.06 $  & $  5.9 \pm  0.7 $  &  1.11 \\ 
 $M_B-5{\rm log}h<-19.5$ & $0.4<z<0.6$    &  5271  & $ 0.51 \pm 0.06 $  & $ 2.04 \pm 0.08 $  & $  5.9 \pm  0.8 $  &  1.02 \\ 
 $M_B-5{\rm log}h<-20.0$ & $0.4<z<0.6$    &  3041  & $ 0.55 \pm 0.07 $  & $ 2.04 \pm 0.10 $  & $  6.1 \pm  0.9 $  &  0.81 \\ 
 $M_B-5{\rm log}h<-20.5$ & $0.4<z<0.6$    &  1396  & $ 0.68 \pm 0.10 $  & $ 2.16 \pm 0.14 $  & $  5.9 \pm  1.1 $  &  1.10 \\ 
\\
 $M_B-5{\rm log}h<-19.0$ & $0.6<z<0.8$    & 11154  & $ 0.28 \pm 0.04 $  & $ 2.07 \pm 0.09 $  & $  4.9 \pm  0.7 $  &  0.73 \\ 
 $M_B-5{\rm log}h<-19.5$ & $0.6<z<0.8$    &  8321  & $ 0.31 \pm 0.04 $  & $ 2.09 \pm 0.08 $  & $  5.0 \pm  0.6 $  &  0.80 \\ 
 $M_B-5{\rm log}h<-20.0$ & $0.6<z<0.8$    &  5290  & $ 0.33 \pm 0.05 $  & $ 2.17 \pm 0.10 $  & $  4.8 \pm  0.6 $  &  0.45 \\ 
 $M_B-5{\rm log}h<-20.5$ & $0.6<z<0.8$    &  2731  & $ 0.41 \pm 0.05 $  & $ 2.21 \pm 0.13 $  & $  5.1 \pm  0.8 $  &  0.74 \\ 
 $M_B-5{\rm log}h<-21.0$ & $0.6<z<0.8$    &  1132  & $ 0.57 \pm 0.12 $  & $ 2.09 \pm 0.20 $  & $  6.4 \pm  1.5 $  &  0.57 \\ 
\\
 $M_B-5{\rm log}h<-19.5$ & $0.8<z<1.0$    & 12398  & $ 0.24 \pm 0.03 $  & $ 2.03 \pm 0.07 $  & $  5.1 \pm  0.5 $  &  0.81 \\ 
 $M_B-5{\rm log}h<-20.0$ & $0.8<z<1.0$    &  8619  & $ 0.24 \pm 0.03 $  & $ 2.08 \pm 0.08 $  & $  4.9 \pm  0.5 $  &  0.79 \\ 
 $M_B-5{\rm log}h<-20.5$ & $0.8<z<1.0$    &  4744  & $ 0.27 \pm 0.04 $  & $ 2.08 \pm 0.13 $  & $  5.1 \pm  0.8 $  &  1.26 \\ 
 $M_B-5{\rm log}h<-21.0$ & $0.8<z<1.0$    &  2085  & $ 0.44 \pm 0.06 $  & $ 1.87 \pm 0.14 $  & $  7.7 \pm  1.1 $  &  1.27 \\ 
\\
\hline \\
\multicolumn{7}{l}{Space density selected samples}\\
\\ \hline
\\
 $10^{-3.0} h^{3} {\rm Mpc^{-3}}$ & $0.2<z<0.4$    &   770  & $ 0.78 \pm 0.20 $  & $ 2.05 \pm 0.20 $  & $  5.8 \pm  1.6 $  &  0.99 \\ 
 $10^{-3.0} h^{3} {\rm Mpc^{-3}}$ & $0.4<z<0.6$    &  1836  & $ 0.57 \pm 0.09 $  & $ 2.08 \pm 0.13 $  & $  5.9 \pm  1.0 $  &  1.44 \\ 
 $10^{-3.0} h^{3} {\rm Mpc^{-3}}$ & $0.6<z<0.8$    &  2813  & $ 0.42 \pm 0.05 $  & $ 2.14 \pm 0.13 $  & $  5.4 \pm  0.8 $  &  0.88 \\ 
 $10^{-3.0} h^{3} {\rm Mpc^{-3}}$ & $0.8<z<1.0$    &  3802  & $ 0.34 \pm 0.04 $  & $ 2.03 \pm 0.14 $  & $  6.0 \pm  0.9 $  &  1.56 \\ 
\\
$10^{-3.5} h^{3} {\rm Mpc^{-3}}$ & $0.6<z<0.8$    &   955  & $ 0.56 \pm 0.14 $  & $ 2.12 \pm 0.27 $  & $  6.2 \pm  1.9 $  &  0.75 \\ 
$10^{-3.5} h^{3} {\rm Mpc^{-3}}$ & $0.8<z<1.0$    &  1233  & $ 0.48 \pm 0.10 $  & $ 1.98 \pm 0.18 $  & $  7.3 \pm  1.4 $  &  0.71 \\ 
\enddata
\end{deluxetable}

\clearpage

\begin{deluxetable}{cccccccccc}
\rotate
\tablewidth{19cm}
\tablecolumns{10}
\tabletypesize{\tiny}
\setlength{\tabcolsep}{0.05in} 
\tablecaption{HOD models of the luminosity and correlation functions of red galaxies.\label{table:hod1}}
\tablehead{ 
  \colhead{Subsample Selection}      & 
  \colhead{Photometric}              & 
  \colhead{Space Density}            & 
  \colhead{${\rm log}~M_{\rm min}$\tablenotemark{a}}   &
  \colhead{$\sigma_{{\rm log}~M}$}   &
  \colhead{${\rm log}~M_0$}          &
  \colhead{${\rm log}~M_1^\prime$}   &
  \colhead{$\alpha$}                 &
  \colhead{$b_g$}                    &
  \colhead{$\chi^2/d.o.f.$}                    \\
  \colhead{Criterion}                          & 
  \colhead{Redshift}                           &
  \colhead{$(h^{3} Mpc^{-3})$}                 &
  \colhead{}     &
  \colhead{}                      &
  \colhead{}                                   &
  \colhead{}                                   &
  \colhead{}                                   &
  \colhead{}                                   &
  \colhead{}                                   
}
\startdata
$ M_B - 5 {\rm log h} < -17.5 $ & $ 0.2 < z < 0.4 $ & $ 6.98 \pm 0.92 \times 10^{-3}$ & $ 11.96 \pm  0.08 $ & $  0.27 \pm  0.19 $ & $  8.57 \pm  2.76 $ & $ 12.78 \pm  0.13 $ & $  1.01 \pm  0.13 $ & $  1.41 \pm  0.06 $ & $  0.77 $  \\ 
$ M_B - 5 {\rm log h} < -18.0 $ & $ 0.2 < z < 0.4 $ & $ 6.11 \pm 0.80 \times 10^{-3}$ & $ 12.00 \pm  0.08 $ & $  0.30 \pm  0.18 $ & $ 10.60 \pm  2.89 $ & $ 12.84 \pm  0.16 $ & $  0.99 \pm  0.15 $ & $  1.40 \pm  0.06 $ & $  0.85 $  \\ 
$ M_B - 5 {\rm log h} < -18.5 $ & $ 0.2 < z < 0.4 $ & $ 5.14 \pm 0.69 \times 10^{-3}$ & $ 12.05 \pm  0.08 $ & $  0.33 \pm  0.18 $ & $  6.91 \pm  4.39 $ & $ 12.98 \pm  0.14 $ & $  1.00 \pm  0.14 $ & $  1.38 \pm  0.06 $ & $  0.93 $  \\ 
$ M_B - 5 {\rm log h} < -19.0 $ & $ 0.2 < z < 0.4 $ & $ 3.96 \pm 0.53 \times 10^{-3}$ & $ 12.11 \pm  0.08 $ & $  0.31 \pm  0.19 $ & $  9.38 \pm  3.62 $ & $ 13.21 \pm  0.13 $ & $  1.06 \pm  0.19 $ & $  1.38 \pm  0.07 $ & $  1.02 $  \\ 
$ M_B - 5 {\rm log h} < -19.5 $ & $ 0.2 < z < 0.4 $ & $ 2.59 \pm 0.36 \times 10^{-3}$ & $ 12.27 \pm  0.08 $ & $  0.28 \pm  0.19 $ & $ 10.73 \pm  2.43 $ & $ 13.46 \pm  0.18 $ & $  1.03 \pm  0.27 $ & $  1.40 \pm  0.08 $ & $  0.96 $  \\ 
$ M_B - 5 {\rm log h} < -20.0 $ & $ 0.2 < z < 0.4 $ & $ 1.42 \pm 0.20 \times 10^{-3}$ & $ 12.50 \pm  0.09 $ & $  0.30 \pm  0.19 $ & $  9.36 \pm  2.64 $ & $ 13.82 \pm  0.18 $ & $  1.15 \pm  0.33 $ & $  1.47 \pm  0.07 $ & $  1.17 $  \\ 
\\ 
$ M_B - 5 {\rm log h} < -18.0 $ & $ 0.4 < z < 0.6 $ & $ 5.75 \pm 0.53 \times 10^{-3}$ & $ 12.02 \pm  0.07 $ & $  0.35 \pm  0.17 $ & $ 11.58 \pm  0.94 $ & $ 12.81 \pm  0.12 $ & $  1.07 \pm  0.10 $ & $  1.57 \pm  0.03 $ & $  0.56 $  \\ 
$ M_B - 5 {\rm log h} < -18.5 $ & $ 0.4 < z < 0.6 $ & $ 4.92 \pm 0.46 \times 10^{-3}$ & $ 12.06 \pm  0.07 $ & $  0.32 \pm  0.19 $ & $ 11.87 \pm  1.26 $ & $ 12.86 \pm  0.25 $ & $  1.06 \pm  0.19 $ & $  1.57 \pm  0.03 $ & $  0.58 $  \\ 
$ M_B - 5 {\rm log h} < -19.0 $ & $ 0.4 < z < 0.6 $ & $ 3.89 \pm 0.37 \times 10^{-3}$ & $ 12.14 \pm  0.07 $ & $  0.32 \pm  0.18 $ & $ 12.38 \pm  0.63 $ & $ 12.86 \pm  0.28 $ & $  0.94 \pm  0.21 $ & $  1.59 \pm  0.03 $ & $  0.36 $  \\ 
$ M_B - 5 {\rm log h} < -19.5 $ & $ 0.4 < z < 0.6 $ & $ 2.71 \pm 0.27 \times 10^{-3}$ & $ 12.29 \pm  0.07 $ & $  0.37 \pm  0.18 $ & $ 10.59 \pm  3.06 $ & $ 13.23 \pm  0.15 $ & $  1.10 \pm  0.16 $ & $  1.60 \pm  0.04 $ & $  1.10 $  \\ 
$ M_B - 5 {\rm log h} < -20.0 $ & $ 0.4 < z < 0.6 $ & $ 1.56 \pm 0.16 \times 10^{-3}$ & $ 12.50 \pm  0.07 $ & $  0.37 \pm  0.17 $ & $ 10.88 \pm  1.86 $ & $ 13.51 \pm  0.14 $ & $  1.18 \pm  0.22 $ & $  1.68 \pm  0.04 $ & $  0.88 $  \\ 
$ M_B - 5 {\rm log h} < -20.5 $ & $ 0.4 < z < 0.6 $ & $ 7.17 \pm 0.78 \times 10^{-4}$ & $ 12.77 \pm  0.07 $ & $  0.30 \pm  0.19 $ & $  4.97 \pm  3.55 $ & $ 13.88 \pm  0.10 $ & $  1.18 \pm  0.23 $ & $  1.83 \pm  0.06 $ & $  1.61 $  \\ 
\\ 
$ M_B - 5 {\rm log h} < -19.0 $ & $ 0.6 < z < 0.8 $ & $ 3.88 \pm 0.29 \times 10^{-3}$ & $ 12.10 \pm  0.07 $ & $  0.33 \pm  0.19 $ & $  8.74 \pm  5.33 $ & $ 13.06 \pm  0.12 $ & $  1.10 \pm  0.15 $ & $  1.64 \pm  0.03 $ & $  0.58 $  \\ 
$ M_B - 5 {\rm log h} < -19.5 $ & $ 0.6 < z < 0.8 $ & $ 2.87 \pm 0.22 \times 10^{-3}$ & $ 12.23 \pm  0.07 $ & $  0.37 \pm  0.19 $ & $ 12.47 \pm  0.54 $ & $ 12.96 \pm  0.25 $ & $  0.82 \pm  0.24 $ & $  1.69 \pm  0.04 $ & $  1.01 $  \\ 
$ M_B - 5 {\rm log h} < -20.0 $ & $ 0.6 < z < 0.8 $ & $ 1.82 \pm 0.15 \times 10^{-3}$ & $ 12.38 \pm  0.06 $ & $  0.26 \pm  0.18 $ & $  5.79 \pm  3.47 $ & $ 13.40 \pm  0.06 $ & $  1.12 \pm  0.12 $ & $  1.77 \pm  0.04 $ & $  0.51 $  \\ 
$ M_B - 5 {\rm log h} < -20.5 $ & $ 0.6 < z < 0.8 $ & $ 9.38 \pm 0.85 \times 10^{-4}$ & $ 12.64 \pm  0.08 $ & $  0.34 \pm  0.18 $ & $  3.96 \pm  4.84 $ & $ 13.69 \pm  0.08 $ & $  1.16 \pm  0.20 $ & $  1.91 \pm  0.05 $ & $  0.66 $  \\ 
$ M_B - 5 {\rm log h} < -21.0 $ & $ 0.6 < z < 0.8 $ & $ 3.89 \pm 0.37 \times 10^{-4}$ & $ 12.93 \pm  0.08 $ & $  0.28 \pm  0.19 $ & $ 10.01 \pm  2.41 $ & $ 14.07 \pm  0.23 $ & $  1.33 \pm  0.52 $ & $  2.15 \pm  0.08 $ & $  0.82 $  \\ 
\\ 
$ M_B - 5 {\rm log h} < -19.5 $ & $ 0.8 < z < 1.0 $ & $ 3.29 \pm 0.24 \times 10^{-3}$ & $ 12.09 \pm  0.06 $ & $  0.24 \pm  0.19 $ & $ 11.21 \pm  1.29 $ & $ 13.13 \pm  0.13 $ & $  1.22 \pm  0.18 $ & $  1.80 \pm  0.03 $ & $  0.82 $  \\ 
$ M_B - 5 {\rm log h} < -20.0 $ & $ 0.8 < z < 1.0 $ & $ 2.26 \pm 0.17 \times 10^{-3}$ & $ 12.28 \pm  0.07 $ & $  0.39 \pm  0.17 $ & $ 10.84 \pm  1.90 $ & $ 13.28 \pm  0.12 $ & $  1.26 \pm  0.18 $ & $  1.85 \pm  0.04 $ & $  1.09 $  \\ 
$ M_B - 5 {\rm log h} < -20.5 $ & $ 0.8 < z < 1.0 $ & $ 1.22 \pm 0.09 \times 10^{-3}$ & $ 12.47 \pm  0.07 $ & $  0.31 \pm  0.18 $ & $  9.96 \pm  1.49 $ & $ 13.57 \pm  0.07 $ & $  1.29 \pm  0.17 $ & $  1.98 \pm  0.05 $ & $  1.12 $  \\ 
$ M_B - 5 {\rm log h} < -21.0 $ & $ 0.8 < z < 1.0 $ & $ 5.34 \pm 0.47 \times 10^{-4}$ & $ 12.77 \pm  0.07 $ & $  0.28 \pm  0.19 $ & $  8.37 \pm  3.29 $ & $ 13.89 \pm  0.12 $ & $  1.29 \pm  0.32 $ & $  2.22 \pm  0.07 $ & $  1.28 $  \\ 
\\ 
$ 10^{-3.0} h^3 {\rm Mpc}^{-3} $ & $ 0.2 < z < 0.4 $ & $ 1.00 \pm 0.14 \times 10^{-3}$ & $ 12.65 \pm  0.09 $ & $  0.31 \pm  0.19 $ & $  4.83 \pm  3.88 $ & $ 14.05 \pm  0.25 $ & $  1.22 \pm  0.56 $ & $  1.53 \pm  0.07 $ & $  1.31 $  \\ 
$ 10^{-3.0} h^3 {\rm Mpc}^{-3} $ & $ 0.4 < z < 0.6 $ & $ 1.00 \pm 0.12 \times 10^{-3}$ & $ 12.64 \pm  0.08 $ & $  0.31 \pm  0.19 $ & $ 12.36 \pm  4.40 $ & $ 13.67 \pm  0.29 $ & $  1.00 \pm  0.36 $ & $  1.75 \pm  0.06 $ & $  1.86 $  \\ 
$ 10^{-3.0} h^3 {\rm Mpc}^{-3} $ & $ 0.6 < z < 0.8 $ & $ 1.00 \pm 0.09 \times 10^{-3}$ & $ 12.59 \pm  0.07 $ & $  0.25 \pm  0.19 $ & $ 10.41 \pm  1.86 $ & $ 13.68 \pm  0.11 $ & $  1.10 \pm  0.25 $ & $  1.90 \pm  0.05 $ & $  0.91 $  \\ 
$ 10^{-3.0} h^3 {\rm Mpc}^{-3} $ & $ 0.8 < z < 1.0 $ & $ 1.00 \pm 0.08 \times 10^{-3}$ & $ 12.55 \pm  0.07 $ & $  0.29 \pm  0.19 $ & $ 10.78 \pm  2.25 $ & $ 13.61 \pm  0.10 $ & $  1.19 \pm  0.27 $ & $  2.04 \pm  0.06 $ & $  1.17 $  \\ 
\\ 
$ 10^{-3.5} h^3 {\rm Mpc}^{-3} $ & $ 0.6 < z < 0.8 $ & $ 3.16 \pm 0.30 \times 10^{-4}$ & $ 12.94 \pm  0.07 $ & $  0.12 \pm  0.16 $ & $ 10.15 \pm  0.80 $ & $ 14.00 \pm  0.15 $ & $  1.89 \pm  0.35 $ & $  2.33 \pm  0.13 $ & $  0.87 $  \\ 
$ 10^{-3.5} h^3 {\rm Mpc}^{-3} $ & $ 0.8 < z < 1.0 $ & $ 3.16 \pm 0.28 \times 10^{-4}$ & $ 12.93 \pm  0.07 $ & $  0.27 \pm  0.18 $ & $  9.06 \pm  3.58 $ & $ 14.20 \pm  0.43 $ & $  1.16 \pm  0.47 $ & $  2.38 \pm  0.08 $ & $  0.63 $  \\ 
\enddata
\tablenotetext{a}{All masses are in units of $h^{-1}~M_\odot$.}
\end{deluxetable}

\clearpage

\begin{deluxetable}{cccccccccc}
\rotate
\tablewidth{19cm}
\tablecolumns{10}
\tabletypesize{\tiny}
\setlength{\tabcolsep}{0.05in} 
\tablecaption{HOD models of the luminosity and correlation functions of red galaxies with $M_{\rm min}$ and $M_1^\prime$ as the only free parameters.\label{table:hod2}}
\tablehead{ 
  \colhead{Subsample Selection}      & 
  \colhead{Photometric}              & 
  \colhead{Space Density}            & 
  \colhead{${\rm log}~M_{\rm min}$\tablenotemark{a}}      &
  \colhead{$\sigma_{{\rm log}~M}$}   &
  \colhead{${\rm log}~M_0$}          &
  \colhead{${\rm log}~M_1^\prime$}   &
  \colhead{$\alpha$}                 &
  \colhead{$b_g$}                    &
  \colhead{$\chi^2/d.o.f.$}                    \\
  \colhead{Criterion}                          & 
  \colhead{Redshift}                           &
  \colhead{$(h^{3} Mpc^{-3})$}                 &
  \colhead{}                                   &
  \colhead{}                                   &
  \colhead{}                                   &
  \colhead{}                                   &
  \colhead{}                                   &
  \colhead{}                                   &
  \colhead{}                                   
}
\startdata
$ M_B - 5 {\rm log h} < -17.5 $ & $ 0.2 < z < 0.4 $ & $ 6.98 \pm 0.92 \times 10^{-3}$ & $ 11.94 \pm  0.05 $ &   0.30 & $ 11.94 \pm  0.05 $ & $ 12.73 \pm  0.08 $ &   1.00 & $  1.44 \pm  0.04 $ & $  0.55 $  \\ 
$ M_B - 5 {\rm log h} < -18.0 $ & $ 0.2 < z < 0.4 $ & $ 6.11 \pm 0.80 \times 10^{-3}$ & $ 11.98 \pm  0.05 $ &   0.30 & $ 11.98 \pm  0.05 $ & $ 12.81 \pm  0.08 $ &   1.00 & $  1.43 \pm  0.04 $ & $  0.61 $  \\ 
$ M_B - 5 {\rm log h} < -18.5 $ & $ 0.2 < z < 0.4 $ & $ 5.14 \pm 0.69 \times 10^{-3}$ & $ 12.03 \pm  0.05 $ &   0.30 & $ 12.03 \pm  0.05 $ & $ 12.93 \pm  0.09 $ &   1.00 & $  1.42 \pm  0.04 $ & $  0.66 $  \\ 
$ M_B - 5 {\rm log h} < -19.0 $ & $ 0.2 < z < 0.4 $ & $ 3.96 \pm 0.53 \times 10^{-3}$ & $ 12.11 \pm  0.05 $ &   0.30 & $ 12.11 \pm  0.05 $ & $ 13.10 \pm  0.10 $ &   1.00 & $  1.40 \pm  0.04 $ & $  0.72 $  \\ 
$ M_B - 5 {\rm log h} < -19.5 $ & $ 0.2 < z < 0.4 $ & $ 2.59 \pm 0.36 \times 10^{-3}$ & $ 12.27 \pm  0.06 $ &   0.30 & $ 12.27 \pm  0.06 $ & $ 13.33 \pm  0.11 $ &   1.00 & $  1.43 \pm  0.04 $ & $  0.75 $  \\ 
$ M_B - 5 {\rm log h} < -20.0 $ & $ 0.2 < z < 0.4 $ & $ 1.42 \pm 0.20 \times 10^{-3}$ & $ 12.50 \pm  0.06 $ &   0.30 & $ 12.50 \pm  0.06 $ & $ 13.72 \pm  0.17 $ &   1.00 & $  1.48 \pm  0.05 $ & $  0.92 $  \\ 
\\ 
$ M_B - 5 {\rm log h} < -18.0 $ & $ 0.4 < z < 0.6 $ & $ 5.75 \pm 0.53 \times 10^{-3}$ & $ 12.00 \pm  0.04 $ &   0.30 & $ 12.00 \pm  0.04 $ & $ 12.74 \pm  0.05 $ &   1.00 & $  1.57 \pm  0.02 $ & $  0.47 $  \\ 
$ M_B - 5 {\rm log h} < -18.5 $ & $ 0.4 < z < 0.6 $ & $ 4.92 \pm 0.46 \times 10^{-3}$ & $ 12.05 \pm  0.03 $ &   0.30 & $ 12.05 \pm  0.03 $ & $ 12.82 \pm  0.05 $ &   1.00 & $  1.58 \pm  0.02 $ & $  0.48 $  \\ 
$ M_B - 5 {\rm log h} < -19.0 $ & $ 0.4 < z < 0.6 $ & $ 3.89 \pm 0.37 \times 10^{-3}$ & $ 12.14 \pm  0.04 $ &   0.30 & $ 12.14 \pm  0.04 $ & $ 12.93 \pm  0.06 $ &   1.00 & $  1.60 \pm  0.02 $ & $  0.32 $  \\ 
$ M_B - 5 {\rm log h} < -19.5 $ & $ 0.4 < z < 0.6 $ & $ 2.71 \pm 0.27 \times 10^{-3}$ & $ 12.26 \pm  0.04 $ &   0.30 & $ 12.26 \pm  0.04 $ & $ 13.14 \pm  0.07 $ &   1.00 & $  1.61 \pm  0.03 $ & $  0.80 $  \\ 
$ M_B - 5 {\rm log h} < -20.0 $ & $ 0.4 < z < 0.6 $ & $ 1.56 \pm 0.16 \times 10^{-3}$ & $ 12.47 \pm  0.04 $ &   0.30 & $ 12.47 \pm  0.04 $ & $ 13.41 \pm  0.08 $ &   1.00 & $  1.69 \pm  0.03 $ & $  0.68 $  \\ 
$ M_B - 5 {\rm log h} < -20.5 $ & $ 0.4 < z < 0.6 $ & $ 7.17 \pm 0.78 \times 10^{-4}$ & $ 12.76 \pm  0.04 $ &   0.30 & $ 12.76 \pm  0.04 $ & $ 13.75 \pm  0.09 $ &   1.00 & $  1.85 \pm  0.04 $ & $  1.19 $  \\ 
\\ 
$ M_B - 5 {\rm log h} < -19.0 $ & $ 0.6 < z < 0.8 $ & $ 3.88 \pm 0.29 \times 10^{-3}$ & $ 12.09 \pm  0.03 $ &   0.30 & $ 12.09 \pm  0.03 $ & $ 12.96 \pm  0.05 $ &   1.00 & $  1.66 \pm  0.02 $ & $  0.47 $  \\ 
$ M_B - 5 {\rm log h} < -19.5 $ & $ 0.6 < z < 0.8 $ & $ 2.87 \pm 0.22 \times 10^{-3}$ & $ 12.21 \pm  0.03 $ &   0.30 & $ 12.21 \pm  0.03 $ & $ 13.09 \pm  0.05 $ &   1.00 & $  1.71 \pm  0.02 $ & $  0.72 $  \\ 
$ M_B - 5 {\rm log h} < -20.0 $ & $ 0.6 < z < 0.8 $ & $ 1.82 \pm 0.15 \times 10^{-3}$ & $ 12.38 \pm  0.03 $ &   0.30 & $ 12.38 \pm  0.03 $ & $ 13.29 \pm  0.06 $ &   1.00 & $  1.79 \pm  0.02 $ & $  0.41 $  \\ 
$ M_B - 5 {\rm log h} < -20.5 $ & $ 0.6 < z < 0.8 $ & $ 9.38 \pm 0.85 \times 10^{-4}$ & $ 12.62 \pm  0.03 $ &   0.30 & $ 12.62 \pm  0.03 $ & $ 13.57 \pm  0.07 $ &   1.00 & $  1.93 \pm  0.03 $ & $  0.52 $  \\ 
$ M_B - 5 {\rm log h} < -21.0 $ & $ 0.6 < z < 0.8 $ & $ 3.89 \pm 0.37 \times 10^{-4}$ & $ 12.93 \pm  0.03 $ &   0.30 & $ 12.93 \pm  0.03 $ & $ 13.97 \pm  0.11 $ &   1.00 & $  2.17 \pm  0.04 $ & $  0.67 $  \\ 
\\ 
$ M_B - 5 {\rm log h} < -19.5 $ & $ 0.8 < z < 1.0 $ & $ 3.29 \pm 0.24 \times 10^{-3}$ & $ 12.11 \pm  0.03 $ &   0.30 & $ 12.11 \pm  0.03 $ & $ 13.03 \pm  0.05 $ &   1.00 & $  1.78 \pm  0.02 $ & $  1.16 $  \\ 
$ M_B - 5 {\rm log h} < -20.0 $ & $ 0.8 < z < 1.0 $ & $ 2.26 \pm 0.17 \times 10^{-3}$ & $ 12.25 \pm  0.03 $ &   0.30 & $ 12.25 \pm  0.03 $ & $ 13.19 \pm  0.06 $ &   1.00 & $  1.86 \pm  0.03 $ & $  1.26 $  \\ 
$ M_B - 5 {\rm log h} < -20.5 $ & $ 0.8 < z < 1.0 $ & $ 1.22 \pm 0.09 \times 10^{-3}$ & $ 12.47 \pm  0.03 $ &   0.30 & $ 12.47 \pm  0.03 $ & $ 13.47 \pm  0.07 $ &   1.00 & $  1.99 \pm  0.03 $ & $  0.86 $  \\ 
$ M_B - 5 {\rm log h} < -21.0 $ & $ 0.8 < z < 1.0 $ & $ 5.34 \pm 0.47 \times 10^{-4}$ & $ 12.77 \pm  0.03 $ &   0.30 & $ 12.77 \pm  0.03 $ & $ 13.80 \pm  0.10 $ &   1.00 & $  2.23 \pm  0.04 $ & $  0.95 $  \\ 
\\ 
$ 10^{-3.0} h^3 {\rm Mpc}^{-3} $ & $ 0.2 < z < 0.4 $ & $ 1.00 \pm 0.14 \times 10^{-3}$ & $ 12.64 \pm  0.06 $ &   0.30 & $ 12.64 \pm  0.06 $ & $ 13.85 \pm  0.18 $ &   1.00 & $  1.56 \pm  0.06 $ & $  0.94 $  \\ 
$ 10^{-3.0} h^3 {\rm Mpc}^{-3} $ & $ 0.4 < z < 0.6 $ & $ 1.00 \pm 0.12 \times 10^{-3}$ & $ 12.64 \pm  0.04 $ &   0.30 & $ 12.64 \pm  0.04 $ & $ 13.63 \pm  0.10 $ &   1.00 & $  1.76 \pm  0.04 $ & $  1.37 $  \\ 
$ 10^{-3.0} h^3 {\rm Mpc}^{-3} $ & $ 0.6 < z < 0.8 $ & $ 1.00 \pm 0.09 \times 10^{-3}$ & $ 12.60 \pm  0.03 $ &   0.30 & $ 12.60 \pm  0.03 $ & $ 13.55 \pm  0.07 $ &   1.00 & $  1.92 \pm  0.03 $ & $  0.67 $  \\ 
$ 10^{-3.0} h^3 {\rm Mpc}^{-3} $ & $ 0.8 < z < 1.0 $ & $ 1.00 \pm 0.08 \times 10^{-3}$ & $ 12.55 \pm  0.03 $ &   0.30 & $ 12.55 \pm  0.03 $ & $ 13.54 \pm  0.08 $ &   1.00 & $  2.05 \pm  0.03 $ & $  0.91 $  \\ 
\\ 
$ 10^{-3.5} h^3 {\rm Mpc}^{-3} $ & $ 0.6 < z < 0.8 $ & $ 3.16 \pm 0.30 \times 10^{-4}$ & $ 13.00 \pm  0.04 $ &   0.30 & $ 13.00 \pm  0.04 $ & $ 14.11 \pm  0.13 $ &   1.00 & $  2.22 \pm  0.04 $ & $  0.66 $  \\ 
$ 10^{-3.5} h^3 {\rm Mpc}^{-3} $ & $ 0.8 < z < 1.0 $ & $ 3.16 \pm 0.28 \times 10^{-4}$ & $ 12.94 \pm  0.03 $ &   0.30 & $ 12.94 \pm  0.03 $ & $ 14.04 \pm  0.13 $ &   1.00 & $  2.40 \pm  0.05 $ & $  0.44 $  \\ 
\enddata
\tablenotetext{a}{All masses are in units of $h^{-1}~M_\odot$.}
\end{deluxetable}

\end{document}